\documentclass[a4paper,11pt]{article}
\usepackage[utf8]{inputenc}
\usepackage{lmodern}
\usepackage[T1]{fontenc}
\usepackage{amsmath,amsthm,amsfonts,amssymb,bm}
\usepackage{dsfont}			
\usepackage{enumitem}
\usepackage{graphicx}
\usepackage{booktabs}
\usepackage{xcolor}
\usepackage[authoryear]{natbib}
\usepackage[right=2.5cm,left=2.5cm,top=2cm,bottom=3cm]{geometry}
\usepackage[font={small,it}]{caption}	
\usepackage[colorlinks,linkcolor=blue,citecolor=blue,urlcolor=blue]{hyperref}
\usepackage{subfig}
\usepackage{mathtools}
\DeclarePairedDelimiter\floor{\lfloor}{\rfloor}
\usepackage{multirow}
\usepackage{makecell}
\usepackage{afterpage}

\newcommand{\independent}{\mathrel{\perp\mspace{-10mu}\perp}}

\newcommand{\x}{\bm{x}}
\newcommand{\MU}{\bm{\mu}}
\newcommand{\SIGMA}{\bm{\Sigma}}
\newcommand{\THETA}{\bm{\Theta}}
\newcommand{\sumi}{\sum_{i=1}^N}
\newcommand{\sumk}{\sum_{k=1}^K}
\newcommand{\GG}{\mathcal{G}}
\newcommand{\A}{\bm{A}}
\newcommand{\vbar}{\,\lvert\,}
\DeclareMathOperator*{\argmax}{arg\,max}

\title{ \textsc{Model-based Clustering with\\ Sparse Covariance Matrices} }
\author{Michael Fop\thanks{School of Mathematics \& Statistics and Insight Research Centre, University College Dublin, Belfield, Dublin 4, Ireland. This work was supported by the Science Foundation Ireland funded Insight Research Centre (SFI/12/RC/2289)} \and Thomas Brendan Murphy\footnotemark[1] \and Luca Scrucca\thanks{Department of Economics, Universit\`a degli Studi di Perugia, Via A. Pascoli 20, 06123 Perugia, Italy}}
\date{}

\begin{document}
\maketitle

\begin{abstract}
Finite Gaussian mixture models are widely used for model-based clustering of continuous data. 
Nevertheless, since the number of model parameters scales quadratically with the number of variables, these models can be easily over-parameterized.
For this reason, parsimonious models have been developed via covariance matrix decompositions or assuming local independence. However, these remedies do not allow for direct estimation of sparse covariance matrices nor do they take into account that the structure of association among the variables can vary from one cluster to the other. To this end, we introduce mixtures of Gaussian covariance graph models for model-based clustering with sparse covariance matrices. A penalized likelihood approach is employed for estimation and a general penalty term on the graph configurations can be used to induce different levels of sparsity and incorporate prior knowledge. 
Model estimation is carried out using a structural-EM algorithm for parameters and graph structure estimation, where two alternative strategies based on a genetic algorithm and an efficient stepwise search are proposed for inference.
With this approach, sparse component covariance matrices are directly obtained. The framework results in a parsimonious model-based clustering of the data via a flexible model for the within-group joint distribution of the variables. Extensive simulated data experiments and application to illustrative datasets show that the method attains good classification performance and model quality. 
\end{abstract}


\section{Introduction}
\label{intro}
Model-based clustering \citep{fraley:raftery:2002,mcnicholas:2016} is a popular and well established framework for clustering multivariate data. In this approach, the data generating process is represented as a finite mixture of probability distributions where each component distribution corresponds to a group. When the observations are measured as continuous variables, it is common to model each component density using a multivariate Gaussian distribution. Hence, the component covariance matrices encode the within-group association structure among the observed variables. In several situations, this association structure may vary between the groups and two (or more) variables correlated within one cluster may be independent in another. In such cases, assuming a model where the variables are all independent and the component covariance matrices are diagonal would be too restrictive. On the other hand, not placing any constraint on the covariance terms would introduce unnecessary parameters when some of the variables have weak or almost null correlation \citep{dempster:1972}. Therefore, sparse covariance matrices can be used to characterize the component densities in order to better model and define a parsimonious representation of the within-group association structure.

Graphical models \citep{whittaker:1990,koller:2009} are widely used to model the relations among a collection of random variables. When the joint distribution of these variables is multivariate Gaussian, a subclass of such models, the Gaussian covariance graph model, defines a correspondence between the graph and the pattern of correlation embedded in the covariance matrix \citep{chaudhuri:etal:2007,richardson:spirtes:2002}. The graph depicts the association structure of the variables and two or more variables are independent if there is no edge joining them. Thus, the marginal independence statements of the graph coincide to zero covariance terms in the covariance matrix.

In this work we develop a framework for model-based clustering with sparse covariance matrices. This framework is built upon the combination of Gaussian mixture models and Gaussian covariance graph models. The component densities are then characterized by graphs representing the structure of association of each cluster and by covariance matrices with zero patterns concomitant to the missing edges of the graphs. The approach results in a parsimonious model-based clustering of the data via a flexible model for the within-group joint distribution of the observed variables.

The article is structured as follows. Section~\ref{mbc} briefly recalls the model-based clustering framework via finite mixture of Gaussian distributions. Section~\ref{covgraph} describes the Gaussian covariance models for modeling the marginal dependences among a collection of random variables. Section~\ref{mixcovgraph} introduces the mixture of Gaussian covariance graph models employed for model-based clustering with sparse covariance matrices. In particular, Section~\ref{spec} focuses on model specification and Section~\ref{estim} on its estimation by means of a penalized log-likelihood. Section~\ref{reg} presents a simple Bayesian regularization approach for avoiding degeneracies of the likelihood. We present and discuss different penalty functions for graph estimation in Section~\ref{pen}. These functions place a direct penalty on the graph structure, hence the problem of structure estimation corresponds to a combinatorial optimization task. 
Section~\ref{ga} describes two alternative strategies for graph structure search and sparse covariance estimation based respectively on genetic algorithm and stepwise search.
In Section~\ref{sim} we assess the proposed method on simulated data experiments and in Section~\ref{data} it is applied to two illustrative data examples. The paper ends with a discussion in Section~\ref{disc}.

\section{Model-based clustering}
\label{mbc}
Let $\mathbf{X}$ be the $N\times V$ data matrix, in which each observation $\x_i$ is a realization of a $V$-dimensional vector of random variables $( X_1, \dots, X_j, \dots, X_V )$. Model-based clustering assumes that the data arise from 
a finite mixture of $K$ distributions, corresponding to the groups. For continuous data, a popular approach is to model each component density by a multivariate Gaussian distribution. Therefore, the density of each data point is given by:
$$
f\left( \x_i \vbar \THETA \right) = \sumk \tau_k\, \phi \left( \x_i \vbar \MU_k, \SIGMA_k \right),
$$
where $\tau_k\,$ are the mixing proportions such that $\sumk \tau_k\, = 1$ and $\tau_k\, > 0$, and $\phi(\cdot)$ is the multivariate Gaussian density with mean vector $\MU_k$ and covariance matrix $\SIGMA_k$, and $\THETA=\left(\tau_1, \dots,\tau_{K-1}, \MU_1, \dots, \MU_K, \SIGMA_1, \dots, \SIGMA_K \right)$ is the vector of model parameters. In this model, the component densities characterize the groups and each observation belongs to the corresponding cluster according to a latent group membership indicator variable $Z_{ik}$, such that $Z_{ik} = 1$ if $\x_i$ arises from the $k$th subpopulation, 0 otherwise. For a fixed number of components $K$, the model is usually estimated using the EM algorithm \citep{dempster:etal:1977}. See \cite{mclachlan:peel:2000} and \cite{fraley:raftery:2002} for further details, and \cite{mcnicholas:2016} for a recent review.

In such setting, the \emph{curse of dimensionality} \citep{bellman:1957} takes the form of a dramatic over-parametrization of the model. Indeed, the number of parameters is of order $\mathcal{O}(KV^2)$ and is mainly led by the number of covariance terms in the matrices $\SIGMA_k$ \citep{bouveyron:2014}. In the literature, different methods and alternative parameterizations of the component densities have been proposed in order to overcome this issue and attain parsimony. For example, \cite{banfield:raftery:1993} and \cite{celeux:govaert:1995} propose a parsimonious parametrization of the covariance matrices based on an eigenvalue decomposition which allows the control of the volume, shape and orientation of the Gaussian ellipsoids; \cite{mcnicholas:murphy:2008} present a factorization of the covariance matrix based on a factor analysis model where parsimony is attained by constraining the loading and noise matrices. \cite{bouveyron:2012} propose a framework for model-based clustering in a low-dimensional subspace of the data; \cite{biernacki:lourme:2014} suggest a decomposition of the covariance based on conditional variance and conditional correlation matrices, different parsimonious models are defined by placing constraints on such matrices. Several other approaches have been presented, and for a review we suggest the excellent survey of \cite{bouveyron:2014}.


Most of the frameworks developed in the literature rely on some sort of matrix decomposition. In fact, they often focus on the geometric properties of the mixture components, rather than the dependence structure between the variables conveyed in the covariance matrices. However, parsimony can also be obtained by direct modelling of such association structure via estimation of sparse covariance matrices, where some covariance terms are set to zero. In this way, parsimonious models can be defined by taking into account the fact that two (or more) variables correlated within a cluster may be independent in another one. Hence, the corresponding covariance parameter should be enforced to zero in order to avoid the estimation of unneeded parameters. Furthermore, this would also enable the definition of a general model where the association structure may vary across the mixture components: capturing this feature with the model can ease the interpretation of the clustering result and can lead to a better representation of the data generating process. 
In the next section we will introduce a tool that allows to estimate sparse covariance matrices and model the relations among variables.

\section{Gaussian covariance graph models}
\label{covgraph}
A graph $\GG$ is a mathematical object denoted as the pair $\GG=(\mathcal{V}, \mathcal{E})$, where $\mathcal{V}$ is the set of vertices (or nodes) and $\mathcal{E}$ the set of edges (or arcs). We denote with $V$ and $E$ the cardinalities of these sets respectively. In the graph, two vertices $j$ and $h$ are adjacent if there is an edge joining them. Edges can be directed, undirected or bi-directed, carrying different interpretations; here we focus on the case of graphs with only bi-directed edges. Such type of graph is denoted as \emph{covariance graph} and can represent the pattern of zeros in a sparse covariance matrix, and consequently the embedded association structure \citep{chaudhuri:etal:2007}.

\begin{figure}[tb]
 \centering
 \includegraphics[scale=0.9]{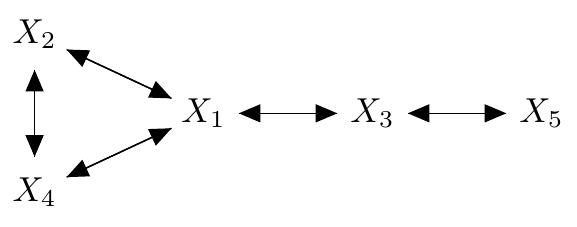}
 \caption{\label{fig_graph_example} The covariance graph corresponding to the covariance matrix presented in \ref{eq:1}.}
\end{figure}

Let us consider a bi-directed graph $\GG$ whose node set $\mathcal{V}$ of dimension $V$ represents a collection of random variables $( X_1,\dots, X_j,\dots,$ $X_V )$ distributed according to a multivariate Gaussian distribution. In this framework there is a one to one correspondence between the graph and the joint distribution of the random variables \citep{koller:2009}. A \emph{Gaussian covariance graph model} is the family of multivariate Gaussian distributions in which the restrictions on the graph hold in the covariance matrix. Thus, a missing edge in the graph between any two nodes is equivalent to the corresponding variables being marginally independent and the following properties hold \citep{edwards:2000}:
$$
(j,h) \notin \mathcal{E} \quad \Leftrightarrow \quad X_j \independent X_h \quad \Leftrightarrow \quad \sigma_{jh} = 0.
$$
For example, the graph in Figure~\ref{fig_graph_example} corresponds to the covariance matrix:
\begin{equation}
\label{eq:1}
  \SIGMA = \begin{bmatrix}
	          \sigma_{1}  & \sigma_{12} & \sigma_{13} & \sigma_{14} &     0        \\
	          \sigma_{12} & \sigma_{2}  &      0       & \sigma_{24} &    0         \\
	          \sigma_{13} &      0       & \sigma_{3}  &      0       & \sigma_{35} \\
	          \sigma_{14} & \sigma_{24} &       0      & \sigma_{4}  &       0      \\
	               0       &       0      & \sigma_{35} &      0       & \sigma_{5}  \\
		\end{bmatrix}.
\end{equation}
Formally, we define a Gaussian covariance graph model as the collection of multivariate Normal distributions:
$$
\lbrace\, \mathcal{N}(\MU, \SIGMA) ~ \colon \SIGMA \in \mathcal{C}^+( \GG) \,\rbrace,
$$
where $\mathcal{C}^+(\GG)$ denotes the cone of $V\times V$ positive definite matrices induced by the graph $\GG$. Note that the model framework is different from the standard Gaussian graphical model where the Normal distribution is parameterized in terms of the precision matrix $\bm{\Omega}=\SIGMA^{-1}$ and an undirected graph is used to represent the relations. When the precision matrix is considered, the graph poses a set of pairwise conditional independences and sparsity in $\SIGMA$ may be obtained only as a by-product of inverting $\bm{\Omega}$, but it is not guaranteed \citep{whittaker:1990,pourahmadi:2011}. Instead, with a covariance graph model, a sparse $\SIGMA$ is obtained directly, since the graph places a sets of marginal independence restrictions on the corresponding pairs of variables. 

Estimation of a covariance graph model refers to two tasks: structure learning, corresponding to the task of inferring a graph structure from the data, and parameter estimation, concerning the estimation of the covariance matrix terms according to the pairwise restrictions of the graph and the constraint of the matrix being positive definite. The aim is closely related to the estimation of a sparse covariance matrix for a vector of random variables, a problem that has been tackled in a plethora of ways in the literature. For example, by using maximum likelihood methods \citep{kauermann:1996,wermuth:etal:2006,chaudhuri:etal:2007}, by using penalized likelihood methods and regularization techniques \citep{huang:etal:2006,zhou:etal:2011,bien:tibshirani:2011,rothman:2012}, or by exploiting a Bayesian framework with shrinkage priors \citep{wang:2015}. 

In particular, in this paper we focus on the work of \cite{chaudhuri:etal:2007}. The authors propose a maximum likelihood method for estimating a positive definite covariance matrix with zero entries given by a fixed graph structure. The method relies on the \emph{Iterative Conditional Fitting} (ICF) algorithm. The procedure estimates the joint distribution of the variables by fixing the marginal distribution of a subset of variables and finding the conditional distribution of a variable given the rest under the constraints induced by the graph. Then the joint distribution is updated by multiplying the two distributions. The method is fast and easy to implement and the covariance matrix obtained is ensured to be positive definite. Appendix~\ref{appendix:a} contains a more detailed description of the algorithm within the context of this work.

A vast amount of literature exists on graphical models, and we conclude this section suggesting some general references on the topic: \cite{whittaker:1990}, \cite{edwards:2000} and \cite{koller:2009} for an in-depth discussion on the subject and \cite{drton:2017} for a recent review on structure learning. Compared to undirected and directed graphs, bi-directed graphs are usually employed for graphical modeling of the marginal dependences of a set of random variables; \cite{richardson:spirtes:2002} contains a review on different graph types and their properties.

\section{Mixtures of Gaussian covariance graph models}
\label{mixcovgraph}
Gaussian covariance graph models determine a framework for estimating multivariate Normal distributions with sparse covariance matrices and for modeling the relations among a set of variables. In this section, we incorporate this framework into model-based clustering to obtain a clustering of the data with sparse covariance matrices and groups with different association patterns.

\subsection{Model specification}
\label{spec}
In a mixture of Gaussian covariance graph models we assume that the density of each data point is defined as follows:
\begin{equation}
\begin{split}
\label{eq:2}
 f\left( \x_i \vbar \THETA, \mathbb{G} \right) &= \sumk \tau_k\, \phi \left( \x_i \vbar \MU_k, \SIGMA_k, \GG_k \right)\\ 
                                               &\quad \text{with}\quad \SIGMA_k \in \mathcal{C}^+( \GG_k),
\end{split}
\end{equation}
where $\mathbb{G} = \lbrace \GG_1, \dots, \GG_k, \dots, \GG_K \rbrace$ is the collection of graphs of mixture components and $\mathcal{C}^+\left(\GG_k \right)$ denotes the cone of positive definite matrices induced by the graph $\GG_k$. Within each component, a graph $\GG_k = (\mathcal{V}, \mathcal{E}_k)$ poses a collection of marginal independence restrictions on the joint distribution of the variables. This results in the corresponding component covariance matrix being sparse with the related covariance terms set to zero. In addition, clusters with differing dependence patterns are described by different sets of edges $\mathcal{E}_k$. Therefore, the model takes into account that groups can be characterized by dissimilar association structures and allows the performing of model-based clustering with sparse covariance matrices.

\subsection{Model estimation}
\label{estim}
For a fixed number of components $K$, model estimation concerns the estimation of mixture parameters $\THETA$ and the selection of graph structures $\mathbb{G}$. To accomplish the task we introduce a structural EM algorithm (S-EM) \citep{friedman:1997,friedman:1998}. The algorithm allows the estimation of model parameters and inferring graph configurations in presence of incomplete data, combining the standard EM algorithm \citep{dempster:etal:1977} and the penalized EM algorithm \citep{green:1990} with a graph structure search. The S-EM algorithm maximizes a penalized version of the log-likelihood, where the penalization term is some function of the graph structure. The penalty term allows the definition of a scoring rule to be used for searching the best graph at each step of the algorithm. The general outline is similar to the conventional EM algorithm, with the relevant exception that we optimize not only parameters, but also graph edge sets.

The set of arcs $\mathcal{E}_k$ defines the structure of graph $\GG_k$. Let us represent it by introducing the symmetric adjacency matrix $\A_k$ such that an entry $a_{jhk}$ is equal to zero if $(j,h) \notin \mathcal{E}_k$, 1 if $(j,h) \in \mathcal{E}_k$; in addition, $\text{diag}\lbrace \A_k \rbrace = \mathbf{0}$. Let us also denote with $\mathbb{A}$ the collection of adjacency matrices representing $\mathbb{G}$. For the model in \eqref{eq:2} we consider the following penalized log-likelihood:
\begin{equation}
\label{eq:3}
 \ell = \sumi \log \left\lbrace \sumk \tau_k\, \phi \left( \x_i \vbar \MU_k, \SIGMA_k, \GG_k \right) \right\rbrace - \sumk Q(\A_k),
\end{equation}
where $Q(\cdot)$ is a function that penalizes the graph complexity. Different choices of $Q(\cdot)$ will be discussed in Section~\ref{pen}. Equation \eqref{eq:3} leads to the following penalized complete log-likelihood:
$$
 \ell_C = \sumi \sum_{k=1}^K z_{ik} \log \left\lbrace \tau_k\, \phi \left( \x_i \vbar \MU_k, \SIGMA_k, \GG_k \right) \right\rbrace - \sumk Q(\A_k),
$$
where we denoted by $z_{ik}$ a realization of $Z_{ik}$.

The S-EM algorithm is used to maximize \eqref{eq:3} with respect to model parameters and graph structures. The algorithm alternates between the two standard steps, E(xpectation) and M(aximization). In addition, the M step includes the structure learning step, the so-called S step, employed to search for the optimal graph configurations within the mixture components. We describe the S-EM algorithm in detail in the following subsections. A description of how the the algorithm is initialized is in Appendix~\ref{appendix:b}

\subsubsection{E step}
\label{estep}
At iteration $t$ of the S-EM algorithm, the estimated a posteriori probabilities $\hat{z}^{(t)}_{ik} = \widehat{\Pr}\left(Z_{ik}=1 \vbar \x_i \right)$ are computed using mixture parameters and graph configurations as follows:
$$
\hat{z}^{(t)}_{ik} = \dfrac{\hat{\tau}^{(t-1)}_k\, \phi (\x_i \vbar \hat{\MU}^{(t-1)}_k, \hat{\SIGMA}^{(t-1)}_k, \hat{\GG}^{(t-1)}_k )}{\sum_{l=1}^K\hat{\tau}^{(t-1)}_l\, \phi (\x_i \vbar \hat{\MU}^{(t-1)}_l, \hat{\SIGMA}^{(t-1)}_l, \hat{\GG}^{(t-1)}_l )},
$$
where $\hat{\tau}^{(t-1)}_k, \hat{\MU}^{(t-1)}_k, \hat{\SIGMA}^{(t-1)}_k, \hat{\GG}^{(t-1)}_k$ are parameters and graph structures estimated in the M and S steps at the previous iteration $(t-1)$.

\subsubsection{M step}
\label{mstep}
In the M step we solve the following maximization problem:
\begin{equation*}
\underset{\THETA,\mathbb{A}}\argmax ~\sumi \sum_{k=1}^K \hat{z}^{(t)}_{ik} \log \left\lbrace \tau_k\, \phi \left( \x_i \vbar \MU_k, \SIGMA_k, \GG_k \right) \right\rbrace- \sumk Q(\A_k). 
\end{equation*}
Note that finding the optimal collection of adjacency matrices $\mathbb{A}$ corresponds to finding the optimal set of graphs $\mathbb{G}$. Since the penalization term does not involve mixing proportions and cluster means, the updating formulas for these parameters are readily given by:
$$
\hat{\tau}^{(t)}_k\ = \dfrac{N^{(t)}_k}{N},	\qquad\qquad	\hat{\MU}^{(t)}_k = \dfrac{1}{N_k}\sumi \hat{z}^{(t)}_{ik}\, \x_i,
$$
where $N^{(t)}_k =\sumi \hat{z}^{(t)}_{ik}$. Estimation of the matrices $\SIGMA_k$ is coupled with the estimation of the graphs $\GG_k$. In fact, $\SIGMA_k$ needs to fulfill the constraint $\SIGMA_k \in \mathcal{C}^+(\GG_k)$. We resort to the subsequent S step to solve the optimization problem.

\subsubsection{S step}
For fixed $(\hat{\MU}^{(t)}_k, \hat{\tau}^{(t)}_k)$, estimates of $\SIGMA_k$ and $\A_k$ are found solving the maximization problem:
\begin{equation}
\begin{split}
\label{eq:4}
\underset{\SIGMA,\mathbb{A}}\argmax ~&- \dfrac{1}{2} \sumk \left\lbrace N^{(t)}_k \left[ \text{tr}(\mathbf{S}^{(t)}_k\SIGMA_k^{-1}) + \log\det \SIGMA_k  \right] \right\rbrace
				     - \sumk Q(\A_k), \\ 
                                     &\quad \text{with}\quad \SIGMA_k \in \mathcal{C}^+\left( \GG_k \right),
\end{split}
\end{equation}
where $\mathbf{S}^{(t)}_k=\frac{1}{N^{(t)}_k}\sumi \hat{z}^{(t)}_{ik} ( \x_i - \hat{\MU}^{(t)}_k )( \x_i - \hat{\MU}^{(t)}_k )^{\!\top}$ is the within-component sample covariance matrix and the objective function above corresponds to the (penalized) profile complete log-likelihood. The solution to \eqref{eq:4} is obtained by solving the problem component-wise with respect to $\SIGMA_k$ and $\A_k$:
\begin{equation}
\begin{split}
\label{eq:5}
\underset{\SIGMA_k,\mathbf{A}_k}\argmax ~&- \dfrac{N^{(t)}_k}{2} \left[ \text{tr}(\mathbf{S}^{(t)}_k\SIGMA_k^{-1}) + \log\det \SIGMA_k  \right]  - Q(\A_k), \\
                                         &\quad \text{with}\quad \SIGMA_k \in \mathcal{C}^+\left( \GG_k \right).
\end{split}
\end{equation}
Here the problem corresponds to the estimation of a covariance graph model. In this case, the structure learning task coincides with a combinatorial optimization problem. Given a proposed graph $\GG^\star_{k}$ represented by $\mathbf{A}^\star_{k}$, the corresponding $\SIGMA^\star_{k}$ is estimated using the ICF algorithm (see Appendix~\ref{appendix:a} for details). Then, the objective function in \eqref{eq:5} is evaluated for $(\mathbf{A}^\star_{k}, \SIGMA^\star_{k})$ and is used to rank different graph structures. Consequently, $\hat{\mathbf{A}}^{(t)}_k$ (thus $\hat{\GG}^{(t)}_k$)  and $\hat{\SIGMA}^{(t)}_k$ are determined as the couple $(\mathbf{A}^\star_{k}, \SIGMA^\star_{k})$ that maximizes this quantity. Carrying out an exhaustive search is infeasible since there are $2^{\binom{V}{2}}$ possible graphs. We propose to efficiently solve the problem by means of two alternative strategies based on genetic algorithm and stepwise search, both described in Section~\ref{ga}.

\subsection{Bayesian regularization}
\label{reg}
The likelihood of a Gaussian mixture model can be prone to degeneracies and singularities, especially related to the covariance matrix estimate \citep{titterington:1985}. Moreover, the ICF algorithm employed to estimate a sparse $\SIGMA_k$ requires the within-component sample covariance matrix $\mathbf{S}_k$ to be strictly positive definite \citep{chaudhuri:etal:2007}. This condition may not be attained in practice if the expected number of observations falling into a cluster is less than the number of variables or because of singularities, due to highly correlated variables, for example. To overcome the issue, we delineate a Bayesian framework for regularization where the maximum likelihood estimate is replaced by the \emph{maximum a posteriori} (MAP) estimate. A similar approach has already been suggested by \cite{ciuperca:2003}, \cite{fraley:2005,fraley:2007} and \cite{baudry:2015}. Here the main purpose is in regularizing the estimate of the covariance parameters, thus we place no prior distributions on the mixing proportions and the component means. We consider exchangeable priors on the covariance matrices, such that the prior factorizes as $\prod_k p(\SIGMA_k)$. Then, the aim is optimizing the following regularized log-likelihood:
\begin{equation}
\label{eq:6}
\ell_R = \sumi \log \left\lbrace \sumk \tau_k\, \phi \left( \x_i \vbar \MU_k, \SIGMA_k, \GG_k \right) \right\rbrace + \sumk \log p(\SIGMA_k)
       - \sumk Q(\A_k),
\end{equation}
where we take $p(\SIGMA_k)$ to be an Inverse Wishart distribution, $IW(\omega, \mathbf{W})$, the standard conjugate prior in this setting.

To maximize \eqref{eq:6} with respect to parameters and graph configurations we use the same S-EM algorithm of the previous section. The E step is unchanged and estimates  $\hat{z}_{ik}$ are given as in Section~\ref{estep}. Also estimates for $\tau_k$ an $\MU_k$ are obtained in the same way as Section~\ref{mstep}. On the other hand, maximization of \eqref{eq:6} for $\SIGMA_k$ and $\A_k$ leads to the following optimization problem:
\begin{equation*}
\begin{split}
\underset{\SIGMA_k,\mathbf{A}_k}\argmax ~ &- \dfrac{\tilde{N}_k}{2} \left[ \text{tr}(\tilde{\mathbf{S}}_k\SIGMA_k^{-1}) + \log\det \SIGMA_k  \right]  - Q(\A_k),\\ 
                                          &\quad \text{with}\quad \SIGMA_k \in \mathcal{C}^+\left( \GG_k \right),
\end{split}
\end{equation*}
where
\begin{equation}
\label{eq:7}
\tilde{N}_k = N_k + \omega + V + 1, \qquad \tilde{\mathbf{S}}_k = \dfrac{1}{\tilde{N}_k} \left[ N_k \mathbf{S}_k + \mathbf{W} \right].
\end{equation}
Numerical solution to this problem is found using the same approach adopted for solving \eqref{eq:5}, this time replacing the maximum likelihood estimate of $\SIGMA_k$ under the covariance graph model with its MAP estimate. Again, the process involves a type of combinatorial optimization, which is solved using two alternative strategies as described in Section~\ref{ga}. To find the MAP estimate of $\SIGMA_k$ given a graph structure, the ICF algorithm is modified consequently. Appendix~\ref{appendix:a} contains further details. 

Using arguments similar to \cite{fraley:2005,fraley:2007} and \cite{baudry:2015}, we set: 
$$
\omega = V + 2, \qquad \mathbf{W} = \dfrac{\mathbf{S}}{\det(\mathbf{S})^{1/V}}\left( \dfrac{c}{K} \right)^{1/V},
$$ 
where $\mathbf{S} = \frac{1}{N}\sumi ( \x_i - \bar{\x} )( \x_i - \bar{\x} )^{\!\top}$, the empirical covariance matrix of all the data, with $\bar{\x}$ the sample mean, $\bar{\x} = \frac{1}{N}\sumi \x_i$. With this choice, $\det(\mathbf{W}) = \frac{c}{K}$ and the tuning parameter $c$ determines the amount of regularization. The parameter allows a weaker regularization than the one suggested in \cite{fraley:2005,fraley:2007}. A small value for $c$ avoids masking the clustering structure and allows to obtain components whose volume is within the volume of the data \citep{baudry:2015}. We set this parameter equal to 0.001 by default. A further discussion on the choice of hyperparameters for the Inverse Wishart distribution in mixture modeling is in \cite{fruhwirth:2006}.

\subsection{Penalty functions}
\label{pen}
The framework for sparse covariance estimation has been rendered as a combinatorial optimization problem, where the function $Q(\cdot)$ in \eqref{eq:3} acts with the purpose of scoring the different graph configurations. This function penalizes the complexity of a graph structure and different specifications of it lead to different modeling strategies and control on the amount of sparsity induced in the component covariance matrices. Furthermore, within the context of maximum penalized likelihood estimation, the choice of $Q(\cdot)$ can also be interpreted as a choice for the prior distribution $p(\A_k)$, thus, indirectly, $p(\GG_k)$; in fact, it may be considered $p(\GG_k)=p(\A_k) \propto e^{-Q(\A_k)}$ \citep{green:1990}. With this view, the decision can be made as to include prior knowledge about the correlation pattern among the variables or to penalize more some structures of association than others. Indeed, specification of the form of $Q(\cdot)$ is context and purpose-dependent. For example, in high-dimensional settings, one may want to have sparser component covariance matrices, opting for a function that penalizes significantly complex association structures; also, if the aim is to derive a graphical model for the within-cluster joint distribution of the variables, a penalty function based on a model selection criterion could be specified. In the subsequent sections we suggest some alternatives that we found to work well in practice; these are tailored to different situations and have a meaningful interpretation. 

\subsubsection{BIC-type}
Within the S-EM algorithm, the structure learning task can be recast as a model selection problem. The set of edges delineates a model for the association among the variables within a mixture component and selection of the optimal structure corresponds to selection of the best model for such association. Let us denote by $E_k = \sum_{j>h} a_{jhk}$ the number of arcs in a graph $\GG_k$, i.e. the number of non-zero off-diagonal entries of $\A_k$, corresponding to the number of covariance parameters of the associated matrix $\SIGMA_k$; let also $T=\binom{V}{2}$, i.e. the total number of covariance terms for a set of $V$ variables. In the context of Gaussian graphical model selection, a natural penalty function is such that the score in \eqref{eq:5} corresponds to the Bayesian Information Criterion \citep[BIC,][]{schwarz:1978,koller:2009} of a Gaussian graph covariance model. In this case the function is given by:
$$
Q_{\text{\tiny BIC}}(\A_k) = \dfrac{1}{2} E_k \log N.
$$
With this choice of $Q(\cdot)$, solving the problem in \eqref{eq:5} is equivalent to selecting the best covariance graph model using the BIC. The score obtained in this way is an approximation to the marginal likelihood of the Gaussian covariance graph model and consistency properties hold \citep{koller:2009}. When $N$ and $V$ are of comparable size, this score may select graphs that are overly complex. In this case, \cite{foygel:2010} suggest an extended Bayesian information criterion (EBIC). The corresponding $Q(\cdot)$ function is given by:
$$
Q_{\text{\tiny EBIC}}(\A_k) = \dfrac{1}{2} E_k \log N + 2 \gamma E_k \log V,
$$
where $0 \leq \gamma \leq 1$. The parameter $\gamma$ downweighs the probability of selecting graphs with a large number of arcs. In the case $\gamma=1$, the probability of selecting a graph with $E_k$ edges is proportional to $\binom{T}{E_k}^{-1}$.  Clearly, for a choice of $\gamma = 0$ the BIC score is recovered. In practice, setting $\gamma=1$ results in very sparse covariance matrices and is particularly suitable when the number of variables is large. We refer to \cite{chen:2008}, \cite{foygel:2010}, and \cite{barber:2015} for more details.

\subsubsection{Erd\H{o}s-R\'enyi}
The Erd\H{o}s-R\'enyi model is a popular model for random graphs. Under this model, the probability of a graph $\GG_k$ with $E_k$ arcs is given by $\alpha^{E_k}(1-\alpha)^{T-E_k}$, where $\alpha$ is the probability that two nodes are associated \citep{erdos:1959,bollobas:2001}. From this quantity, the following penalty function can be derived:
$$
Q_{\text{\tiny ER}}(\A_k) = -E_k \log\alpha - (T-E_k)\log(1-\alpha).
$$
The parameter $\alpha$ controls the connectivity of a graph. In particular, \cite{erdos:1960} derived a tight bound on the density of a graph in relation to the value of $\alpha$. For values of this parameter less than $\log V/V$, the graph will be almost surely disconnected as $V \rightarrow \infty$, i.e. there exists two nodes such that there is no path in the graph joining them  \citep{edwards:2000}. Therefore, for small values of $\alpha$ the penalization tends to favor situations where the component joint distribution decomposes into the product of independent blocks, which contain correlated variables. We suggest setting $\alpha=\log V/T$, a value such that the expected number of arcs is equal to $\log V$ and such that the graph will almost surely have disconnected components of size larger than $\mathcal{O}(\log V)$ \citep{bollobas:2001}.

\subsubsection{Power law}
\begin{figure}[tb]
 \centering
 \subfloat[][]
 {\includegraphics[scale=0.9]{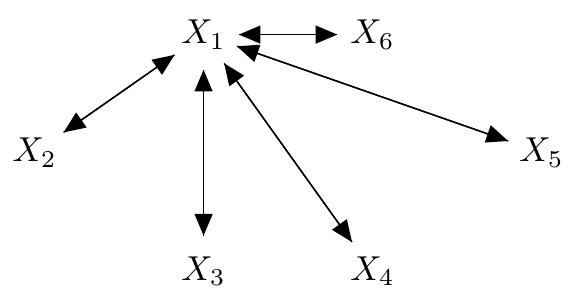}} \qquad 
 \subfloat[][]
 {\includegraphics[scale=0.9]{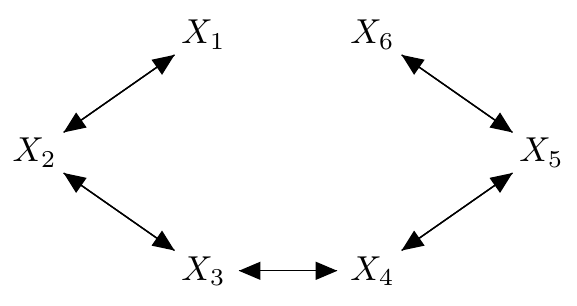}}
 \caption{\label{fig:pl} Example of the $Q_{\text{\tiny PL}}$ penalty function. The two graphs have the same number of edges while they differ in the association structure. The graph in \emph{(a)} corresponds to a matrix where $X_1$ is a central variable, while the graph in \emph{(b)} to a case where the variables are pairwise correlated. For \emph{(a)}, $Q_{\text{\tiny PL}} = 5.08~\!\beta$, while for \emph{(b)}, $Q_{\text{\tiny PL}} = 6.60~\!\beta $. }
\end{figure}

The previous $Q(\cdot)$ functions penalize in the same way graphs with equal number of edges but dissimilar configurations. However, in some situations some form of association structures may be preferred to others a priori. To assign different penalization to different structures defined on the same number of arcs, we consider the following penalty function:
$$
Q_{\text{\tiny PL}}(\A_k) = \beta \sum_{j=1}^V \log( d_{kj} + 1 ),
$$
where $\beta$ is a weighting coefficient and $d_{kj} = \sum_{h=1}^V a_{jhk}$, the degree of node $j$ in graph $\GG_k$, i.e. the number of nodes connected to it. The penalty is derived from a power law on the nodes of a graph of the form $\prod_j(d_{kj}+1)^{\beta}$. With this function, for a fixed number of edges, structures of association characterized by few hub variables correlated to the others are preferred. Figure~\ref{fig:pl} contains an explicit example.
With the choice $\beta = \log (N V)$, the penalty function can be rewritten as $Q_{\text{\tiny PL}}(\A_k) = \sum_{j=1}^V \log( d_{kj} + 1 )\log N + \sum_{j=1}^V \log( d_{kj} + 1 )\log V$, and thus its magnitude is approximately on a similar scale as BIC and EBIC penalizations. However, contrary to $Q_{\text{\tiny EBIC}}$ and $Q_{\text{\tiny BIC}}$ functions, it is not linear in the number of parameters and denser graphs will tend to be less penalized.

\subsection{Solving the optimization problem in the S step}
\label{ga}
We resort to two alternative strategies in order to solve the optimization problem in the S-step and obtain estimates of the graph structures and the corresponding covariance matrices. The first is based on a genetic algorithm, while the second is based on a stepwise search. We note that both strategies allow parallelization of the search procedure, leading to a notable reduction of computing time.

\subsubsection{Genetic algorithm}
Genetic algorithms (GAs) are stochastic optimization algorithms based on concepts and operators of biological evolution and natural selection \citep{goldberg:1989, holland:1992}. These algorithms have been applied in numerous fields of statistics \citep[see][for example]{chatterjee:1996,bozdogan:2004,galimberti:2017}, in graphical modeling \citep{poli:1998,roverato:2004}, and Gaussian mixture model estimation \citep{martinez:2000,pernkopf:2005}.

Compared to standard stepwise search strategies, GAs are less prone to be trapped in local optima, but may not scale well to problems with a large space of possible solutions. Although GAs require some parameter tuning, the stochastic evolutionary nature of these algorithms make the final solution less sensitive to initialization \citep{goldberg:1989}. Furthermore, convergence results have been derived \citep[][for example]{greenhalgh:2000,sharapov:2006}.

A GA is started with a population of randomly generated individuals or solutions. The fitness of every individual in the population is evaluated and a new population is formed by applying genetic operators. In our framework, a graph is encoded through its adjacency matrix as a binary string indicating the presence or absence of an arc between any pair of variables. For example, the graph represented in Figure~\ref{fig_graph_example} is encoded as the binary vector represented in Table~\ref{tab:bin}, where the pairs of variables coincide to the off-diagonal elements of the related adjacency matrix.

\begin{table*}[b]
\centering\footnotesize
\begin{tabular}{cccccccccc}
\toprule
 $X_1\!-\!X_2$ & $X_1\!-\!X_3$ & $X_1\!-\!X_4$ & $X_1\!-\!X_5$ & $X_2\!-\!X_3$ & $X_2\!-\!X_4$ & $X_2\!-\!X_5$ & $X_3\!-\!X_4$ & $X_3\!-\!X_5$ & $X_4\!-\!X_5$\\
\midrule
 1 & 1 & 1 & 0 & 0 & 1 & 0 & 0 & 1 &  0\\
\bottomrule
\end{tabular}
\caption{\label{tab:bin} The binary string representing the adjacency matrix corresponding to the graph in Figure~\ref{fig_graph_example}.}
\end{table*}

In our setting, a population corresponds to a collection of possible graphs. Then the fitness of each individual is evaluated according to the objective function of Equation \eqref{eq:5} (or Equation \eqref{eq:7} in case of regularization). At each iteration, a new population is generated by means of the following operators:
\begin{itemize}[noitemsep]
 \item {\bf Selection}: this step involves selecting a subset of graphs for breeding. A weighted rank selection scheme is used to assign a weight between 0 and 1 to each graph structure based on its fitness value. Consequently, a new population is randomly sampled with such computed weights. Thus, better models of association have higher chance of being included in the next generation.
 \item {\bf Crossover}: two strings (parents) are selected at random with probability 0.8 (by default) and re-combined in order to produce two different strings (offspring). Single-point crossover selects a point at random and the resulting graph is then obtained copying one parent from beginning to the crossover point and the rest is  from the second parent.
 \item {\bf Mutation}: a random mutation is introduced in the population to ensure that the searching process does not get trapped in some local optima of the searching space. With probability 0.2 (by default), an arc is either introduced or removed from the graph.
 \item {\bf Elitism}: the graph structure with the largest fitness value is retained at each iteration of the genetic algorithm. Moreover, in order to maintain the general monotonicity property, elitism is performed also between each iteration of the S-EM algorithm. Therefore, the optimal graph structure selected in the S-step at the previous iteration is included in the starting population of the S-step at the following iteration.
\end{itemize}

For a population of graphs, the related sparse covariance matrices are estimated using the ICF algorithm and the optimal couple $(\SIGMA_k, \A_k)$ is selected as the one with the largest fitness value in the population. At each iteration of the GA, the evolutionary scheme is repeated and the aim is to generate novel population members that gradually improve their average fitness value. The GA stops when there are no further improvements in the fitness value of the optimal couple $(\SIGMA_k, \A_k)$ of a population for a fixed number of consecutive iterations. By default, we set this number of consecutive iterations equal to 100, a value ensuring that a stable solution has been reached without slowing down the procedure. Due to the elitism operator, the general S-EM algorithm is in the class of \emph{generalized} EM algorithms and generates a sequence of values of $\ell$ (or $\ell_R$) that monotonically converges to a stationary point \citep{wu:1983,green:1990,friedman:1997,friedman:1998}. 

The genetic algorithm is implemented in practice using the \texttt{R} \citep{R} package \texttt{GA} \citep{scrucca:2017,scrucca:2013}. The implementation allows parallelization of the search procedure. Moreover, the nature of the optimization problem allows to discard solutions already evaluated during the previous iterations. This results in a considerable reduction of the amount of computing time.

\subsubsection{Stepwise search}
Although less prone to be trapped in a local optimum, GAs can be computationally intensive and not suited for high-dimensional problems. Despite being sub-optimal, stepwise searching strategies are standard procedures for combinatorial model search \citep{miller:2002,wiegand:2010} and can scale better in higher dimensions. Here we propose a stepwise search particularly suited to the case when the number of variables $V$ is large.

Let $O(\SIGMA_k, \A_k)$ be the value of the objective function in \eqref{eq:5} (or \eqref{eq:7} in case of regularization) at a given iteration of the S-EM algorithm (note we omit the iteration superscript $t-1$ for ease of notation). Let also denote with $\mathcal{A}_k^+$ the collection of adjacency matrices where an edge has been added to $\A_k$, and with $\mathcal{A}_k^-$ the collection of adjacency matrices where an edge has been removed from $\A_k$. We indicate with $e$ a generic edge, thus $\A_k^e$ is the adjacency matrix whose edge $e$ has been added or removed. For matrix $\A_k^e$, the corresponding sparse covariance matrix $\SIGMA_k^e$ is estimated using the ICF algorithm. We alternate the following steps.
\begin{itemize}
 \item \textbf{Addition} - Add one edge:
 \begin{enumerate}
  \item For $\A_k^e \in \mathcal{A}_k^+$, estimate $\SIGMA_k^e$ using the ICF algorithm given $\A_k^e$ and compute $O(\SIGMA^e_k, \A^e_k)$;
  \item Find the couple ($\SIGMA_k^{\star}, \A_k^{\star}) = \underset{\A_k^e \in \mathcal{A}_k^+}{\argmax}\lbrace O(\SIGMA^e_k, \A^e_k) \rbrace$
  \item If $O(\SIGMA^\star_k, \A^{\star}_k) > O(\SIGMA_k, \A_k)$, an edge is added to $\A_k$, thus set $\SIGMA_k = \SIGMA^{\star}_k$, $\A_k = \A^{\star}_k$ and $O(\SIGMA_k, \A_k) = O(\SIGMA^{\star}_k, \A^\star_k)$.
 \end{enumerate}\vspace*{0.8cm}
 \item \textbf{Removal} - Remove one edge:
 \begin{enumerate}
  \item For $\A_k^e \in \mathcal{A}_k^-$, estimate $\SIGMA_k^e$ using the ICF algorithm given $\A_k^e$ and compute $O(\SIGMA^e_k, \A^e_k)$;
  \item Find the couple $(\SIGMA_k^{\star}, \A_k^{\star}) = \underset{\A_k^e \in \mathcal{A}_k^-}{\argmax}\lbrace O(\SIGMA^e_k, \A^e_k) \rbrace$
  \item If $O(\SIGMA^\star_k, \A^{\star}_k) \geq O(\SIGMA_k, \A_k)$, an edge is removed from $\A_k$, therefore set $\SIGMA_k = \SIGMA^{\star}_k$, $\A_k = \A^{\star}_k$ and $O(\SIGMA_k, \A_k) = O(\SIGMA^{\star}_k, \A^\star_k)$.
 \end{enumerate}
\end{itemize}
The procedure is repeated until no edges are added or removed for two consecutive addition and removal steps. 

In some situations, at a given addition or removal step, the number of adjacency matrices in the collection $\mathcal{A}_k$ to be considered as a potential solution can still be remarkably large. Moreover, the addition or removal of some edges will give a value of the objective function significantly smaller than the current optimum. Therefore, it is reasonable to reduce the space of candidate adjacency matrices at the subsequent step by discarding those with a value of the objective function $O(\SIGMA_k^{e}, \A_k^{e})$ which is too distant from the current optimal value $O(\SIGMA_k^{\star}, \A_k^{\star})$. To this purpose, after we found the current optimal couple $(\SIGMA_k^{\star}, \A_k^{\star})$, we compute the differences:
$$
D_e = O(\SIGMA_k^{\star}, \A_k^{\star}) - O(\SIGMA_k^{e}, \A_k^{e}).
$$
Then, at the next addition or removal step, we only consider the set of adjacency matrices such that:
$$
\lbrace{\A^e_k \in \mathcal{A}_k \colon D_e \leq C} \rbrace,
$$
where $C$ is a constant to be specified. In this way, only candidate solutions whose value of $O(\SIGMA_k^{e}, \A_k^{e})$ is within the interval 
$[~O(\SIGMA_k^{\star}, \A_k^{\star}) -C~~;~~O(\SIGMA_k^{\star}, \A_k^{\star})+C ~]$
will be evaluated. The rationale is that possible solutions which give a value of the objective function too small compared to the current best are unlikely to be good candidates at the next step and should no longer be considered. 

The idea is closely connected to the Occam's window of \cite{madigan:1992} \citep[see also][]{hoeting:1999} and greatly reduces the number of adjacency matrices to be taken into account at each step of the stepwise search. In this context, the quantity $C$ can be interpreted as the maximum log-odds ratio value between the likelihood of the current best graphical model and the likelihood of a candidate graphical model, both weighted by the corresponding graph structure prior (i.e. the penalty term). Selection of $C$ is context dependent and represents a trade-off between speed and quality of the solution. Smaller values shrink the searching space around the current optimum, thus the algorithm runs faster, but the search could be more prone to reach a sub-optimal solution; larger values allows the algorithm to better explore the space of association structures, but at the price of an higher computational cost. In practice and simulated data experiments we found setting $C=50$ to provide a good balance between quality of the solution and efficiency, especially in high-dimensional settings. 

The overall stepwise strategy is particularly easy to implement, less computationally intensive than a genetic algorithm and allows parallelization of the search procedure as well. Furthermore, also in this case, since the optimal solution is carried to the next iteration, the general S-EM algorithm with stepwise search is in the class of generalized EM algorithms and generates a sequence of log-likelihood values that monotonically converges to a stationary point.

\subsection{Model selection and cluster assignment}
The number of mixture components is often unknown and needs to be inferred from the data. Here we make use of the Bayesian information criterion (BIC) for choosing the number of clusters and performing model selection
$$
\text{BIC} = 2\sumi \log f( \x_i \vbar \hat{\THETA}, \hat{\mathbb{G}} ) - \nu\, \log N,
$$
where $\hat{\THETA}$ and $\hat{\mathbb{G}}$ are the estimated mixture parameters and graphs, and $\nu$ is the number of non-zero parameters. We remark that, differently from penalized model-based clustering with lasso penalty, where \emph{parameters are first estimated and then shrunk to zero}, in our framework covariance parameters corresponding to zero entries in the graph are exactly zero and not estimated, therefore $\nu$ coincides to the actual number of parameters and degrees of freedom \citep[see][]{xie:2008,yuan:2007,pan:shen:2007,zou:2007,pan:2006}. The best model is the one that maximizes the BIC. Also resampling model selection methods (such as cross-validation) could be employed, however BIC has the advantage of being less computationally demanding \citep{shen:2002,ruan:2011}. Several other methods for model selection in mixture models have been proposed in the literature; for an in depth review we recommend \cite{mcLachlan:rathnayake:2014}. Moreover, we point that BIC can be used to compare different sparse covariance models once they have been estimated using different penalty functions in the S-EM algorithm. Note that, in doing so, BIC shall not be used to choose the type of penalization function and state its superiority over the others. Rather, the selection of the penalty function $Q(\cdot)$ is context and purpose dependent. Nevertheless, different penalty functions may lead to different mixtures of Gaussian covariance graph models with different general structures of association, and BIC can be employed to compare these models.

After estimating parameters, graph configurations, and selecting the number of components, each observation $\x_i$ is assigned to the corresponding cluster using the maximum a posteriori rule. The rule assigns an observation to the cluster $k$ if
$$
k = \underset{l}{\operatorname{argmax}} ~ \lbrace \hat{z}_{i1},\, \dots,\, \hat{z}_{il},\, \dots,\, \hat{z}_{iK} \rbrace,
$$
where $\hat{z}_{il}$ are the posterior probabilities as estimated in the E step of Section~\ref{estep}.

\section{Simulated data experiments}
\label{sim}
In this section we assess the proposed sparse modeling approach through different simulated data scenarios. The objective is to evaluate the ability of the mixture of Gaussian covariance graph models framework of discovering the group structure in the data, as well as its ability in modeling the within-cluster associations among the variables. We test the method by considering various configurations of sample size, number of variables and dependence patterns.

For each simulated dataset, we fit a mixture of Gaussian covariance graph models using the four penalizations described in Section~\ref{pen}: BIC-type, EBIC-type, Erd\H{o}s-R\'enyi and power law; we will refer to the sparse covariance models with \texttt{mgc} and to the different penalizations with \texttt{BIC},  \texttt{EBIC},  \texttt{ER} and  \texttt{PL}, respectively. For each penalization, we will estimate the model using both the genetic algorithm and the stepwise search, denoted with \texttt{Ga} and \texttt{Step} respectively. Hence, for example, a sparse covariance model estimated using the stepwise search and the EBIC-type penalization will be indicated by \texttt{mgcStepEBIC}.

For comparison, we also apply the well known model-based clustering approach of \cite{banfield:raftery:1993} and \cite{celeux:govaert:1995}, implemented in the widely popular \texttt{R} package \texttt{mclust} \citep{scrucca:etal:2016}. The approach is based on a family of 14 parsimonious models defined imposing constraints on the covariance matrix eigendecomposition $\SIGMA_k = \lambda_k\mathbf{V}_k \mathbf{D}_k \mathbf{V}_k^{\!\top}$. The models characterize the geometric properties of the clusters, however, with regards to the association structure between the variables, they can only convey two alternatives: diagonal covariance, where all the variables are independent (the eigenvectors $\mathbf{V}_k$ are constrained to correspond to the standard basis vectors), or full covariance, where all the variables are allowed to be dependent (no constraints on $\mathbf{V}_k$); see \cite{scrucca:etal:2016} and \cite{celeux:govaert:1995} for details. In using the package, we let it automatically select the best covariance decomposition model (among the available 14), and we simply use the umbrella term \texttt{mclust} to denote the package, the methodology and the corresponding results.

To evaluate the ability of recovering the generating graphs, we consider the false positive rate (proportion of incorrectly identified edges) and the false negative rate (proportion of incorrectly missed edges). To overcome the problem of label matching and the fact that the selected number of clusters $\hat{K}$ may differ from the data generating one, we compute the following indexes:
$$
\text{FPR} = \dfrac{1}{\hat{K}} \sum_{g=1}^{\hat{K}} \text{FPR}^*_g, \qquad \text{FNR} = \dfrac{1}{\hat{K}} \sum_{g=1}^{\hat{K}} \text{FNR}^*_g,
$$
where
\begin{align*}
\text{FPR}^*_g &= \text{min}\lbrace\, \text{FPR}^{(1)}_g,\, \dots\, \text{FPR}^{(k)}_g\, \dots\, \text{FPR}^{(K)}_g \,\rbrace, \\
\text{FNR}^*_g &= \text{min}\lbrace\, \text{FNR}^{(1)}_g,\, \dots\, \text{FNR}^{(k)}_g\, \dots\, \text{FNR}^{(K)}_g \,\rbrace,
\end{align*}
with $\text{FPR}^{(k)}_g$ and $\text{FNR}^{(k)}_g$ the false positive rate and the false negative rate computed between the estimated graph of component $g$ and the graph corresponding to group $k$. As usual, to evaluate the classification performance we make use of the Adjusted Rand Index \citep[ARI,][]{hubert:arabie:1985}.

We consider four scenarios differentiated by the association structures and the sparsity rates corresponding to the group covariance matrices:

\begin{description}[noitemsep]
 \item[\normalfont \emph{Scenario 1.}] Alternated-blocks covariance matrices. 
 \item[\normalfont \emph{Scenario 2.}] Sparse at random covariance matrices.
 \item[\normalfont \emph{Scenario 3.}] Random hubs covariance matrices.
 \item[\normalfont \emph{Scenario 4.}] Mixed type covariance matrices.
\end{description}

\begin{figure}[!h]
\centering
 \includegraphics[scale=0.25]{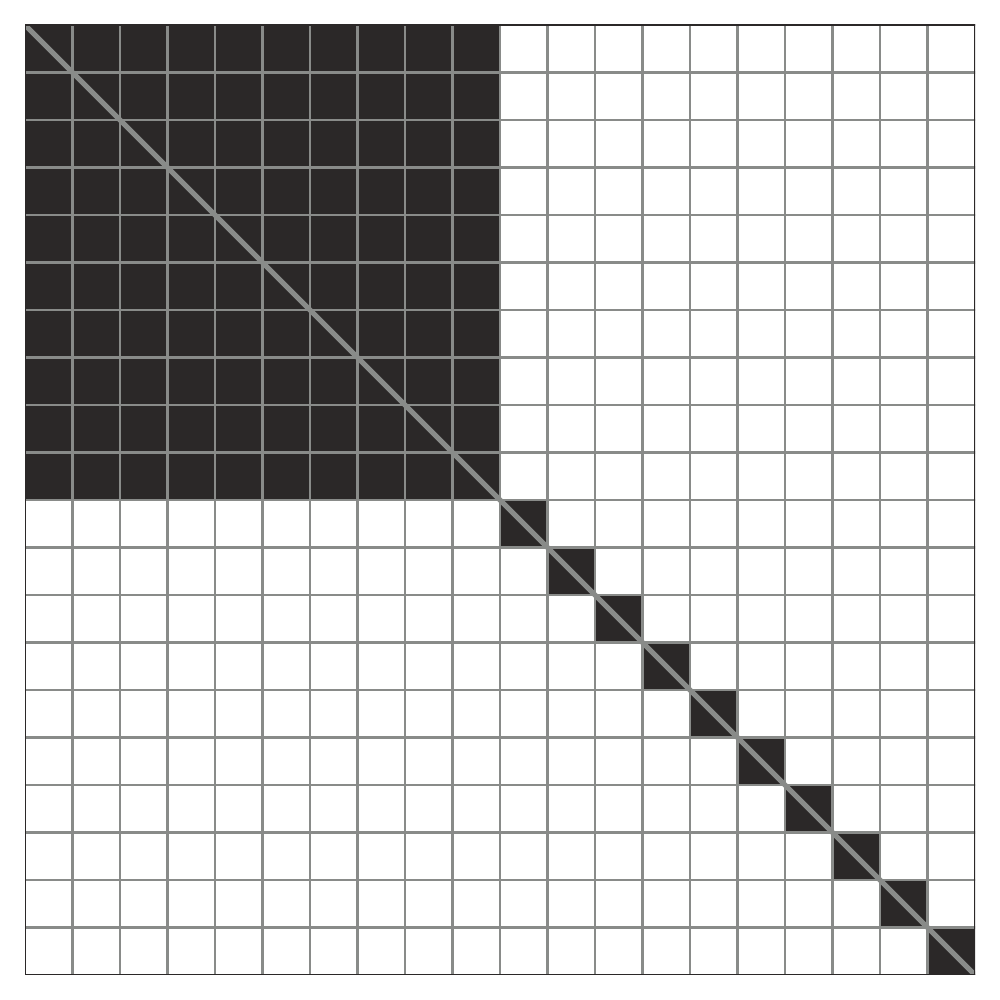}~~\includegraphics[scale=0.25]{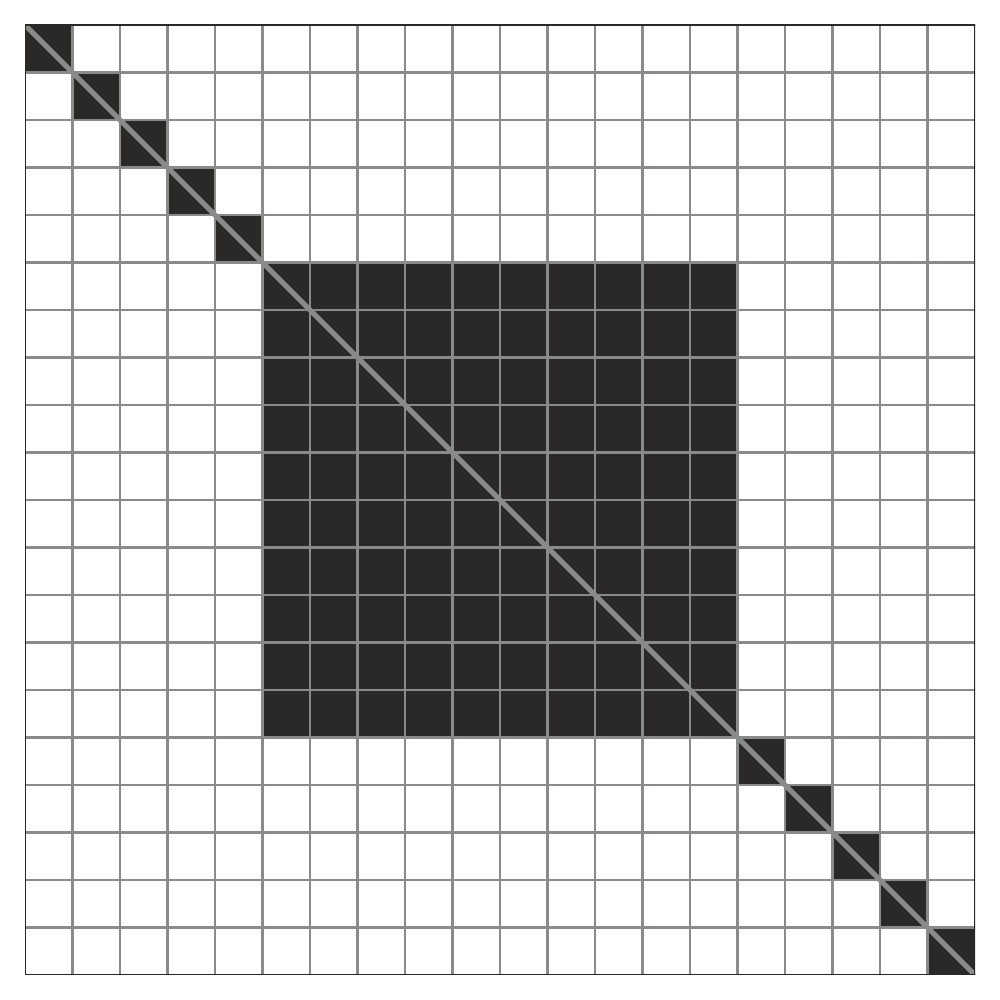}~~\includegraphics[scale=0.25]{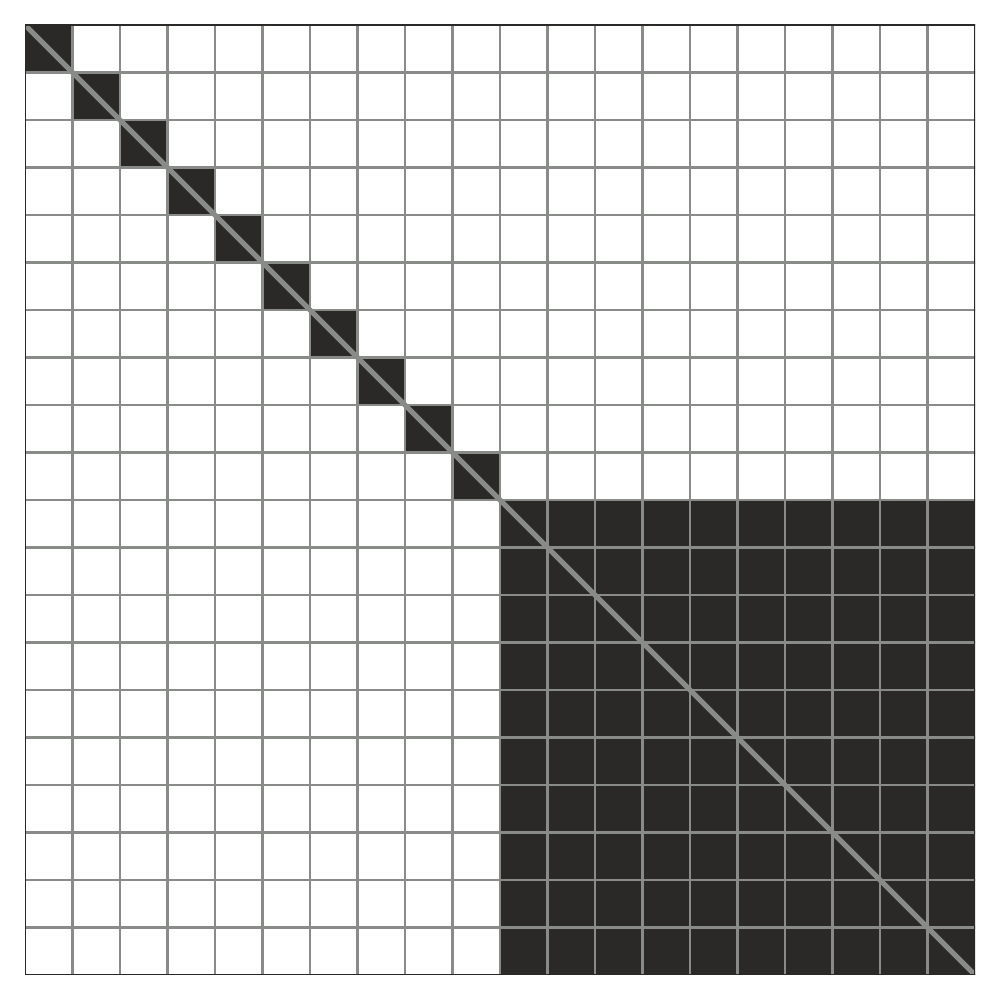}
 \caption{\label{fig:sim:1} Example of simulation setting 1 - Alternated-blocks covariance matrices.}\bigskip
%
\centering
 \includegraphics[scale=0.25]{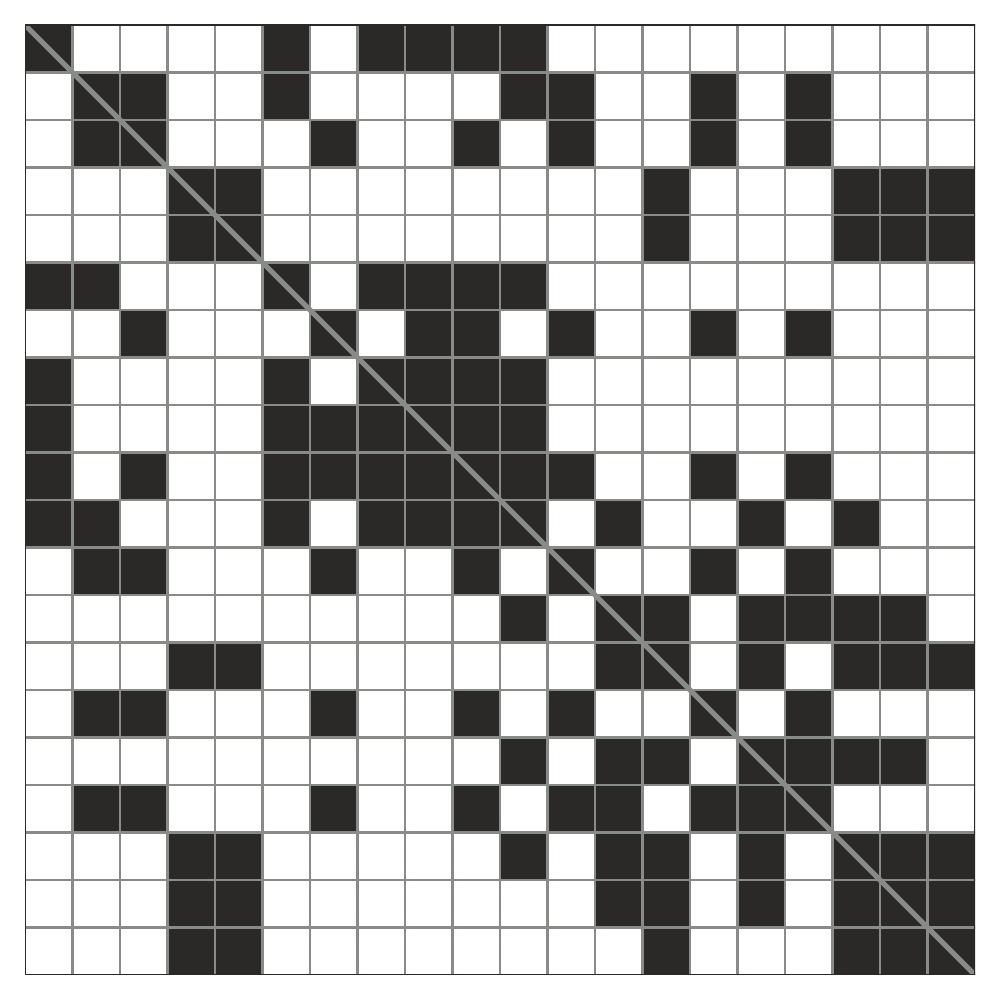}~~\includegraphics[scale=0.25]{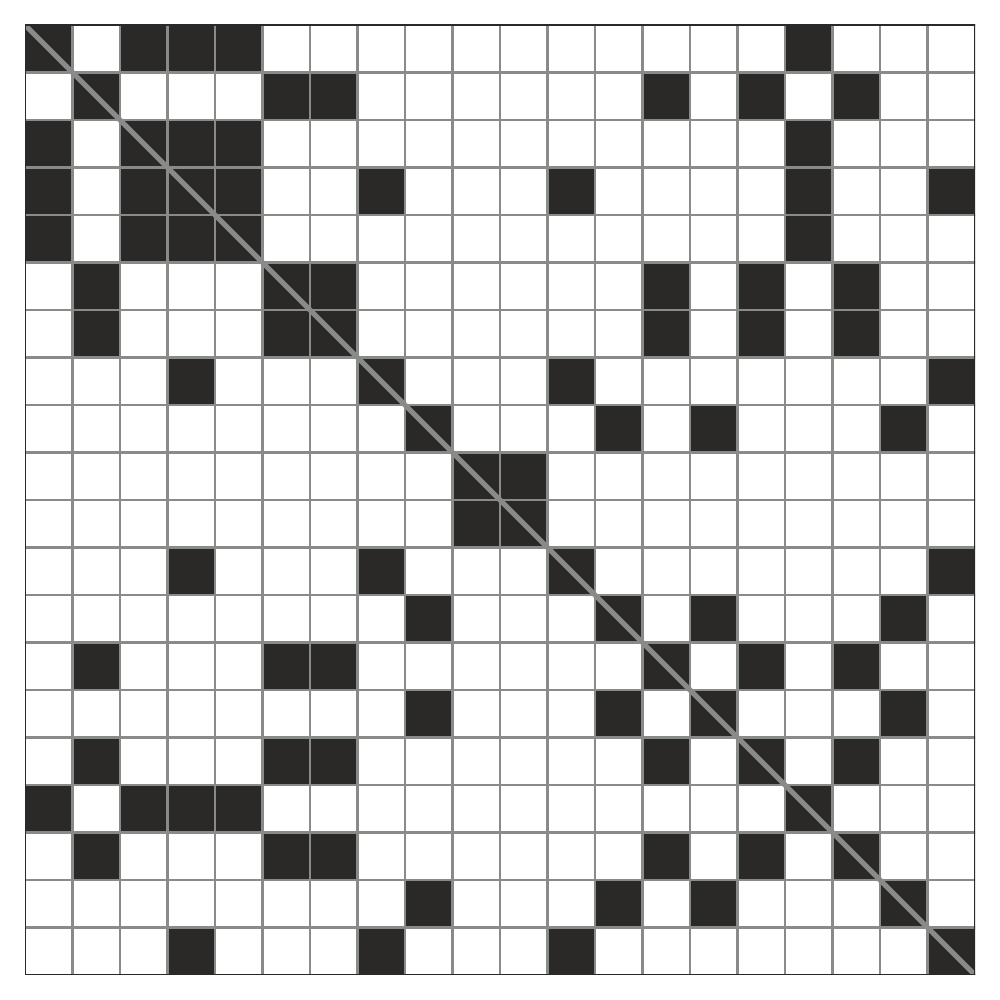}~~\includegraphics[scale=0.25]{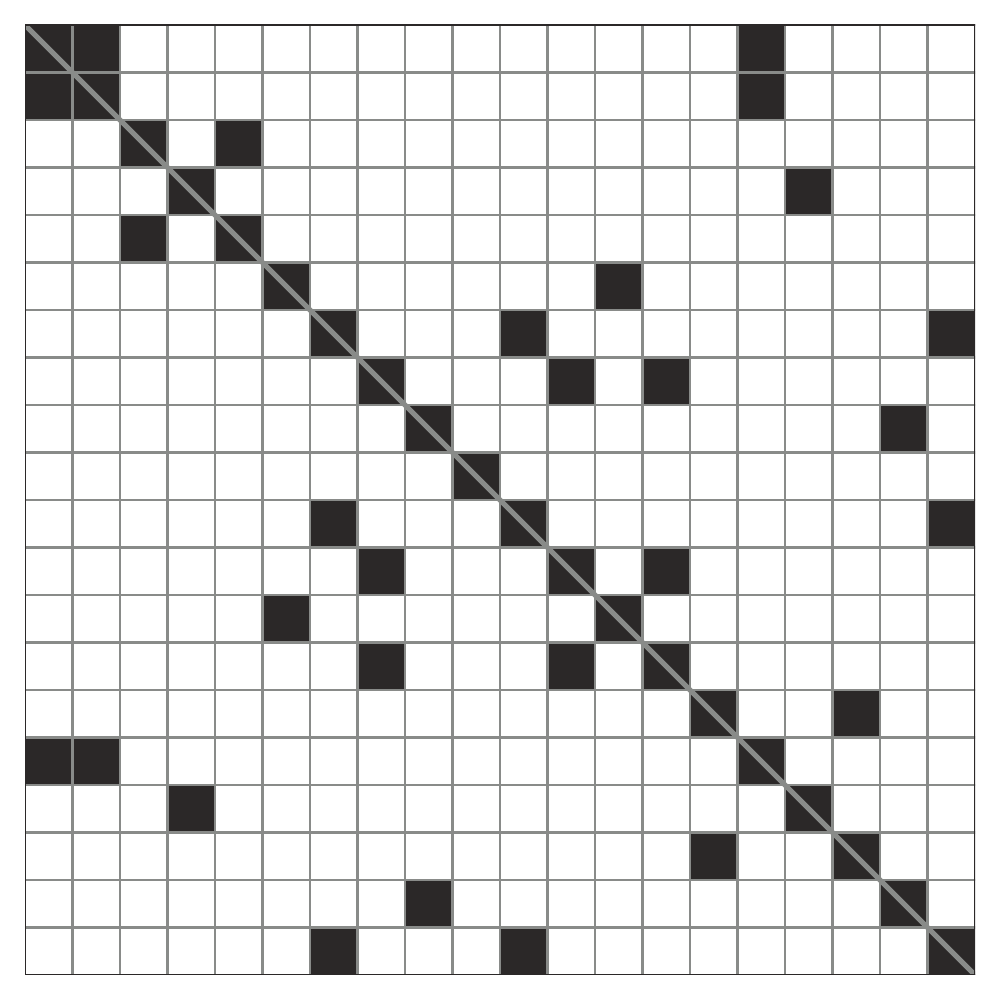}
 \caption{\label{fig:sim:2} Example of simulation setting 2 - Sparse at random covariance matrices.}\bigskip
%
\centering
 \includegraphics[scale=0.25]{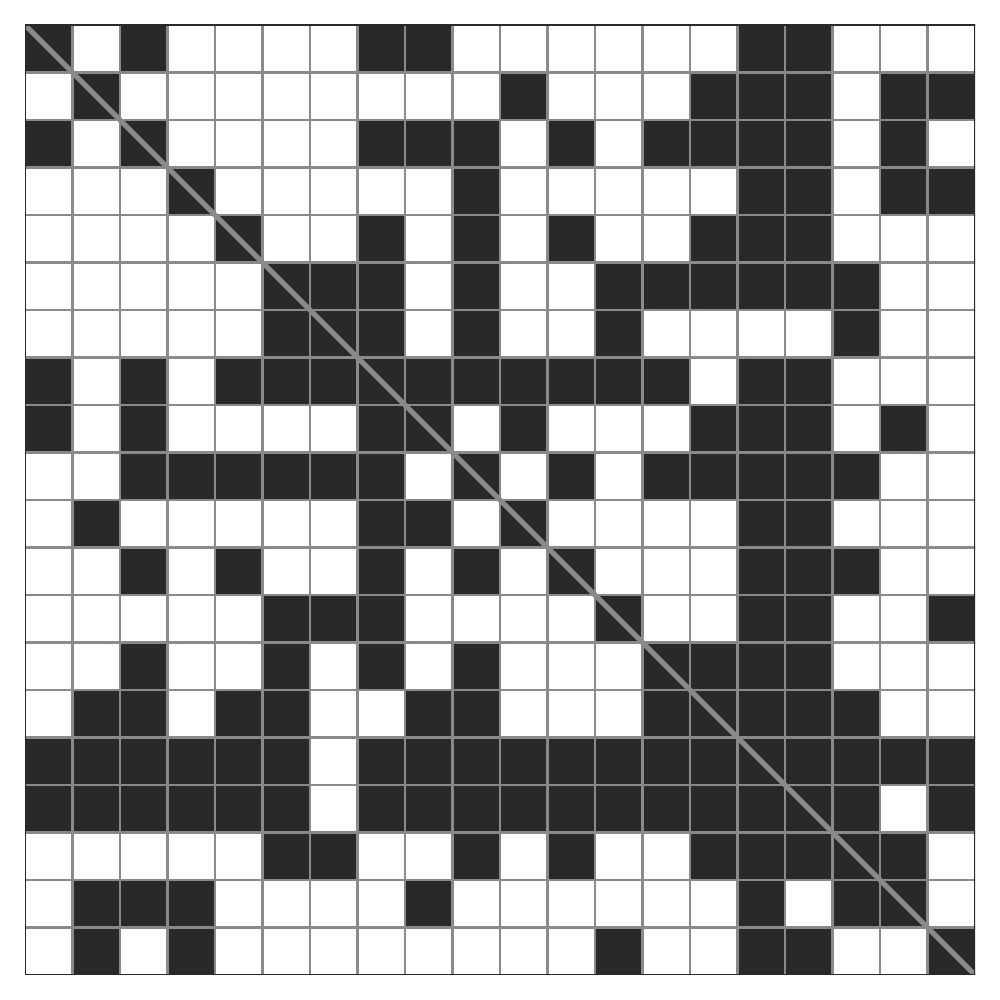}~~\includegraphics[scale=0.25]{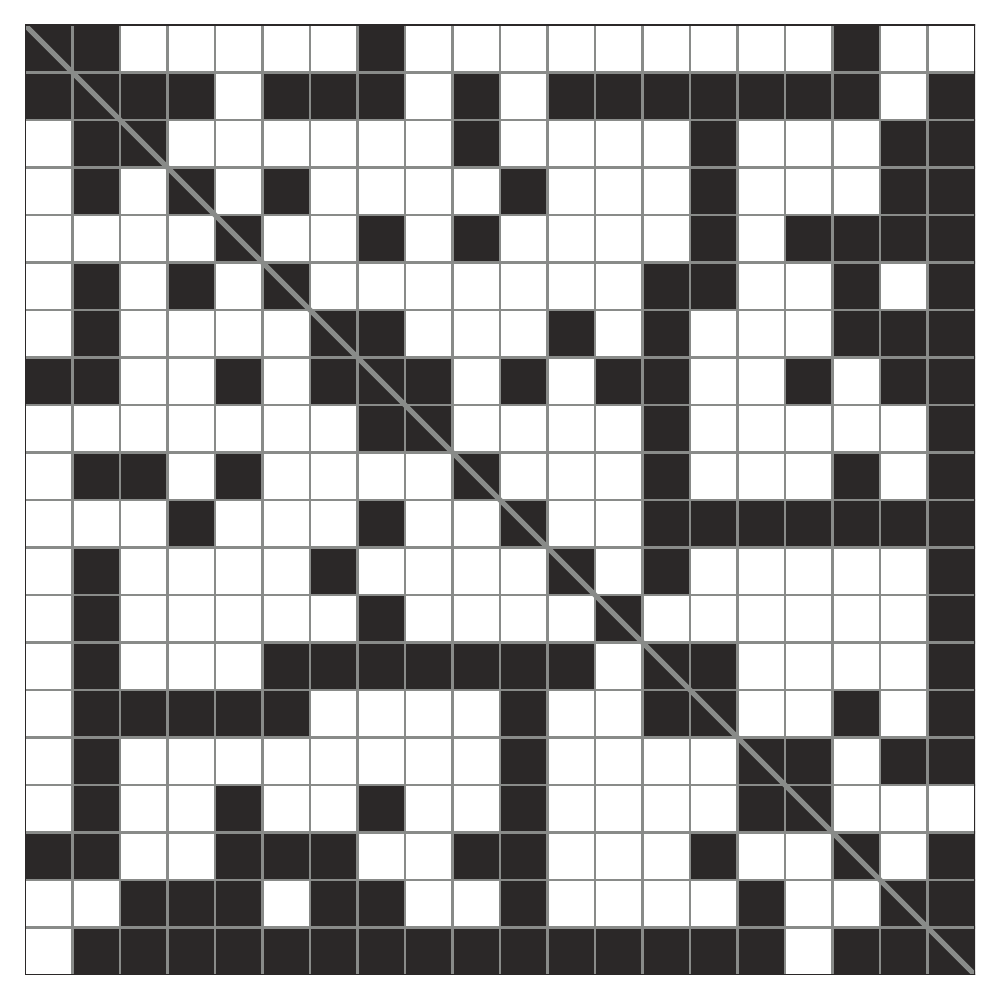}~~\includegraphics[scale=0.25]{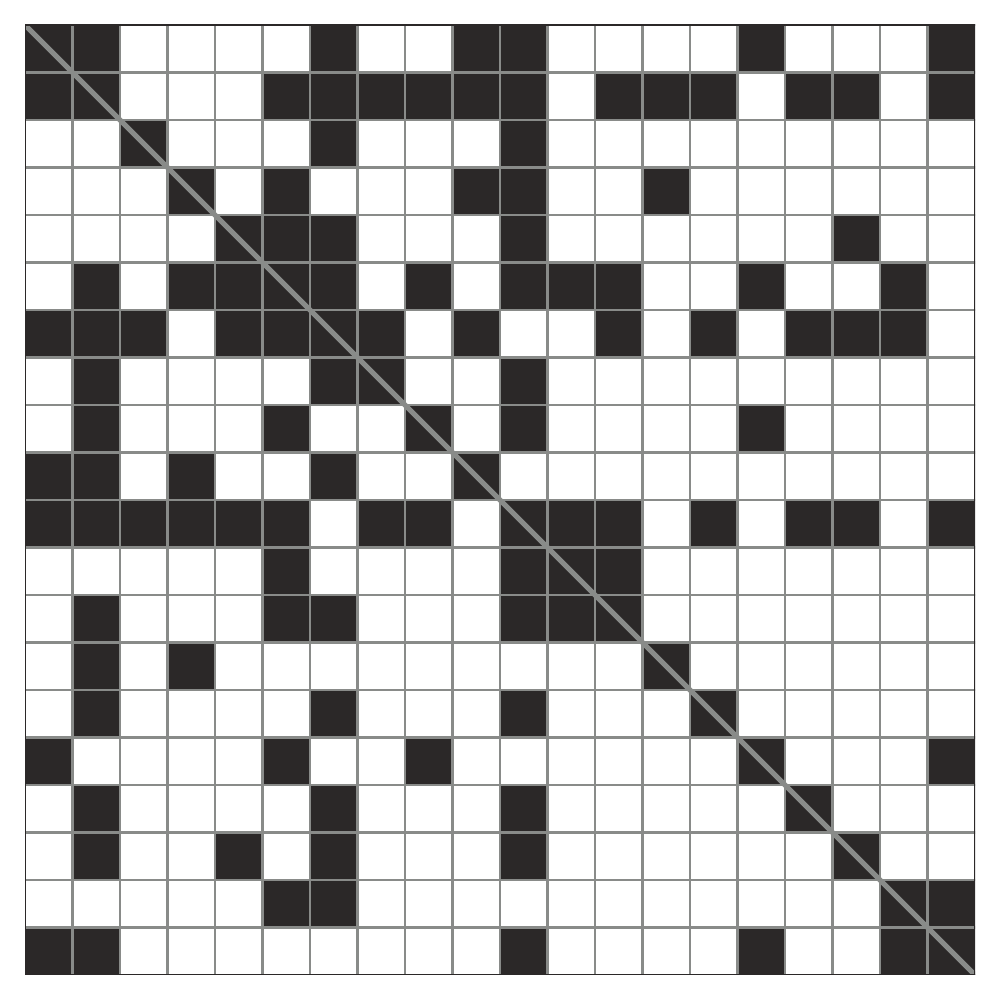}
 \caption{\label{fig:sim:3} Example of simulation setting 3 - Random hubs covariance matrices.}\bigskip
%
\centering
 \includegraphics[scale=0.25]{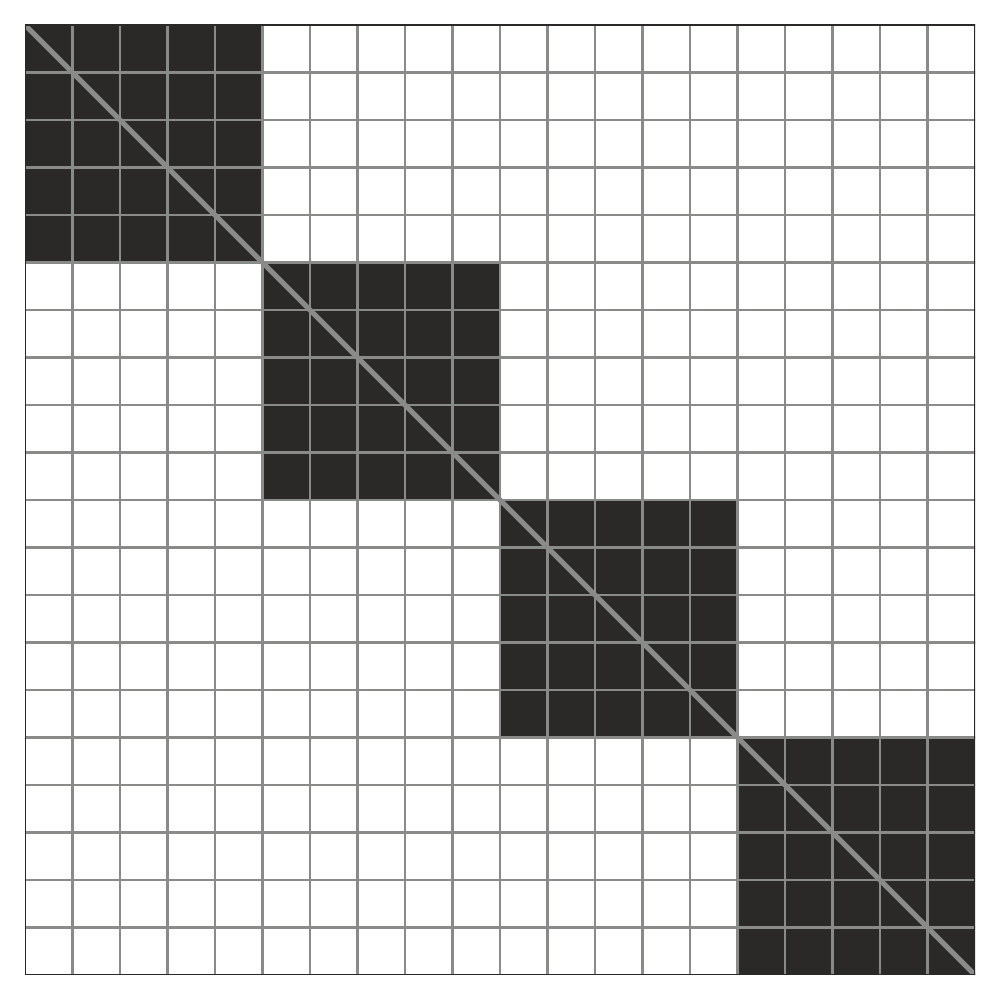}~~\includegraphics[scale=0.25]{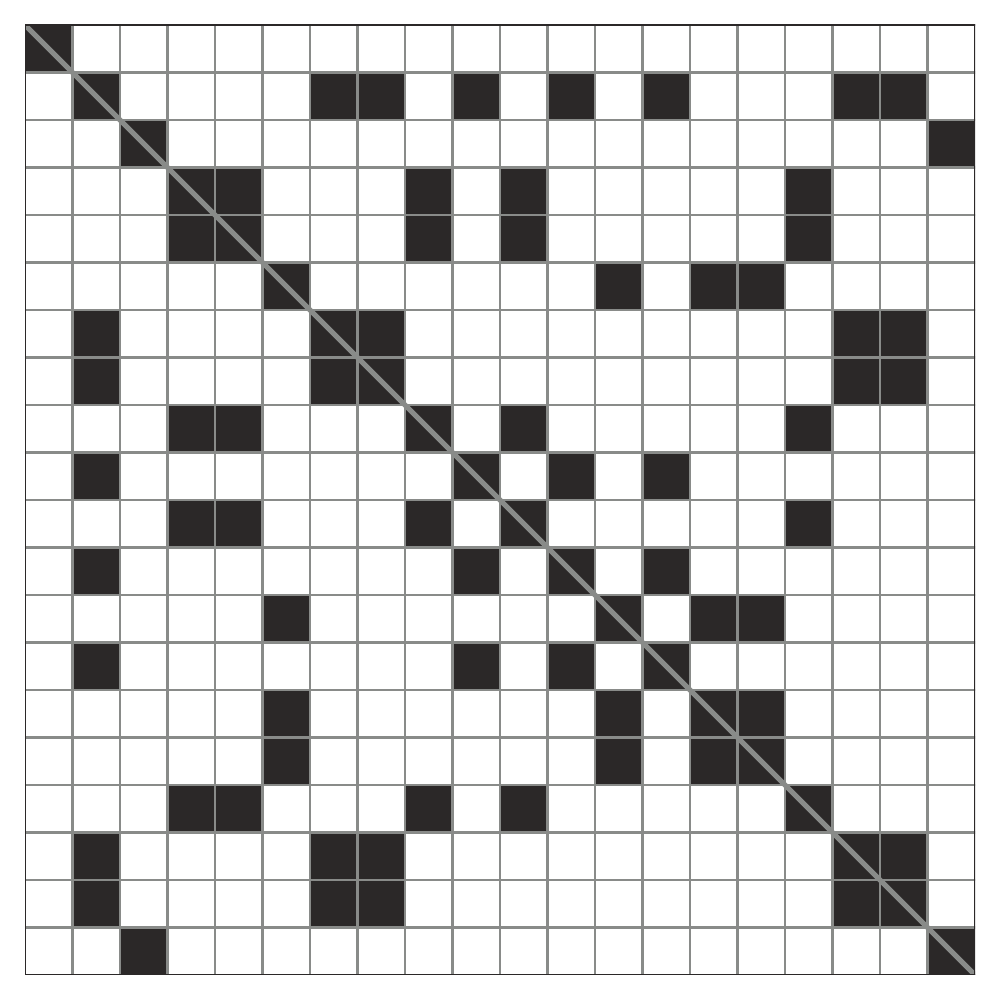}~~\includegraphics[scale=0.25]{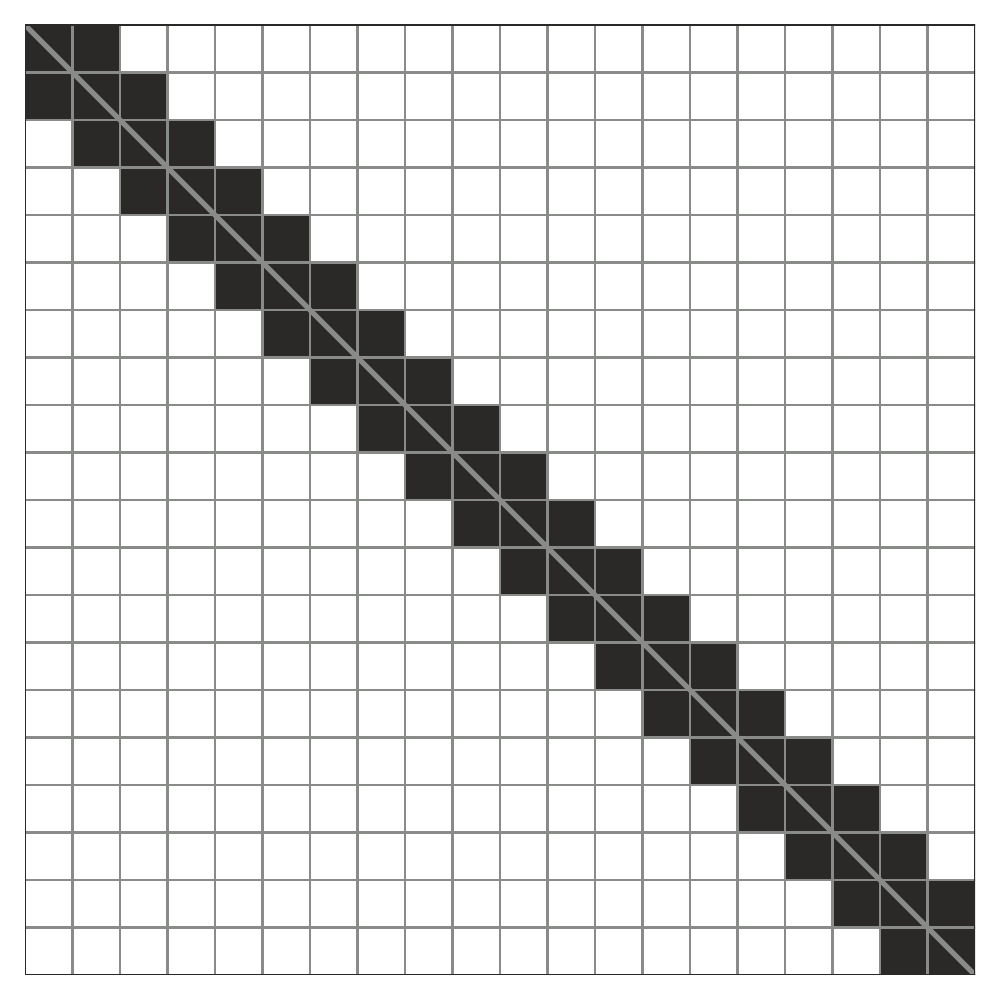}
 \caption{\label{fig:sim:4} Example of simulation setting 4 - Mixed type covariance matrices.}\bigskip
\end{figure}

Examples of the different scenarios are depicted in Figures \ref{fig:sim:1}, \ref{fig:sim:2}, \ref{fig:sim:3} and \ref{fig:sim:4}. In the figures, each large square represents the association structure corresponding to a component covariance matrix. Within each large square, a small black square denotes the presence of an edge between a pair of variables, thus a non-zero covariance term. Appendix~\ref{appendix:c} contains details of the four situations. For each scenario, we simulate from a mixture of $K=3$ multivariate Gaussian distributions with mixing proportions $\bm{\tau} = (0.2, 0.5, 0.3)$. Mean parameters are randomly selected in $(-1,1)$, $(-2,2)$ and $(-3,3)$, respectively.

We will report results concerning BIC, ARI, FPR, FNR, estimated number of clusters and relative computing time with respect to the time taken by \texttt{mclust}. In all cases, we will estimate models considering values of $K$ ranging from 1 to 4. All the experiments are run on a computer cluster with 24 processors Intel Ivybridge E5-2620 @2GHz. Some considerations about computing time evaluation are in Appendix~\ref{appendix:d}.

Different experiments and settings are presented in the following three parts.

\subsection{Part I}
In this section we generate random datasets for different combinations of sample sizes and numbers of variables, $N = (100, 200)$ and $V = (10, 20, 30)$. For every combination of $N$ and $V$ and each scenario we replicate the experiment 100 times. The results are reported in Tables \ref{tab:sim:1}, \ref{tab:sim:2}, \ref{tab:sim:3} and \ref{tab:sim:4}. 

\begin{table*}[]
\centering
\caption{Simulated data \emph{Scenario 1}. The table reports the values of BIC, ARI, FPR, FNR, selected number of clusters and relative time for each method averaged over 100 replicates of the experiment. The relative time is computed with respect to \texttt{mclust}.}
\label{tab:sim:1}
\tiny
\begin{tabular}{@{}lllllllllllllll@{}}
\toprule
        & \multicolumn{6}{c}{$N$ = 100}          &  & \multicolumn{6}{c}{$N$ = 200}         &                     \\[0.2em]
        & \bf BIC  & $K$ & \bf FPR  & \bf FNR  & \bf ARI & Rel. time  &  & \bf BIC  & $K$ & \bf FPR  & \bf FNR  & \bf ARI & Rel. time  &  \\
\midrule
\texttt{mclust}      & -1567 & 3.02 & --- & ---  & 0.99 & 1 &  & -2899 & 3.20 &---  &---  & 0.95 & 1 &  \multirow{9}{*}{$V = 10$}\\[0.5em]   
\texttt{mcgGaBIC}    & \bf -1399 & 2.97 & 0.05 & \bf 0.01 & 0.99 & 56 &  & \bf -2587 & 3.01 & 0.02 & \bf 0.00 & 1.00 & 80 &  \\ 
\texttt{mcgGaEBIC}   & -1453 & 2.97 & \bf 0.01 & 0.28 & 0.99 & 49 &  & -2596 & 3.00 & \bf 0.00 & 0.02 & 1.00 & 68 &  \\ 
\texttt{mcgGaER}     & -1401 & 2.98 & 0.03 & 0.02 & 0.99 & 58 &  & \bf -2587 & 3.00 & 0.02 & \bf 0.00 & 1.00 & 86 &  \\ 
\texttt{mcgGaPL}     & -1417 & 2.97 & \bf 0.00 & 0.07 & 0.99 & 50 &  & -2592 & 3.00 & \bf 0.00 & \bf 0.00 & 1.00 & 76 &  \\[0.5em]  
\texttt{mcgStepBIC}  &  -1400 & 2.97 & 0.05 & \bf 0.01 & 0.99 & 1 &  & \bf -2587 & 3.00 & 0.02 & \bf 0.00 & 1.00 & 2 &  \\ 
\texttt{mcgStepEBIC} & -1435 & 2.97 & \bf 0.01 & 0.16 & 0.99 & 1 &  & -2593 & 3.00 & \bf 0.00 & 0.01 & 1.00 & 2 &  \\ 
\texttt{mcgStepER}   & -1403 & 2.97 & 0.03 & \bf 0.01 & 0.99 & 1 &  & \bf -2587 & 3.00 & 0.02 & \bf 0.00 & 1.00 & 2 &  \\ 
\texttt{mcgStepPL}   & -1417 & 2.97 & 0.01 & 0.06 & 0.99 & 1 &  & -2591 & 3.00 & \bf 0.00 & \bf 0.00 & 1.00 & 2 &  \\
\midrule
\texttt{mclust}    & -3226 & 2.91 & --- & --- & 0.88 & 1 &  & -5046 & 3.12 &---  &---  & 0.97 & 1 & \multirow{9}{*}{$V = 20$} \\[0.5em]    
\texttt{mcgGaBIC}  & -2881 & 2.71 & 0.13 & \bf 0.06 & 0.93 & 162 &  & -4429 & 3.00 & 0.02 & \bf0.00 & 1.00 & 31 &  \\ 
\texttt{mcgGaEBIC} & -3088 & 2.89 & \bf 0.01 & 0.56 & 0.98 & 61 &  & -4634 & 3.00 & \bf0.00 & 0.34 & 1.00 & 16 &  \\ 
\texttt{mcgGaER}   & -2901 & 2.73 & 0.05 & 0.24 & 0.94 & 101 &  & -4444 & 3.00 & \bf0.00 & 0.01 & 1.00 & 26 &  \\ 
\texttt{mcgGaPL}   & -2965 & 2.71 & 0.09 & 0.08 & 0.93 & 117 &  & -4455 & 3.00 & \bf0.00 & \bf0.00 & 1.00 & 22 &  \\[0.5em]    
\texttt{mcgStepBIC}& \bf -2870 & 2.71 & 0.13 & \bf 0.06 & 0.93 & 11 &  & \bf -4418 & 3.00 & 0.03 & \bf0.00 & 1.00 & 2 &  \\ 
\texttt{mcgStepEBIC} & -3064 & 2.92 & \bf 0.01 & 0.54 & 0.98 & 3 &  & -4594 & 3.00 & \bf0.00 & 0.28 & 1.00 & 1 &  \\ 
\texttt{mcgStepER}   & -2877 & 2.76 & 0.05 & 0.21 & 0.95 & 6 &  & -4431 & 3.00 & 0.01 & 0.01 & 1.00 & 1 &  \\ 
\texttt{mcgStepPL}   & -2917 & 2.77 & 0.06 & 0.07 & 0.95 & 5 &  & -4448 & 3.00 & \bf0.00 & \bf0.00 & 1.00 & 1 &  \\
\midrule
\texttt{mclust}   & -5212 & 3.55 &  --- &---  & 0.75 & 1 &  & \bf -7405 & 2.99 &---  &---  & 0.99 & 1 &  \multirow{9}{*}{$V = 30$}\\[0.5em]    
\texttt{mcgGaBIC} & -5764 & 2.72 & 0.31 & 0.10 & 0.92 & 1595 &  & -7792 & 3.00 & 0.09 & 0.04 & 1.00 & 131 &  \\ 
\texttt{mcgGaEBIC}& -5048 & 3.00 & \bf0.01 & 0.73 & 1.00 & 361 &  & -7864 & 3.00 & \bf0.00 & 0.65 & 1.00 & 69 &  \\ 
\texttt{mcgGaER}  & -5047 & 3.00 & 0.05 & 0.47 & 1.00 & 693 &  & -7628 & 3.00 & 0.02 & 0.29 & 1.00 & 80 &  \\ 
\texttt{mcgGaPL}  & -6364 & 1.22 & 0.81 & 0.02 & 0.16 & 1685 &  & -7952 & 3.00 & 0.07 & \bf0.01 & 1.00 & 119 &  \\[0.5em]    
\texttt{mcgStepBIC}  & -5565 & 2.84 & 0.25 & 0.10 & 0.96 & 242 &  & -7751 & 3.00 & 0.10 & 0.04 & 1.00 & 18 &  \\ 
\texttt{mcgStepEBIC} & -5031 & 3.00 & \bf0.01 & 0.74 & 1.00 & 38 &  & -7845 & 3.00 & \bf0.00 & 0.65 & 1.00 & 5 &  \\ 
\texttt{mcgStepER}  & \bf -5012 & 3.00 & 0.06 & 0.46 & 1.00 & 92 &  & -7583 & 3.00 & 0.03 & 0.29 & 1.00 & 10 &  \\ 
\texttt{mcgStepPL}  & -6414 & 1.35 & 0.81 & \bf0.00 & 0.25 & 15 &  & -7664 & 3.00 & 0.03 & 0.03 & 1.00 & 8 &  \\
\bottomrule
\end{tabular}\bigskip\bigskip
\caption{Simulated data \emph{Scenario 2}. The table reports the values of BIC, ARI, FPR, FNR, selected number of clusters and relative time for each method averaged over 100 replicates of the experiment. The relative time is computed with respect to \texttt{mclust}.}
\label{tab:sim:2}
\begin{tabular}{@{}lllllllllllllll@{}}
\toprule
        & \multicolumn{6}{c}{$N$ = 100}          &  & \multicolumn{6}{c}{$N$ = 200}         &                     \\[0.2em]
        & \bf BIC  & $K$ & \bf FPR  & \bf FNR  & \bf ARI & Rel. time  & & \bf BIC &$K$ & \bf FPR  & \bf FNR  & \bf ARI & Rel. time  &  \\
\midrule
\texttt{mclust}      & -2678 & 3.02 &---  &---  & 1.00 & 1 & & -5066 & 3.00 &---  &---  & 1.00 & 1 & \multirow{9}{*}{$V = 10$} \\[0.5em]   
\texttt{mcgGaBIC}    &   \bf -2441 & 3.00 & 0.06 & \bf 0.03 & 1.00 & 85 & & \bf -4673 & 3.00 & 0.04 & \bf 0.01 & 1.00 & 84 &  \\ 
\texttt{mcgGaEBIC}   &   -2491 & 3.00 & \bf 0.00 & 0.23 & 1.00 & 67 & & -4699 & 3.00 & \bf 0.01 & 0.07 & 1.00 & 81 &  \\ 
\texttt{mcgGaER}     &   -2443 & 3.00 & 0.04 & 0.04 & 1.00 & 79 & & \bf -4673 & 3.00 & 0.04 & \bf 0.01 & 1.00 & 87 &  \\ 
\texttt{mcgGaPL}     &   -2476 & 3.00 & 0.02 & 0.20 & 1.00 & 65 & & -4681 & 3.00 & 0.02 & 0.03 & 1.00 & 79 &  \\[0.5em]    
\texttt{mcgStepBIC}  &   -2445 & 3.00 & 0.06 & \bf 0.03 & 1.00 & 2 & & -4676 & 3.00 & 0.05 & \bf 0.01 & 1.00 & 2 &  \\ 
\texttt{mcgStepEBIC} &   -2492 & 3.00 & 0.01 & 0.22 & 1.00 & 1 & & -4699 & 3.00 & \bf 0.01 & 0.06 & 1.00 & 2 &  \\ 
\texttt{mcgStepER}   &   -2447 & 3.01 & 0.04 & 0.04 & 1.00 & 2 & & -4677 & 3.00 & 0.04 & \bf 0.01 & 1.00 & 2 &  \\ 
\texttt{mcgStepPL}   &   -2479 & 3.00 & 0.02 & 0.21 & 1.00 & 1 & & -4684 & 3.00 & 0.02 & 0.04 & 1.00 & 2 &  \\
\midrule
\texttt{mclust}      & -4922 & 2.59 &---  &---  & 0.90 & 1 & & -8490 & 3.03 &---  &---  & 0.99 & 1 &  \multirow{9}{*}{$V = 20$}\\[0.5em]   
\texttt{mcgGaBIC}    &   -4477 & 2.69 & 0.21 & \bf 0.08 & 0.93 & 124 & & -7503 & 3.00 & 0.04 & \bf 0.02 & 1.00 & 57 &  \\ 
\texttt{mcgGaEBIC}   &   -4572 & 2.99 & \bf 0.01 & 0.37 & 1.00 & 55 & & -7749 & 3.00 & \bf 0.00 & 0.23 & 1.00 & 43 &  \\ 
\texttt{mcgGaER}     &   -4439 & 2.92 & 0.07 & 0.16 & 0.98 & 96 & & -7542 & 3.00 & 0.01 & 0.06 & 1.00 & 49 &  \\ 
\texttt{mcgGaPL}     &   -4623 & 2.59 & 0.28 & 0.15 & 0.90 & 118 & & -7542 & 3.00 & 0.05 & 0.06 & 1.00 & 46 &  \\[0.5em]    
\texttt{mcgStepBIC}  &   -4458 & 2.71 & 0.19 & \bf 0.08 & 0.94 & 7 & & \bf -7495 & 3.00 & 0.04 & \bf 0.02 & 1.00 & 2 &  \\ 
\texttt{mcgStepEBIC} &   -4550 & 3.00 & \bf 0.01 & 0.36 & 1.00 & 3 & & -7734 & 3.00 & \bf 0.00 & 0.22 & 1.00 & 2 &  \\ 
\texttt{mcgStepER}   &   \bf -4429 & 2.90 & 0.07 & 0.17 & 0.98 & 6 & & -7535 & 3.00 & 0.02 & 0.05 & 1.00 & 2 &  \\ 
\texttt{mcgStepPL}   &   -4700 & 2.59 & 0.31 & 0.15 & 0.89 & 3 & & -7535 & 3.00 & 0.05 & 0.06 & 1.00 & 2 &  \\
\midrule
\texttt{mclust}      & -7488 & 2.69 &---  &---  & 0.93 & 1 & & -11712 & 2.95 &---  &---  & 0.99 & 1 &  \multirow{9}{*}{$V = 30$}\\[0.5em]    
\texttt{mcgGaBIC}    &   -7613 & 2.14 & 0.49 & 0.08 & 0.65 & 4075 & & -10387 & 2.99 & 0.05 & \bf 0.05 & 1.00 & 663 &  \\ 
\texttt{mcgGaEBIC}   &   -7370 & 2.93 & \bf 0.02 & 0.36 & 0.99 & 1098 & & -11058 & 3.00 & \bf 0.00 & 0.41 & 1.00 & 155 &  \\ 
\texttt{mcgGaER}     &   -7162 & 2.79 & 0.11 & 0.18 & 0.96 & 2435 & & -10553 & 3.00 & 0.01 & 0.20 & 1.00 & 306 &  \\ 
\texttt{mcgGaPL}     &   -7844 & 1.31 & 0.80 & 0.03 & 0.24 & 4499 & & -10642 & 2.95 & 0.14 & 0.07 & 0.99 & 554 &  \\[0.5em]     
\texttt{mcgStepBIC}  &   -7455 & 2.50 & 0.38 & 0.08 & 0.83 & 437 & & \bf -10335 & 2.99 & 0.06 & 0.06 & 1.00 & 49 &  \\
\texttt{mcgStepEBIC} &   -7317 & 2.86 & \bf 0.02 & 0.38 & 0.97 & 84 & & -10987 & 3.00 & \bf 0.00 & 0.41 & 1.00 & 14 &  \\ 
\texttt{mcgStepER}   &   \bf -7093 & 2.76 & 0.10 & 0.20 & 0.95 & 270 & & -10489 & 3.00 & 0.01 & 0.20 & 1.00 & 28 &  \\ 
\texttt{mcgStepPL}   &   -7964 & 1.30 & 0.85 & \bf 0.01 & 0.23 & 0 & & -10725 & 2.93 & 0.17 & 0.08 & 0.98 & 8 &  \\
\bottomrule
\end{tabular}
\end{table*}

\begin{table*}[]
\centering
\caption{Simulated data \emph{Scenario 3}. The table reports the values of BIC, ARI, FPR, FNR, selected number of clusters and relative time for each method averaged over 100 replicates of the experiment. The relative time is computed with respect to \texttt{mclust}.}
\label{tab:sim:3}
\tiny
\begin{tabular}{@{}lllllllllllllll@{}}
\toprule
        & \multicolumn{6}{c}{$N$ = 100}          &  & \multicolumn{6}{c}{$N$ = 200}         &                     \\[0.2em]
        & \bf BIC  & $K$ & \bf FPR  & \bf FNR  & \bf ARI & Rel. time  &  & \bf BIC  & $K$ & \bf FPR  & \bf FNR  & \bf ARI & Rel. time  &  \\
\midrule
\texttt{mclust}      & -1565 & 3.00 &---  &---  & 1.00 & 1 &  & -2796 & 3.01 &---  &---  & 1.00 & 1 & \multirow{9}{*}{$V = 10$} \\[0.5em]     
\texttt{mcgGaBIC}    &   \bf -1352 & 3.00 & 0.05 & \bf 0.36 & 1.00 & 83 &  & \bf -2448 & 3.00 & 0.03 & \bf 0.27 & 1.00 & 91 &  \\ 
\texttt{mcgGaEBIC}   &   -1432 & 3.00 & \bf 0.02 & 0.67 & 1.00 & 78 &  & -2506 & 3.00 & \bf 0.01 & 0.49 & 1.00 & 85 &  \\ 
\texttt{mcgGaER}     &   -1355 & 3.01 & 0.03 & 0.41 & 1.00 & 83 &  & -2449 & 3.00 & 0.02 & 0.29 & 1.00 & 96 &  \\ 
\texttt{mcgGaPL}     &   -1410 & 3.00 & 0.04 & 0.58 & 1.00 & 68 &  & -2480 & 3.00 & 0.02 & 0.42 & 1.00 & 75 &  \\[0.5em]     
\texttt{mcgStepBIC}  &   -1358 & 3.00 & 0.06 & \bf 0.36 & 1.00 & 2 &  & -2453 & 3.00 & 0.04 & \bf 0.27 & 1.00 & 3 &  \\ 
\texttt{mcgStepEBIC} &   -1430 & 3.00 & \bf 0.02 & 0.65 & 1.00 & 2 &  & -2507 & 3.00 & \bf 0.01 & 0.48 & 1.00 & 2 &  \\ 
\texttt{mcgStepER}   &   -1362 & 3.00 & 0.03 & 0.41 & 1.00 & 2 &  & -2452 & 3.00 & 0.03 & 0.29 & 1.00 & 3 &  \\ 
\texttt{mcgStepPL}   &   -1416 & 3.00 & 0.04 & 0.57 & 1.00 & 2 &  & -2483 & 3.00 & 0.02 & 0.41 & 1.00 & 2 &  \\
\midrule
\texttt{mclust}      & -3150 & 2.74 &---  &---  & 0.92 & 1 &  & -5165 & 3.00 &---  &---  & 1.00 & 1 &  \multirow{9}{*}{$V = 20$}\\ [0.5em]    
\texttt{mcgGaBIC}    &   -2634 & 2.69 & 0.18 & \bf 0.46 & 0.93 & 129 &  & -3966 & 3.00 & 0.03 & \bf 0.51 & 1.00 & 44 &  \\ 
\texttt{mcgGaEBIC}   &   -2756 & 2.99 & \bf 0.01 & 0.79 & 1.00 & 72 &  & -4284 & 3.00 & \bf 0.00 & 0.77 & 1.00 & 40 &  \\ 
\texttt{mcgGaER}     &   -2621 & 2.79 & 0.06 & 0.63 & 0.96 & 115 &  & -4012 & 3.00 & 0.01 & 0.60 & 1.00 & 39 &  \\ 
\texttt{mcgGaPL}     &   -2853 & 2.69 & 0.23 & 0.51 & 0.93 & 142 &  & -4070 & 3.00 & 0.01 & 0.62 & 1.00 & 34 &  \\[0.5em]     
\texttt{mcgStepBIC}  &   -2629 & 2.71 & 0.17 & 0.48 & 0.94 & 11 &  & \bf -3958 & 3.00 & 0.03 & \bf 0.51 & 1.00 & 3 &  \\ 
\texttt{mcgStepEBIC} &   -2738 & 2.96 & \bf 0.01 & 0.79 & 0.99 & 6 &  & -4267 & 3.00 & \bf 0.00 & 0.76 & 1.00 & 2 &  \\ 
\texttt{mcgStepER}   &   \bf -2611 & 2.79 & 0.06 & 0.64 & 0.96 & 9 &  & -4007 & 3.00 & 0.01 & 0.60 & 1.00 & 3 &  \\ 
\texttt{mcgStepPL}   &   -2860 & 2.69 & 0.23 & 0.51 & 0.93 & 7 &  & -4061 & 3.00 & 0.01 & 0.61 & 1.00 & 2 &  \\
\midrule
\texttt{mclust}      & -4949 & 2.97 &---  &---  & 0.83 & 1 &  & -7450 & 2.97 &---  &---  & 0.99 & 1 &  \multirow{9}{*}{$V = 30$} \\[0.5em]    
\texttt{mcgGaBIC}    &   -4752 & 2.55 & 0.31 & 0.45 & 0.89 & 1833 &  & -5192 & 2.99 & 0.03 & \bf 0.63 & 1.00 & 79 &  \\ 
\texttt{mcgGaEBIC}   &   -4672 & 2.74 & \bf 0.02 & 0.83 & 0.95 & 453 &  & -5816 & 3.00 & \bf 0.00 & 0.86 & 1.00 & 42 &  \\ 
\texttt{mcgGaER}     &   -4428 & 2.62 & 0.08 & 0.71 & 0.92 & 1025 &  & -5365 & 3.00 & \bf 0.00 & 0.75 & 1.00 & 60 &  \\ 
\texttt{mcgGaPL}     &   -5262 & 1.41 & 0.71 & 0.19 & 0.32 & 2122 &  & -5407 & 2.97 & 0.03 & 0.71 & 0.99 & 65 &  \\[0.5em]     
\texttt{mcgStepBIC}  &   -4625 & 2.59 & 0.28 & 0.48 & 0.91 & 307 &  & \bf -5159 & 2.99 & 0.04 & 0.64 & 1.00 & 9 &  \\ 
\texttt{mcgStepEBIC} &   -4601 & 2.71 & \bf 0.02 & 0.83 & 0.94 & 57 &  & -5738 & 3.00 & \bf 0.00 & 0.85 & 1.00 & 4 &  \\ 
\texttt{mcgStepER}   &   \bf -4353 & 2.60 & 0.07 & 0.72 & 0.92 & 123 &  & -5328 & 3.00 & 0.01 & 0.75 & 1.00 & 9 &  \\ 
\texttt{mcgStepPL}   &   -5360 & 1.41 & 0.75 & \bf 0.16 & 0.32 & 4 &  & -5387 & 2.97 & 0.03 & 0.71 & 0.99 & 3 &  \\
\bottomrule
\end{tabular}\bigskip\bigskip
\caption{Simulated data \emph{Scenario 4}. The table reports the values of BIC, ARI, FPR, FNR, selected number of clusters and relative time for each method averaged over 100 replicates of the experiment. The relative time is computed with respect to \texttt{mclust}.}
\label{tab:sim:4}
\begin{tabular}{@{}lllllllllllllll@{}}
\toprule
        & \multicolumn{6}{c}{$N$ = 100}          &  & \multicolumn{6}{c}{$N$ = 200}         &                     \\[0.2em]
        & \bf BIC  & $K$ & \bf FPR  & \bf FNR  & \bf ARI & Rel. time  &  & \bf BIC  & $K$ & \bf FPR  & \bf FNR  & \bf ARI & Rel. time  &  \\
\midrule
\texttt{mclust}      & -1362 & 3.02 & --- & --- & 0.99 & 1 &  & -2452 & 3.04 & --- & --- & 0.99 & 1 &  \multirow{9}{*}{$V = 10$}\\[0.5em]
\texttt{mcgGaBIC}    &   \bf -1184 & 3.02 & 0.06 & 0.06 & 1.00 & 79 &  & \bf -2125 & 3.00 & 0.03 & \bf 0.01 & 1.00 & 83 &  \\ 
\texttt{mcgGaEBIC}   &   -1271 & 3.00 & 0.06 & 0.3 & 1.00 & 68 &  & -2143 & 3.00 & \bf 0.01 & 0.05 & 1.00 & 67 &  \\ 
\texttt{mcgGaER}     &   -1186 & 2.99 & \bf 0.04 & 0.06 & 1.00 & 79 &  & \bf -2125 & 3.00 & 0.03 & \bf 0.01 & 1.00 & 85 &  \\ 
\texttt{mcgGaPL}     &   -1248 & 2.99 & 0.13 & 0.2 & 1.00 & 70 &  & -2152 & 3.01 & 0.04 & 0.07 & 1.00 & 53 &  \\[0.5em]
\texttt{mcgStepBIC}  &   -1188 & 2.99 & 0.07 & \bf 0.05 & 1.00 & 2 &  & -2134 & 3.00 & 0.05 & \bf 0.01 & 1.00 & 2 &  \\ 
\texttt{mcgStepEBIC} &   -1259 & 2.99 & 0.07 & 0.25 & 1.00 & 2 &  & -2153 & 3.00 & \bf 0.01 & 0.07 & 1.00 & 2 &  \\ 
\texttt{mcgStepER}   &   -1191 & 3.01 & 0.05 & 0.07 & 1.00 & 2 &  & -2133 & 3.00 & 0.05 & \bf 0.01 & 1.00 & 2 &  \\ 
\texttt{mcgStepPL}   &   -1254 & 2.99 & 0.14 & 0.21 & 1.00 & 2 &  & -2170 & 3.00 & 0.06 & 0.09 & 1.00 & 2 &  \\
\midrule
\texttt{mclust}      & -2748 & 2.68 & --- & --- & 0.92 & 1 &  & -4128 & 3.02 & --- & --- & 0.99 & 1 & \multirow{9}{*}{$V = 20$} \\[0.5em]
\texttt{mcgGaBIC}    &   -2225 & 2.71 & 0.18 & \bf 0.06 & 0.93 & 140 &  & -3031 & 3.00 & 0.04 & \bf 0.01 & 1.00 & 30 &  \\ 
\texttt{mcgGaEBIC}   &   -2384 & 2.96 & \bf 0.03 & 0.33 & 0.99 & 85 &  & -3185 & 3.00 & 0.01 & 0.13 & 1.00 & 22 &  \\ 
\texttt{mcgGaER}     &   -2211 & 2.85 & 0.07 & 0.11 & 0.97 & 121 &  & -3049 & 3.00 & 0.01 & 0.02 & 1.00 & 25 &  \\ 
\texttt{mcgGaPL}     &   -2449 & 2.69 & 0.30 & 0.14 & 0.93 & 144 &  & -3132 & 3.00 & 0.05 & 0.09 & 1.00 & 22 &  \\[0.5em]
\texttt{mcgStepBIC}  &   -2221 & 2.72 & 0.17 & \bf 0.06 & 0.94 & 9 &  & \bf -3022 & 3.00 & 0.03 & \bf 0.01 & 1.00 & 1 &  \\ 
\texttt{mcgStepEBIC} &   -2363 & 2.95 & \bf 0.03 & 0.31 & 0.99 & 5 &  & -3173 & 3.00 & \bf 0.00 & 0.12 & 1.00 & 1 &  \\ 
\texttt{mcgStepER}   &   \bf -2208 & 2.84 & 0.07 & 0.11 & 0.97 & 10 &  & -3037 & 3.00 & 0.01 & \bf 0.01 & 1.00 & 1 &  \\ 
\texttt{mcgStepPL}   &   -2483 & 2.74 & 0.32 & 0.12 & 0.94 & 6 &  & -3139 & 3.00 & 0.06 & 0.09 & 1.00 & 1	 &  \\
\midrule
\texttt{mclust}      & -4565 & 2.61 & --- & --- & 0.91 & 1 &  & -6075 & 2.96 & --- & --- & 0.99 & 1 &  \multirow{9}{*}{$V = 30$}\\[0.5em]
\texttt{mcgGaBIC}    &   -4263 & 2.59 & 0.32 & 0.06 & 0.90 & 2597 &  & -4082 & 2.99 & 0.05 & \bf 0.01 & 1.00 & 171 &  \\ 
\texttt{mcgGaEBIC}   &   -4252 & 2.81 & \bf 0.02 & 0.3 & 0.96 & 723 &  & -4680 & 3.00 & \bf 0.00 & 0.28 & 1.00 & 92 &  \\ 
\texttt{mcgGaER}     &   -3953 & 2.62 & 0.12 & 0.18 & 0.92 & 1761 &  & -4152 & 3.00 & 0.01 & 0.03 & 1.00 & 107 &  \\ 
\texttt{mcgGaPL}     &   -4786 & 1.39 & 0.77 & 0.04 & 0.30 & 2907 &  & -4380 & 2.97 & 0.08 & 0.13 & 0.99 & 165 &  \\[0.5em] 
\texttt{mcgStepBIC}  &   -4163 & 2.59 & 0.30 & 0.06 & 0.90 & 343 &  & \bf -4032 & 2.98 & 0.05 & \bf 0.01 & 1.00 & 20 &  \\ 
\texttt{mcgStepEBIC} &   -4176 & 2.86 & 0.03 & 0.27 & 0.97 & 72 &  & -4584 & 3.00 & \bf 0.00 & 0.24 & 1.00 & 7 &  \\ 
\texttt{mcgStepER}   &   \bf -3902 & 2.62 & 0.13 & 0.18 & 0.92 & 279 &  & -4097 & 3.00 & 0.01 & 0.03 & 1.00 & 13 &  \\ 
\texttt{mcgStepPL}   &   -4872 & 1.39 & 0.8 & \bf 0.01 & 0.30 & 4 &  & -4388 & 2.96 & 0.09 & 0.14 & 0.99 & 7 &  \\
\bottomrule
\end{tabular}
\end{table*}

For sample size $N=100$, as the number of variables increases, the mixture of covariance graph models with different penalization terms tends to outperform \texttt{mclust}, both in terms of classification and identification of the correct number of clusters, and also in terms of BIC. Nevertheless, models with \texttt{PL} penalty perform surprisingly badly in all the simulation settings for $V=30$. The fact suggests that the power law penalty function may be particularly sensitive to the tuning parameter if $N$ is not decisively larger than $V$. For sample size $N = 200$, all the methods tend to attain an almost perfect classification of the data and consistently select the correct number of groups. However, compared to the \texttt{mclust} results, the BIC for the mixture of Gaussian covariance models is higher on average. In particular, models \texttt{ER} are on average almost always preferred to the others in terms of BIC, for both sample sizes and different dimensions. Regardless of the covariance eigen-decomposition, \texttt{mclust} can estimate either full covariance matrices or diagonal ones and is not capable of recovering the underlying association structure within the clusters. With respect to the ability of selecting the correct graph structures, the EBIC-type penalty tends to select very sparse graphs, while the power law favors denser graphs, especially for sample size equal to 100. Moreover, the BIC-type penalization selects graph with spurious associations in some cases. Models estimated using the Erd\H{o}s-R\'enyi penalty function outperforms the others on average in terms of dependence structure detection. Note that in \emph{Scenario 3} it is particularly difficult to infer the underlying within group correlation structures, and all the models estimated by the different penalizations have an high mean false negative rate. In fact, the method used to simulate the sparse covariance matrices often downweighs some of the covariance terms even when the variables are connected in the corresponding graph (see Appendix~\ref{appendix:c}). Overall, sparse covariance mixture models \texttt{ER} and \texttt{BIC} with Erd\H{o}s-R\'enyi and BIC-type penalty have on average the better performance in terms of classification, graph structure detection and BIC. 

For the sparse covariance models \texttt{mgc}, in all situations the stepwise search \texttt{Step} is remarkably faster than the evolutionary search \texttt{Ga}, especially as the number of variables increases. Surprisingly, on average the stepwise search tends also to provide models of slightly better quality in terms of BIC and accuracy in graph structure detection, although in general the results between \texttt{Step} and \texttt{Ga} are quite similar. We attribute the inferior performance of \texttt{Ga} search to the fact of setting the maximum number of iterations equal to 100, while in higher dimension a larger value would have been beneficial, but at an additional computational cost.

\subsection{Part II}

\begin{figure}[!tb]
\centering
\subfloat[][\emph{Scenario 1}.]
{\includegraphics[scale=0.43]{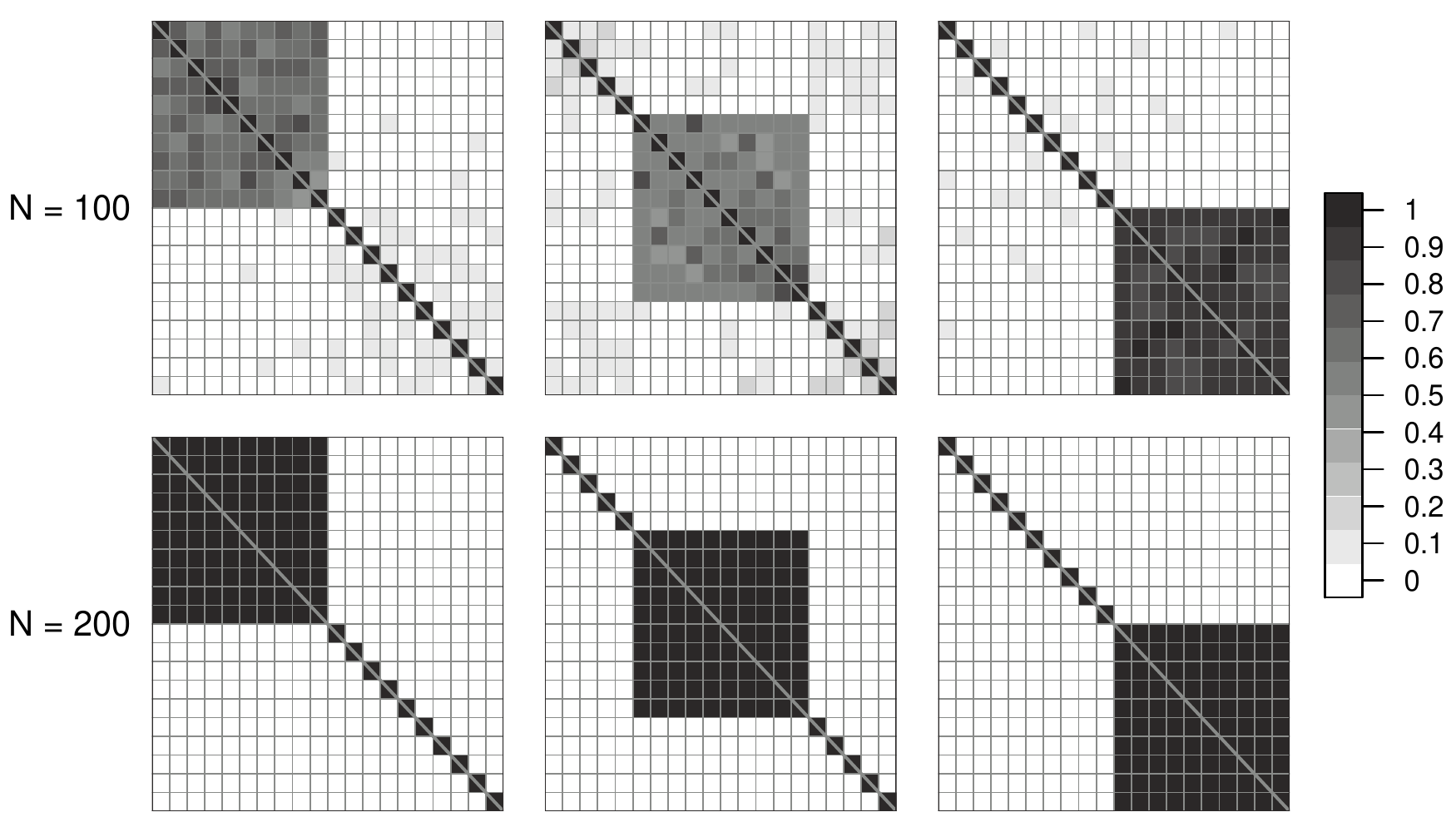}} \quad
\subfloat[][\emph{Scenario 2}.]
{\includegraphics[scale=0.43]{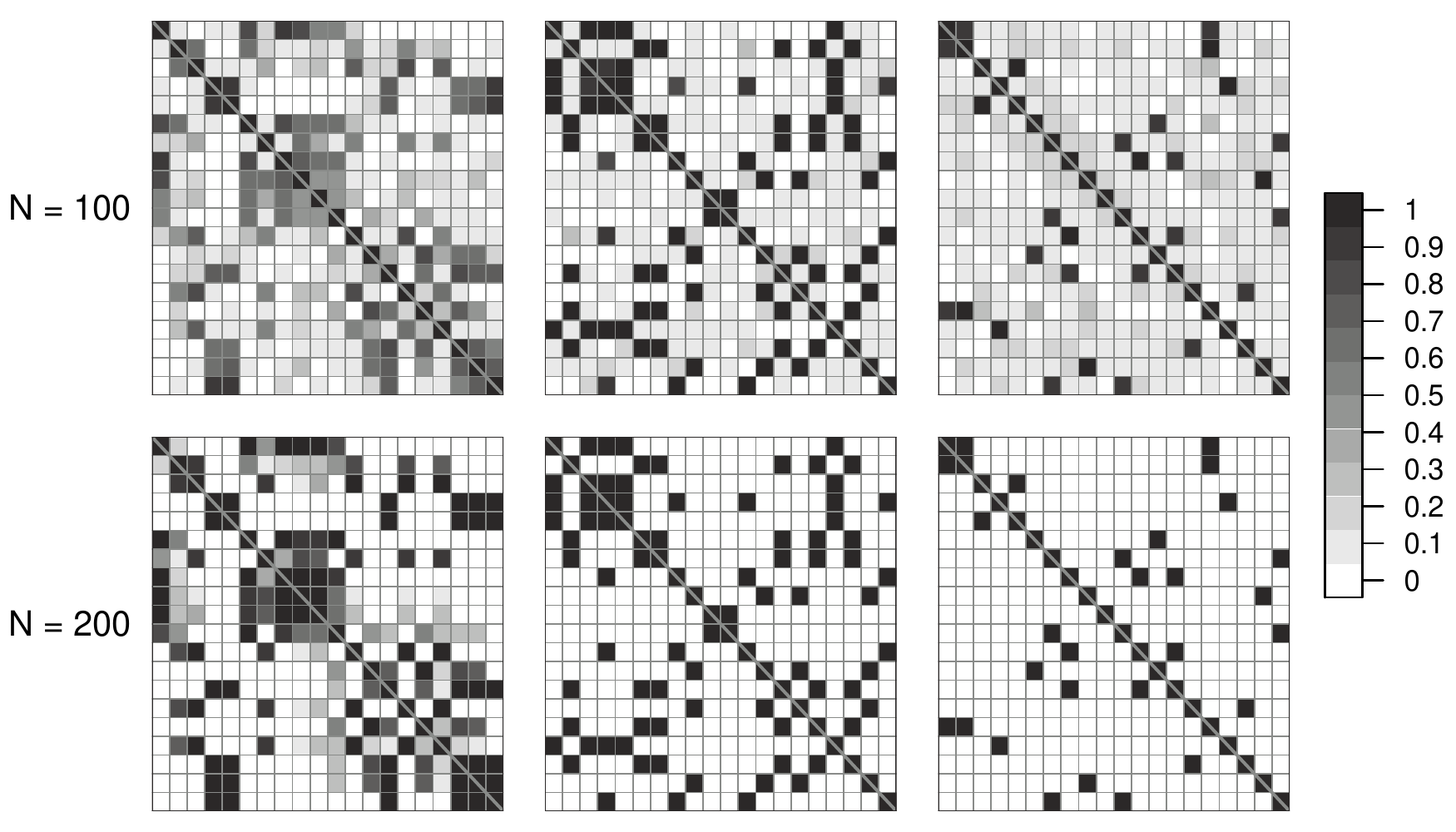}} \\
\subfloat[][\emph{Scenario 3}.]
{\includegraphics[scale=0.43]{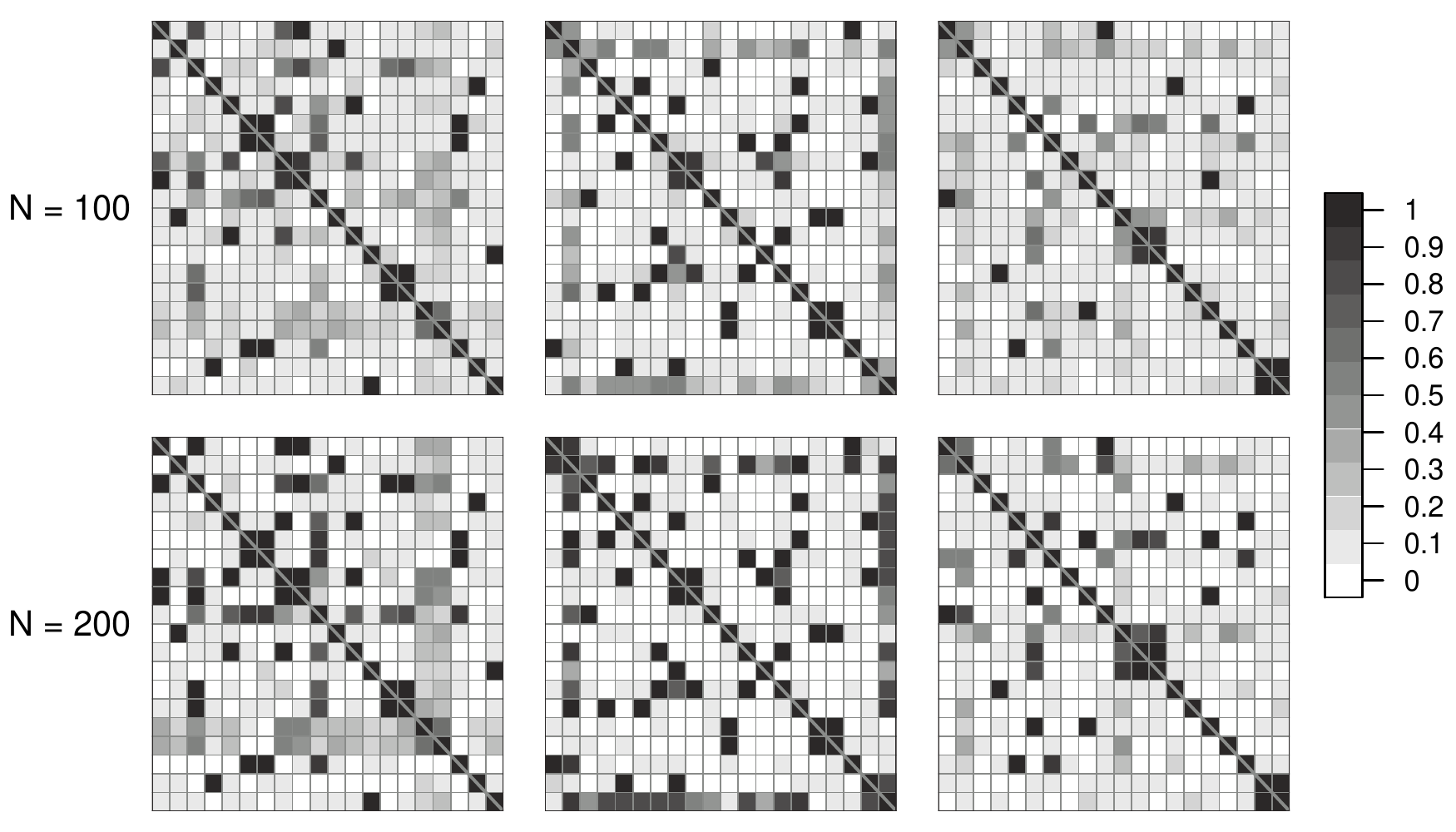}} \quad
\subfloat[][\emph{Scenario 4}.]
{\includegraphics[scale=0.43]{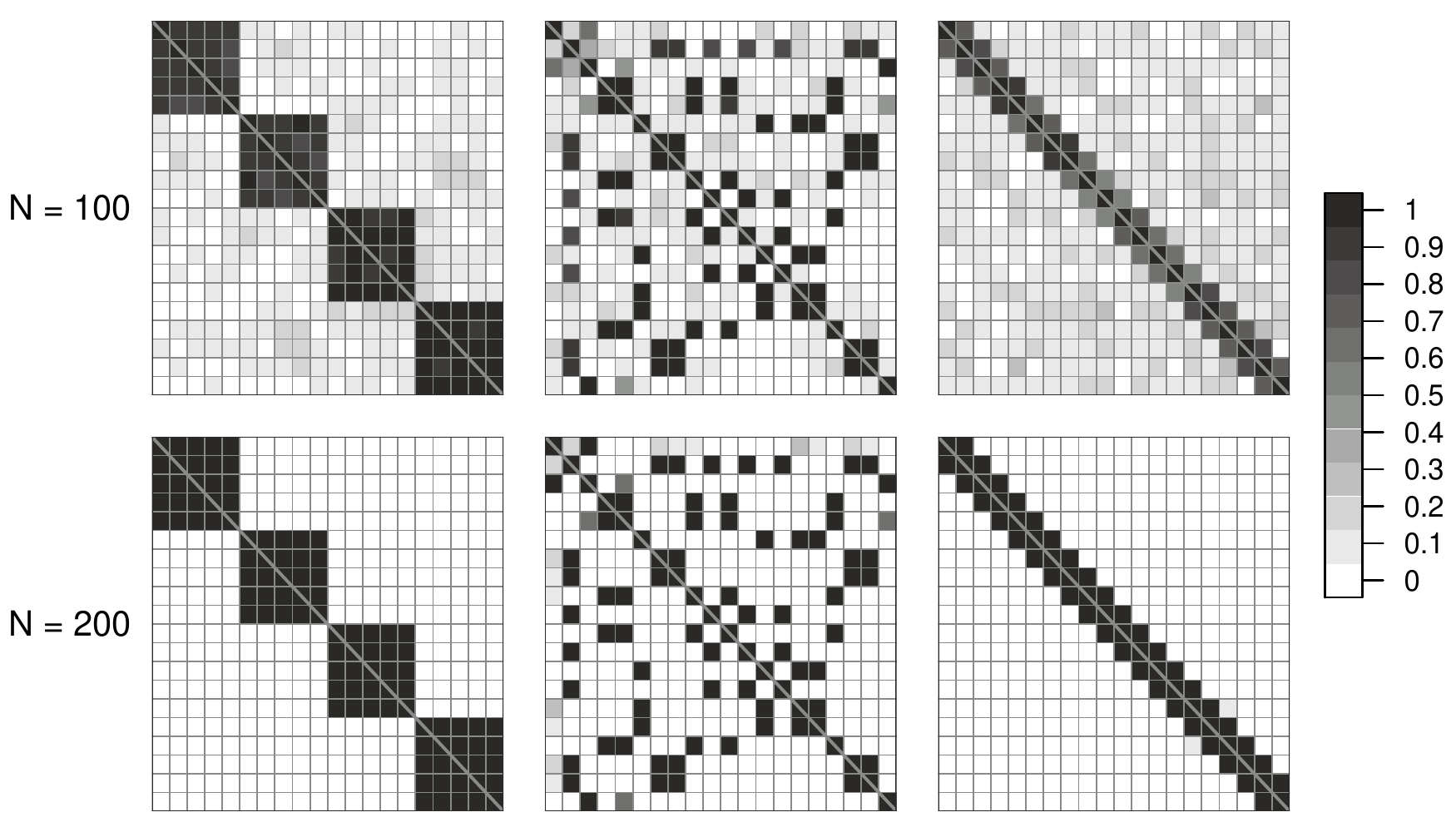}}
\caption{ Heat-map plot of the proportion of times an arc has been estimated between a pair of variables for the simulation setting examples as in Figures~\ref{fig:sim:1} (a), \ref{fig:sim:2} (b), \ref{fig:sim:3} (c), \ref{fig:sim:4} (d)}
\label{fig:subfig}
\end{figure}

In this section we set up another simulated data experiment in order to further investigate the ability of the Erd\H{o}s-R\'enyi penalty function to recover the underlying graph structure. Using the same parameters as before, we generate data according to the association structures depicted in the Figures \ref{fig:sim:1}, \ref{fig:sim:2}, \ref{fig:sim:3} and \ref{fig:sim:4}, for $V=20$ and sample sizes $N = (100, 200)$. On such simulated data, we fit the mixture of Gaussian covariance graph models with Erd\H{o}s-R\'enyi penalty, using the genetic algorithm search \texttt{Ga} to infer the association structures, and with $K$ fixed to 3 in advance. For each association structure and sample size, we replicate the experiment 50 times and we record the inferred graph configurations. Figure~\ref{fig:subfig} reports the proportion of times an edge has been estimated between a pair of variables. The darker the color, the larger the frequency of two variables being declared as associated. Overall, models \texttt{mcg} with Erd\H{o}s-R\'enyi penalty show a good performance in detecting the underlying graph configurations, especially on the association structures related to \emph{Scenario 1}, \emph{2}, and \emph{4}, and as the sample size increases. We point out again that the association structure related to \emph{Scenario 3} is particularly difficult to infer, since some of the correlations are downweighted by the method used to generate the covariance matrices. For this reason, in this scenario sparse covariance mixture models with \texttt{ER} penalty tend to miss some arcs, resulting in larger false negative rates.


\subsection{Part 3}
In this part we evaluate the performance of the \texttt{mgc} models in a high-dimensional setting. We generate data with $N=1000$ and $V=100$ according to the association structures of \emph{Scenario 2} and \emph{Scenario 4}. We replicate the experiment 50 times and we estimate the sparse covariance models for the different penalty functions \texttt{BIC}, \texttt{EBIC}, \texttt{ER}, and \texttt{PL}. We only consider the stepwise search for graph structure inference, as more suited in practice for such a large number of variables. In fact, in this setting for fixed $K$ there are $K2^{4950} \approx K10^{1650}$ possible association structures. To generate the data we use the same parameters (cluster means and mixing proportions) described in Section~\ref{sim}. For this experiment we also compute the difference between the number of estimated parameters and the number of actual parameters of the data generating model. Results are reported in Tables \ref{tab:sim:5} and \ref{tab:sim:6}. 

\begin{table*}[bt]
\centering
\caption{\label{tab:sim:5} Simulated data, high-dimensional setting, \emph{Scenario 2}. The table reports the values of BIC, ARI, FPR, FNR, selected number of clusters, difference between number of estimated and number of actual parameters, and relative time for each method averaged over 50 replicates of the experiment. The relative time is computed with respect to \texttt{mclust}.}
\begin{tabular}{lrrrrrrr}
\toprule
 & \bf BIC  & $K$ & \bf FPR  & \bf FNR  & \bf ARI & Diff. & Rel. time\\
\midrule
\texttt{mclust}	    &  -191763 & 2.60 &---  &---  & 0.78 & 3956.62 & 1 \\[0.5em]
\texttt{mcgStepBIC} & -151948 & 3.00 & 0.11 & 0.04 & 0.95 & 136.18 & 102 \\ 
\texttt{mcgStepEBIC}& -152221 & 3.00 & 0.00 & 0.07 & 0.95 & -0.35 & 52 \\ 
\texttt{mcgStepER}  & -152773 & 2.98 & 0.06 & 0.03 & 0.94 & 23.05 & 63 \\ 
\texttt{mcgStepPL}  & -152120 & 3.00 & 0.13 & 0.08 & 0.95 & 95.93 & 48 \\ 
\bottomrule
\end{tabular}
\end{table*}
\begin{table*}[bt]
\centering
\caption{\label{tab:sim:6} Simulated data, high-dimensional setting, \emph{Scenario 4}. The table reports the values of BIC, ARI, FPR, FNR, selected number of clusters, difference between number of estimated and number of actual parameters, and relative time for each method averaged over 50 replicates of the experiment. The relative time is computed with respect to \texttt{mclust}.}
\begin{tabular}{lrrrrrrr}
\toprule
 & \bf BIC  & $K$ & \bf FPR  & \bf FNR  & \bf ARI & Diff. & Rel. time\\
\midrule
\texttt{mclust}	    & -190438 & 2.79 & --- &---  & 0.79 & 2451.79 & 1 \\ [0.5em]
\texttt{mcgStepBIC} &   -157662 & 3.00 & 0.04 & 0.04 & 0.95 & -19.79 & 92 \\ 
\texttt{mcgStepEBIC}&   -158800 & 3.00 & 0.00 & 0.19 & 0.95 & -379.23 & 42 \\ 
\texttt{mcgStepER}  &   -157702 & 3.00 & 0.04 & 0.03 & 0.96 & -89.11 & 34 \\ 
\texttt{mcgStepPL}  &   -157714 & 3.00 & 0.05 & 0.08 & 0.95 & -68.96 & 50 \\ 
\bottomrule
\end{tabular}
\end{table*}

Overall, the sparse covariance models outperform \texttt{mclust} in terms of model quality, selected number of cluster and classification. In particular, in such high-dimensional situation, despite the parsimonious covariance eigendecomposition, on average the models of \texttt{mclust} family can be largely over-parameterized compared to the \texttt{mgc} models. In fact, the average difference between estimated and effective number of parameters for \texttt{mclust} is significantly larger than for the \texttt{mgc} models. For most of the times, \texttt{mclust} selected the \texttt{VVE} model, for which 5552 mixture parameters need to be estimated for $K=3$, while the actual number of mixture parameters is on average 1137 and 1775 for \emph{Scenario 2} and \emph{Scenario 4} respectively. The remaining times, \texttt{mclust} preferred either the diagonal model \texttt{EEI}, with 402 mixture parameters for $K=3$, or the diagonal model \texttt{VII} with $K=1$ and 101 parameters: both too restrictive and completely missed the presence of association between some of the variables.

\section{Illustrative datasets}
\label{data}
In this section we consider two illustrative data examples. As in the previous section, we fit the mixture of Gaussian covariance graph models using the different penalty functions described in Section~\ref{pen} and using the stepwise and genetic algorithm search for graph configuration inference. Again, the results are compared to \texttt{mclust}. In both examples, the classification of the observations is known and the ARI is used to evaluate the quality of the clustering. 

\subsection{Thyroid gland data}
The data consist of five laboratory tests:
\begin{itemize}[noitemsep]
 \item \textbf{T4}, total Serum thyroxin as measured by the isotopic displacement method.
 \item \textbf{T3}, total serum triiodothyronine as measured by radioimmuno assay.
 \item \textbf{RT3U}, T3-resin uptake test (percentage).
 \item \textbf{TSH}, basal thyroid-stimulating hormone as measured by radioimmuno assay.
 \item \textbf{DTSH}, maximal absolute difference of TSH value after injection of 200 micro grams of thyrotropin-releasing hormone as compared to the basal value.
\end{itemize}
These tests are administered to a sample of 215 patients to assess whether a subject's thyroid gland can be classified as \emph{euthyroidism} (normal functioning, N = 150), \emph{hypothyroidism} (underactive gland not producing enough hormone, N = 30) or \emph{hyperthyroidism} (overactive thyroid producing excessive amounts of the free thyroid hormones T3 and/or thyroxine T4, N = 35). Each patient was assigned to one of the three classes according to a complete medical assessment \citep{coomans:1983}.

\begin{table}[bt]
\centering
\caption{Clustering results for the thyroid gland data: BIC, estimated number of clusters, number of estimated parameters, ARI, and relative time. Relative time is computed with respect to the \texttt{mclust} best model.}
\label{tab:ex:1}
\begin{tabular}{lrrcrr}
\toprule
 & \bf BIC & $K$ & \bf N. par. & \bf ARI & Rel. time\\ 
\midrule
\texttt{mclust-VVV}  & -4810 & 3  & 62 & 0.86 &--- \\
\texttt{mclust}  & -4778 & 3  & 32 & 0.89 & 1\\[0.4em] 

\texttt{mgcGaBIC}  & -4725 & 3  & 41 & 0.86  & 561 \\ 
\texttt{mgcGaEBIC} & -4739 & 3 & 35 & 0.88   & 830 \\ 
\texttt{mgcGaER}   & -4729 & 3   & 44 & 0.86 & 1127 \\ 
\texttt{mgcGaPL}   & -4758 & 3   & 33 & 0.89 & 821 \\[0.4em]

\texttt{mgcStepBIC}  & -4751 & 3  & 47 & 0.86 & 15\\ 
\texttt{mgcStepEBIC} & -4747 & 3 & 37 & 0.88  & 10\\ 
\texttt{mgcStepER}   & -4766 & 3   & 53 & 0.86& 12\\ 
\texttt{mgcStepPL}   & -4759 & 3   & 37 & 0.88&  8\\ 
\bottomrule
\end{tabular}
\end{table}

Table~\ref{tab:ex:1} reports the clustering results. All the methods correctly identify the number of groups and attain a good classification of the patients. \texttt{mclust} selects a \texttt{VVI} model, corresponding to a model where all the variables are independent within each cluster, i.e. the component covariance matrices are diagonal (see \cite{scrucca:etal:2016} for details). However, this could be a restrictive assumption as, for example, hormones T3 and T4 are typically correlated \citep{kumar:1977}. For comparison, we also report the \texttt{mclust} model \texttt{VVV} (\texttt{mclust-VVV}), which places no constraints on the covariance matrices and allows all the variables to be correlated. However, this model is clearly over-parameterized and attains the lowest BIC. Indeed, the models with sparse covariance matrices allow \emph{some} of the variables to be associated in different clusters. All of the sparse covariance mixture models have a larger number of parameters than the model with diagonal covariance matrices, but with a higher BIC value than the one of \texttt{mclust}. Sparse covariance models whose structure of association was estimated using stepwise search give comparable results to those models employing the genetic algorithm, and with a reduced relative computing time.

\begin{figure}[tb]
 \centering
 \subfloat[][\em Cluster 1]
 {\includegraphics[scale=0.39]{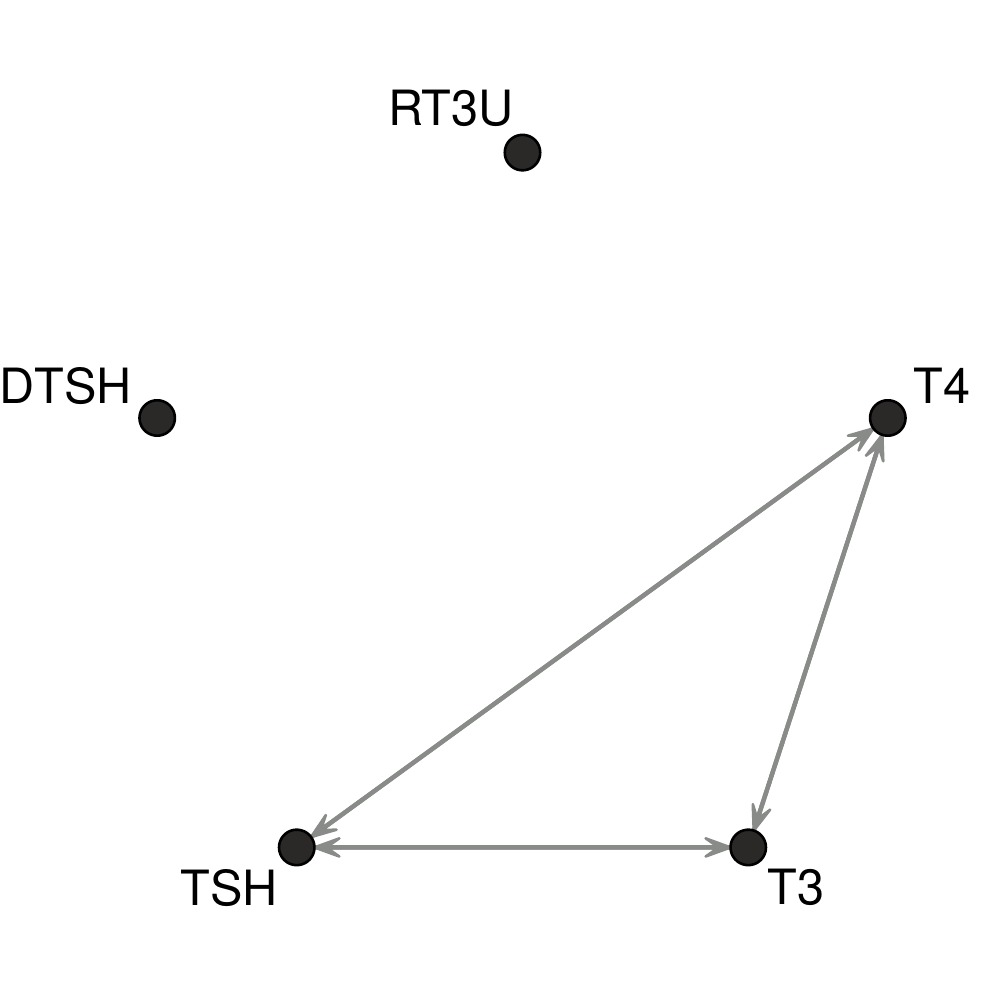}} \quad
 \subfloat[][\em Cluster 2]
 {\includegraphics[scale=0.39]{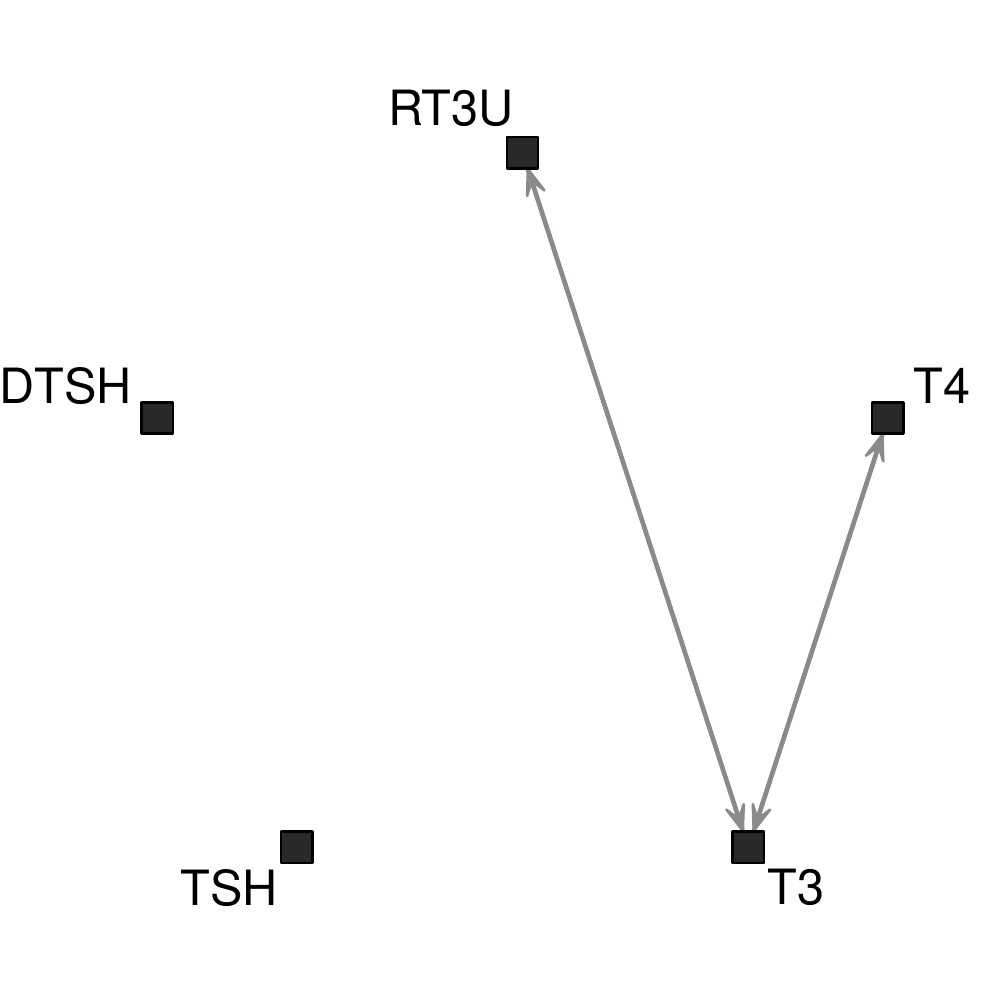}} \quad 
 \subfloat[][\em Cluster 3]
 {\includegraphics[scale=0.39]{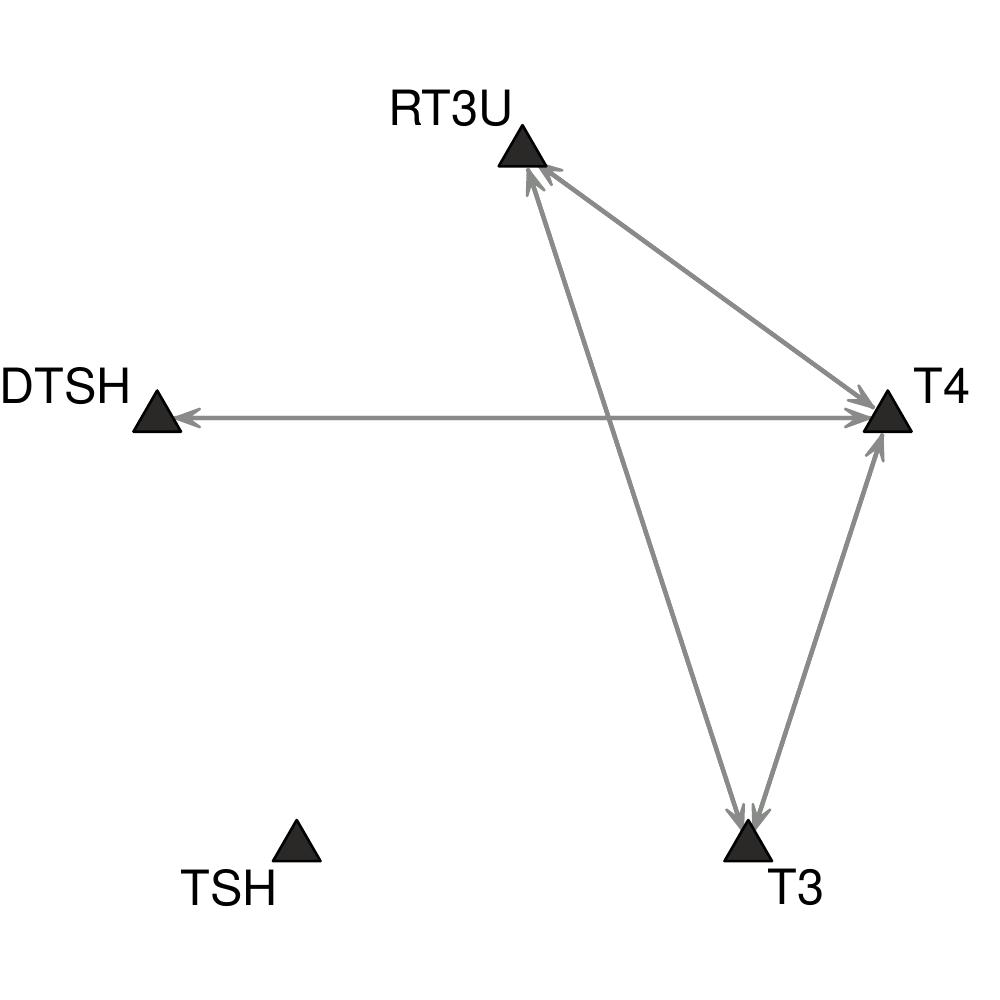}}
 \caption{\label{fig:thyr} Graphs inferred by \texttt{mgcGaBIC} for the thyroid gland data.}
\end{figure}

Among the sparse covariance mixture models, \texttt{mgcGaBIC} is the one with the highest BIC value and Table~\ref{tab:ex:2} presents the cross-tabulation between the patients classification and the estimated partition. There is good agreement between the two partitions and we can match the three diagnosis to the clusters. Figure~\ref{fig:thyr} represents the inferred graphs. Hormones T4 and T3 are associated in all three clusters and overall the correlation structures differ across the groups. In particular, the graph for Cluster 1 is characterized by the relation between T3, T4 and TSH. This cluster is predominantly composed of subjects affected by hypothyroidism and the disease is usually identified by an inverse association between TSH and (T3, T4) \citep{kumar:1977,thyro}. 

\begin{table}[bt]
\centering
\caption{Cross-tabulation between the patients classification and the classification estimated by \texttt{mgcBIC} for the thyroid gland data.}
\label{tab:ex:2}
\begin{tabular}{rccc}
\toprule
& \multicolumn{3}{c}{\bf Cluster}\\
 & 1 & 2 & 3 \\ 
\midrule
Hypothyroidism &  26 &   4 &    \\ 
Euthyroidism &   2 & 145 &   3 \\ 
Hyperthyroidism &    &    &  35 \\ 
\bottomrule
\end{tabular}
\end{table}

\subsection{Italian wine data}
\label{wine}
The data consist of $27$ chemical measurements from $N = 178$ wine samples from Piedmont region, in Italy \citep{forina:1986}. The samples derive from three different cultivars: Barbera $(N = 48)$, Grignolino $(N=71)$ and Barolo $(N = 59)$. Table~\ref{tab:wine} contains the names of the measured variables. 

\begin{table*}[tb]
\small
\centering
\caption{\label{tab:wine} Variables in the Italian wine dataset.}
\begin{tabular}{lll}
\toprule
1. Alcohol & 10. Potassium & 19. Color Intensity \\ 
2. Sugar-free Extract & 11. Calcium & 20. Hue \\ 
3. Fixed Acidity & 12. Magnesium & 21. OD280/OD315 of Diluted Wines \\ 
4. Tartaric Acid & 13. Phosphate & 22. OD280/OD315 of Flavanoids \\ 
5. Malic Acid & 14. Chloride & 23. Glycerol \\ 
6. Uronic Acids & 15. Total Phenols & 24. 2-3-Butanediol \\ 
7. pH & 16. Flavanoids & 25. Total Nitrogen \\ 
8. Ash & 17. Non-flavanoid Phenols & 26. Proline \\ 
9. Alcalinity of Ash & 18. Proanthocyanins & 27. Methanol \\
\bottomrule
\end{tabular}
\end{table*}

\begin{table}[bt]
\centering
\caption{\label{tab:ex:3} Clustering results for the Italian wine data: BIC, estimated number of clusters, number of estimated parameters, ARI, and relative time. Relative time is computed with respect to \texttt{mclust}.}
\begin{tabular}{lrrcrrrr}
\toprule
 &  \bf BIC & $K$ &  \bf N. par. & \bf  ARI & Rel. time \\ 
\midrule
\texttt{mclust-VVV}  & -24254 & 1 & 405 & 0.00 & --- \\
\texttt{mclust}  & -23954 & 3 & 162 & 0.83 & 1\\[0.5em] 

\texttt{mgcGaBIC}  & -23217 & 2 & 248 & 0.48  & 128\\ 
\texttt{mgcGaEBIC} & -23185 & 3 & 189 & 0.88 & 57\\ 
\texttt{mgcGaER}   & -22965 & 3 & 231 & 0.89  & 79\\ 
\texttt{mgcGaPL}   & -23451 & 4 & 240 & 0.83 & 92\\[0.5em] 

\texttt{mgcStepBIC}  & -23485 & 2 & 273& 0.41  & 38\\ 
\texttt{mgcStepEBIC} & -23208 & 3 & 186 & 0.88 & 8\\ 
\texttt{mgcStepER}   & -23042 & 3 & 233 & 0.81  & 17\\ 
\texttt{mgcStepPL}   & -23429 & 4 & 241 & 0.83 & 27\\
\bottomrule
\end{tabular}
\end{table}

\begin{table}[bt]
\centering
\caption{Cross-tabulation between the actual classification and the classification estimated by \texttt{mgcGaER} for the Italian wine data.}
\label{tab:ex:4}
\begin{tabular}{rccc}
\toprule
& \multicolumn{3}{c}{ \bf Cluster}\\
 & 1 & 2 & 3 \\ 
\midrule
Barolo &  59 &    &    \\ 
Grignolino &   6 & 65 &    \\ 
Barbera &    &    &  48 \\ 
\bottomrule
\end{tabular}
\end{table}

The clustering results are reported in Table~\ref{tab:ex:3}. Apart from the sparse covariance model estimated with \texttt{BIC} penalty function, all the models obtain good clustering results, even though the BIC of the sparse covariance model with \texttt{PL} penalization preferred a mixture distribution with 4 components. \texttt{mclust} selects an \texttt{EVI} model, corresponding to graphs where all the variables are independent. However, the assumption could be too restrictive as the characteristics of the wine types are naturally defined by the different relations among the chemical components \citep{wine}. Note that the number of parameters reported in the table for the \texttt{mclust} model is related to the corresponding covariance matrix decomposition, where the volume of the clusters is constrained to be equal across the mixture components (see \cite{scrucca:etal:2016}). The table also reports the \texttt{mclust} model \texttt{VVV} (\texttt{mclust-VVV}), with no restrictions on the component covariance matrices. The VVV model is largely over-parameterized. In fact, for this data, if no constraints are imposed on the covariance matrices, a large number of parameters need to be estimated as the number of component increases, thus resulting in the selection of a mixture model with only one component by BIC. Regarding the sparse covariance mixture models, also in this example the number of estimated parameters is larger than the number of parameters of the model preferred by \texttt{mclust}. Indeed, these models pose less restrictive assumptions on the structure of dependence and allow some of the chemical quantities to be associated in different ways within the clusters. Again, despite the higher number of estimated parameters, the sparse covariance mixture models outperform the \texttt{mclust} model in terms of BIC. Also here, the stepwise search \texttt{Step} provides results comparable to those obtained employing the evolutionary search \texttt{Ga} and with a smaller relative computing time.

\begin{figure}[tb]
 \centering
 \subfloat[][\em Cluster 1]
 {\includegraphics[scale=0.39]{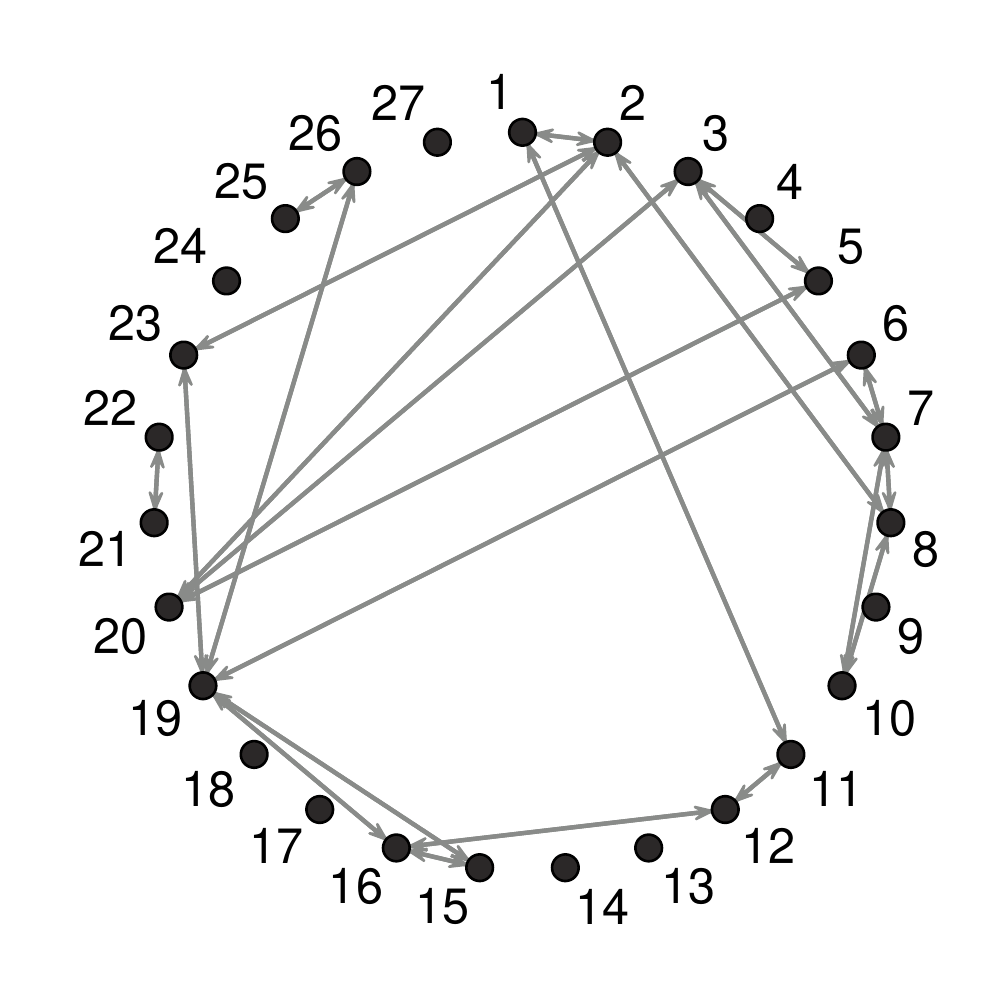}} \quad
 \subfloat[][\em Cluster 2]
 {\includegraphics[scale=0.39]{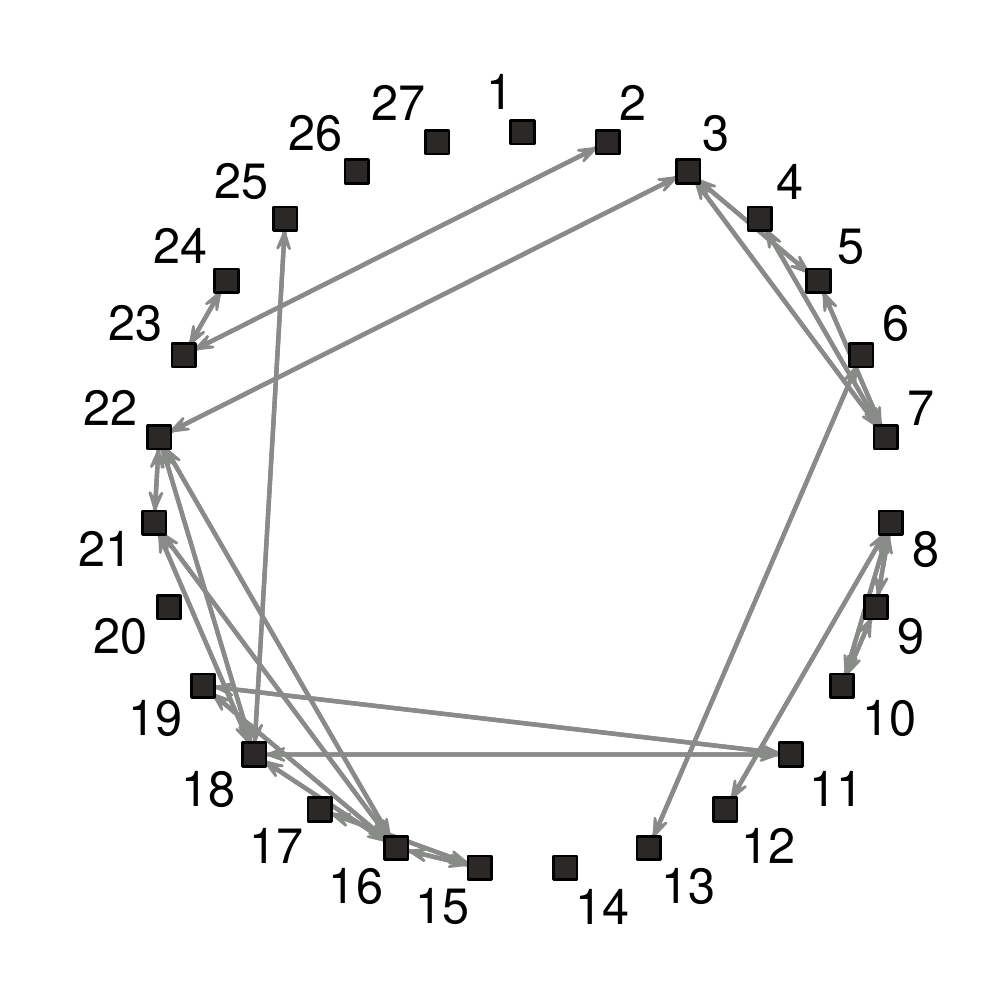}} \quad 
 \subfloat[][\em Cluster 3]
 {\includegraphics[scale=0.39]{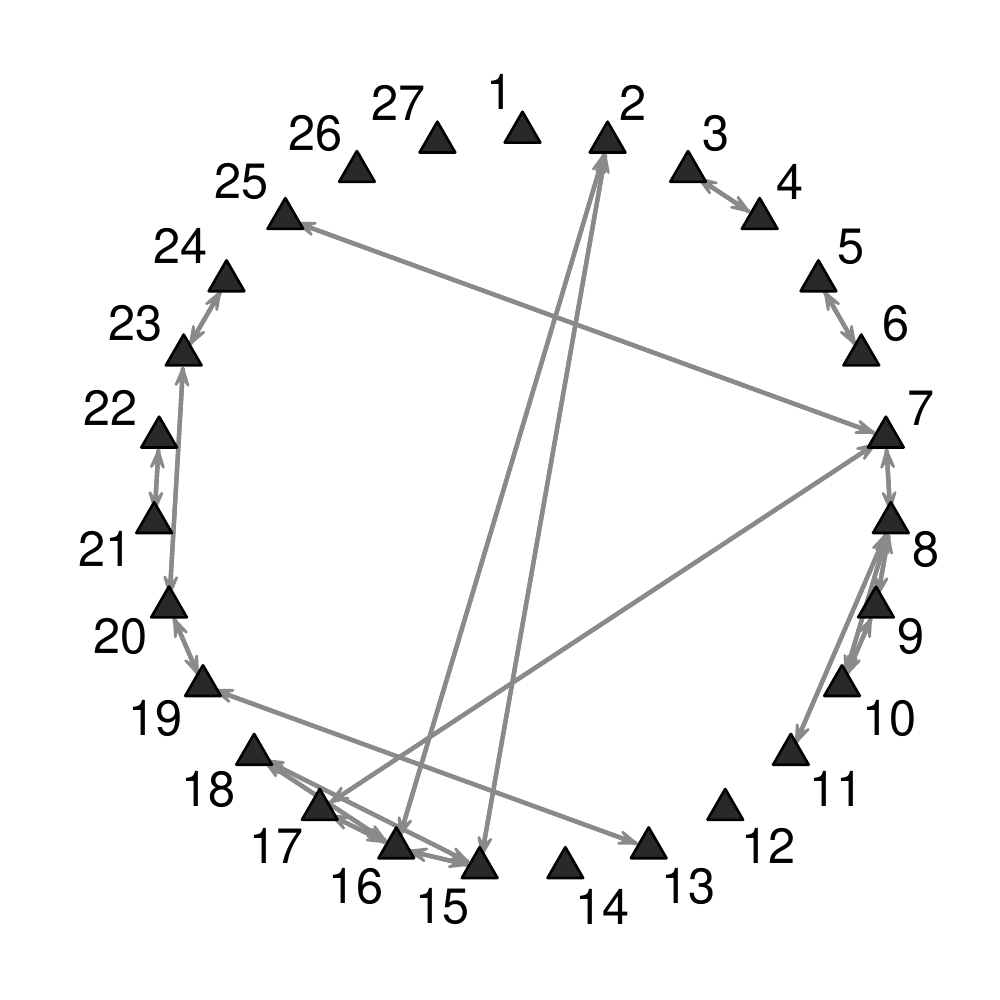}}
 \caption{\label{fig:wine} Graphs inferred by \texttt{mgcGaER} for the Italian wine data. The numbers correspond to the variable names of Table~\ref{tab:wine}.}
\end{figure}

\begin{figure}[tb]
 \centering
 \subfloat[][\em Clusters 1 - 2]
 {\includegraphics[scale=0.45]{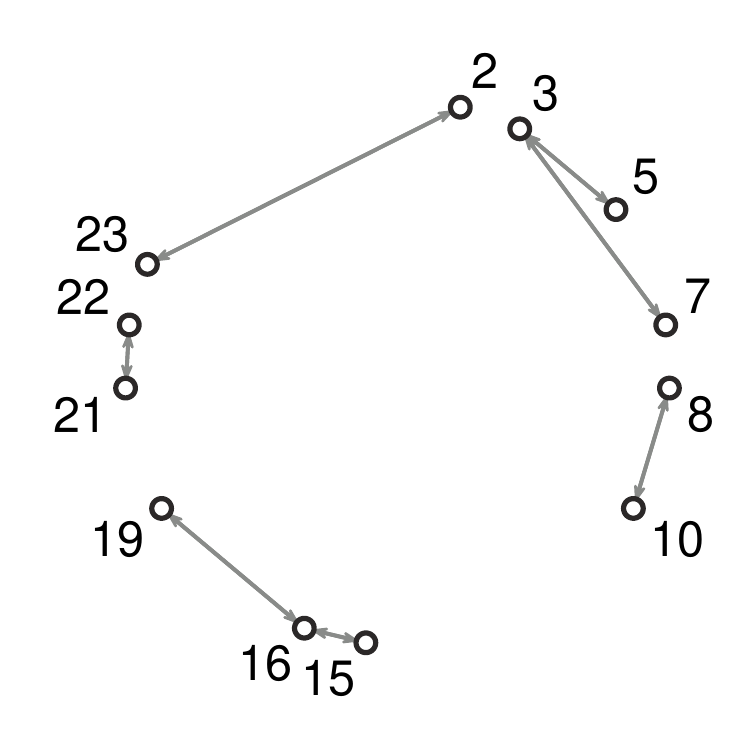}} \qquad 
 \subfloat[][\em Clusters 1 - 3]
 {\vspace*{1cm}\includegraphics[scale=0.45]{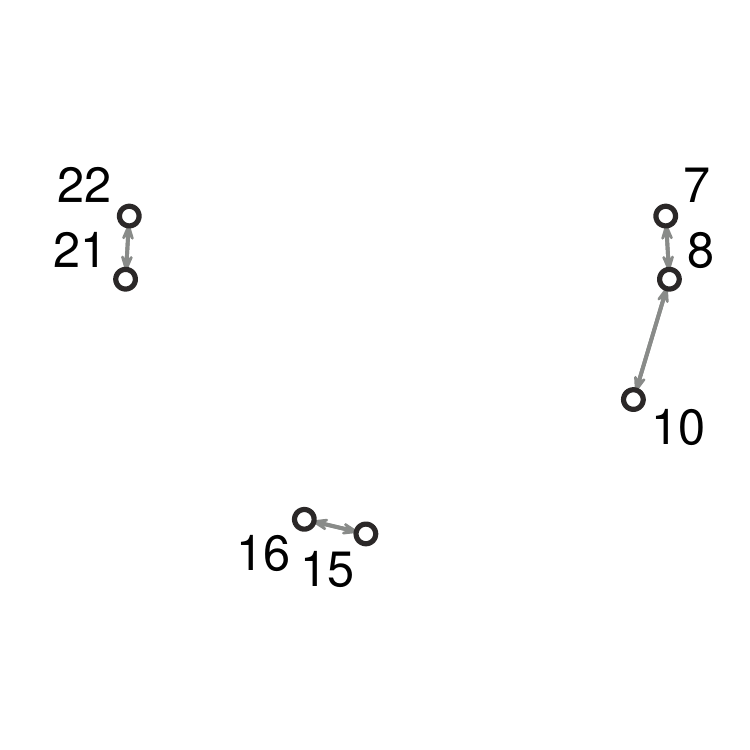}} \qquad 
 \subfloat[][\em Clusters 2 - 3]
 {\includegraphics[scale=0.45]{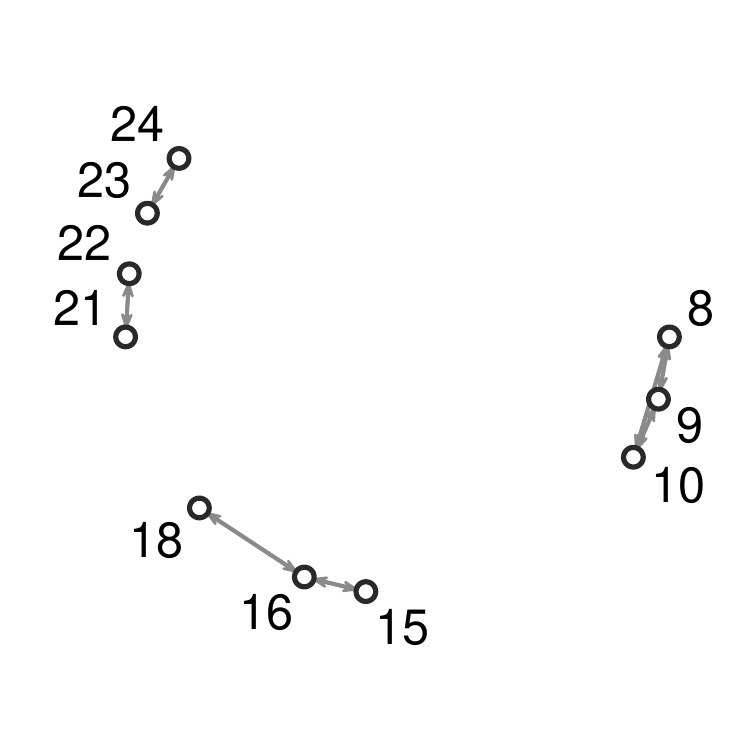}}
 \caption{\label{fig:asswine} Arcs present in each pair of clusters for the Italian wine data as estimated by \texttt{mgcGaER}. Isolated nodes are not shown. The numbers correspond to the variable names of Table~\ref{tab:wine}.}
\end{figure}

In this case,  \texttt{mgcGaER} is the sparse covariance model with the largest BIC and Table~\ref{tab:ex:4} contains the cross-tabulation between the actual classification of the samples and the estimated partition. The clustering shows good agreement to the wine types and only in Cluster 1 there is overlapping between Barolo and Grignolino samples. Figure~\ref{fig:wine} depicts the estimated graphs. The correlation structures among the chemical measurements differ from cluster to cluster, however some similarities are also present. Figure~\ref{fig:asswine} displays the variables that are associated in each pair of clusters. In particular, edges (8, 10), (15, 16), and (21, 22) are present in all three groups. The chemical compounds corresponding to (3, 5) regulate the acidity of the wine and are thus related to the pH measured by variable 7. The subset of variables (15, 16, 17, 18, 19, 20, 21, 22) tend to be particularly connected in the three clusters, with different set of edges. Specifically, variables (15, 16, 17, 18) are related to the phenolic content and are responsible for the coloration of the wine \citep{harbertson}, expressed by the variables (19, 20).

\section{Discussion}
\label{disc}
We present a framework for model-based clustering with sparse covariance matrices. This framework is based on a mixture of Gaussian covariance graph models where the component densities are characterized by bi-directed graphs and corresponding sparse covariance matrices. The approach results in a flexible model that can accommodate association structures among the variables that vary from cluster to cluster. Estimation is accomplished via maximization of a penalized likelihood by means of a structural EM algorithm. Two alternative strategies based on genetic algorithm and stepwise search are suggested to solve the optimization problem related to graph and sparse covariance matrix estimation. In order to introduce different degree of sparsity in the covariance matrices, we proposed a general penalization term on the graph structures that allows for various specifications of the penalty function. 

The presented work is related to the estimation of a Gaussian graphical model when the observed sample arises from an heterogeneous population. Recently, the problem has attracted increasing attention, with particular focus on parameterization of the multivariate Gaussian distribution via its precision matrix. In a supervised context, where the classification of the data points is known in advance, different approaches have been suggested in the literature. \cite{bala:2014} and \cite{peterson:2015} propose full Bayesian frameworks with different parameterizations and priors for the precision matrix. With the aim of joint estimation of multiple graphical models sharing common characteristics among the classes, \cite{guo:2011}, \cite{mohan:2012}, \cite{mohan:2014}, \cite{danaher:2014}, \cite{zhu:2014} and \cite{ma:2016} propose penalized likelihood approaches that place a penalty on the entries of the precision matrix and are inspired by the graphical lasso \citep{friedman:etal:2008}. 

Within the context of clustering and Gaussian mixture models, seminal work can be found in \cite{thiesson::1997}, where the authors parameterize each component density in terms of conditional distributions and a related directed acyclic graph \citep[DAG, ][]{whittaker:1990}. Recent work on mixtures of DAGs is in \cite{chalmond:2015}. \cite{rodriguez:2011} and \cite{talluri:2014} develop a Bayesian framework for estimating infinite mixtures of sparse Gaussian graphical models where different prior distributions on the inverse covariance are employed. \cite{krishnamurthy:2011}, \cite{lotsi:2013}, \cite{gao:2016}, and \cite{lee:2017} present methods for estimating mixture models with sparse precision matrices via penalized likelihood estimation and lasso-type penalty functions. Compared to these approaches, in our proposed framework we parameterize the mixture of Gaussians directly in terms of the component covariance matrices. This has the main advantage of obtaining sparse component covariance matrices immediately, and not as a by-product of inverting the corresponding precision matrices. Indeed, a sparse precision matrix does not guarantee a sparse covariance \citep{whittaker:1990,pourahmadi:2011}. Moreover, zero covariance terms between any pair of variables can be easily understood in terms of marginal independence, instead of conditional independence \citep{whittaker:1990,edwards:2000}, leading to a simpler interpretation of the clustering results. With regard to Gaussian mixture models parameterized by the covariance matrix, \cite{galimberti:2013} propose an approach where the vector of variables is partitioned into subsets that are conditionally independent within the clusters. As a consequence, the component covariance matrices are sparse and have a block diagonal structure. The framework we propose in this paper is more general, since no structure is assumed and the variables are allowed to be dependent also between blocks.

The topic covered in the paper overlaps also with the framework of sparse Gaussian mixture models estimation. The problem was originally introduced by \cite{pan:shen:2007} with the aim of variable selection for clustering, although the authors did not deal with estimation of sparse component covariance matrices, which were assumed all equal and diagonal. Subsequently, in the context of high-dimensional data and regularization, \cite{zhou:2009}, \cite{azizyan:2015}, and \cite{ruan:2011} extended the approach to incorporate sparse inverse covariance estimation via lasso-type penalization. Within a Bayesian framework, \cite{malsiner:etal:2016} propose to use a shrinkage prior on the component means, although no shrinkage prior is considered on the component covariance matrices.

Compared to lasso-type penalizations and the prior distributions employed in the Bayesian frameworks (such as the G-Wishart distribution, \cite{roverato:2002}, for example), we proposed a general penalty term placed on the collection of graph structures. This type of penalty is flexible and allows for any form of functional specification. We discussed some alternatives that are tailored to different situations and objectives. The BIC-type penalty function can be employed when the purpose is to delineate a model for the within-cluster association among the variables. With this aim, graph structure estimation is actually a model selection problem and consistency results of BIC for Gaussian graphical models apply \citep{koller:2009}. In settings with a large number of variables, the BIC could prefer overly complex association structures. In these contexts, the EBIC-type penalty can be considered, since it induces a larger penalization than the BIC-type one, favoring models with sparser covariance matrices. Moreover, consistency results are available in the case where sample size and dimensions of the data are comparable \citep{foygel:2010}. 
The power law penalty function tends to penalize less situations where few nodes in the graph have high degree than situations where all the nodes have comparable degree values (see Figure~\ref{fig:pl}). Thus, it is suitable if the clusters are believed to be characterized by structures of association where a small number of hub variables are correlated to the others. Furthermore, the tuning parameter $\beta$ allows to control the amount of sparsity induced. The Erd\H{o}s-R\'enyi penalty function favors graphs with disconnected components and can be used in situations where the within-cluster joint distribution is believed to decompose into the product of independent blocks containing associated variables. Also for this function, a tuning parameter allows to control the degree of sparsity of the inferred covariance matrices.

By placing a penalty function on the within-component association structures embedded in the adjacency matrices, optimization over the graph space is recast as a combinatorial problem. We propose two alternative strategies based on genetic algorithm and stepwise search to effectively solve the task. In both cases, the nature of the optimization problem allows for parallelization of the computations. In particular, the genetic algorithm extensively explores the space of solutions, but it could be slow and require a substantial number of iterations and computing time to attain convergence to a stable solution when clustering data recorded on a large number of variables. On the other hand, although sub-optimal, the stepwise search is significantly faster and provides models of comparable quality. 

Current and future work focuses on computational improvements and the extension of the methodology to model-based clustering and sparse modeling of categorical and mixed-type data.

The general framework for model-based clustering with sparse covariance matrices is implemented in the \texttt{R} package \texttt{mixGGraph} that will be soon available on CRAN.


\appendix
\section{Iterative conditional fitting algorithm}
\label{appendix:a}
The ICF algorithm \citep{chaudhuri:etal:2007} is employed to estimate a sparse covariance matrix given a certain structure of association. In this appendix, we present the algorithm in application to Gaussian mixture model estimation and we extend it to allow for Bayesian regularization of the covariance matrix.

Given a graph $\GG_k = (\mathcal{V}, \mathcal{E}_k)$, to find the corresponding sparse covariance matrix under the constraint of being positive definite we need to maximize the objective function:
$$
-\dfrac{N_k}{2} \left[ \text{tr}(\mathbf{S}_k\SIGMA_k^{-1}) + \log\det \SIGMA_k  \right] \quad \text{with}\quad \SIGMA_k \in \mathcal{C}^+\left( \GG_k \right).
$$
Let us make use of the following conventions: subscript $[j,h]$ denotes element $(j,h)$ of a matrix, a negative index such as $-j$ denotes that row or column $j$ has been removed, subscript $[\,,j]$ (or $[j,\,]$) denotes that column (or row) $j$ has been selected. Moreover, we denote with $s(j)$ the set of indexes corresponding to the variables connected to variable $X_j$ in the graph, i.e. the positions of the non zero entries in the covariance matrix for $X_j$. Following \cite{chaudhuri:etal:2007}, the ICF algorithm is implemented as follows:
\begin{enumerate}[noitemsep]
 \item Set the iteration counter $r=0$. Initialize the covariance matrix $\hat{\SIGMA}^{(0)}_k = \text{diag}(\mathbf{S}_k)$.
 \item For $j = (1,\, \dots,\, V)$
 \begin{enumerate}
  \item[2.a)] compute $\bm{\Omega}_k^{(r)} = (\hat{\SIGMA}^{(r)}_{k[-j,-j]})^{-1}$
  \item[2.b)] compute the covariance terms estimates
  $$
  \hspace*{-1cm}\hat{\SIGMA}^{(r)}_{k[j,s(j)]} = \left( \mathbf{S}_{k[j,-j]}\,\bm{\Omega}^{(r)}_{k[\,,s(j)]} \right)\,\!\! \left( \bm{\Omega}^{(r)}_{k[s(j),\,]} \mathbf{S}_{k[-j,-j]} \bm{\Omega}^{(r)}_{k[\,,s(j)]} \right)
  $$
  \item[2.c)] compute $\lambda_j = \mathbf{S}_{k[j,j]} - \hat{\SIGMA}^{(r)}_{k[j,s(j)]} \left( \mathbf{S}_{k[j,-j]}\,\bm{\Omega}^{(r)}_{k[\,,s(j)]} \right)^{\!\top}$
  \item[2.d)] compute the variance term estimate
  $$
  \hat{\SIGMA}^{(r)}_{k[j,j]} = \lambda_j + \hat{\SIGMA}^{(r)}_{k[j,s(j)]} \bm{\Omega}^{(r)}_{k[s(j),s(j)]} \hat{\SIGMA}^{(r)}_{k[s(j),j]}
  $$
 \end{enumerate}
 \item Set $\hat{\SIGMA}^{(r+1)}_k = \hat{\SIGMA}_k^{(r)}$, increment $r = r + 1$ and return to (2).
\end{enumerate}
The algorithm stops when the increase in the objective function is less than a pre-specified tolerance. The covariance matrix in output has zero entries corresponding to the graph structure and is guaranteed of being positive definite.

In the case of Bayesian regularization, the objective function becomes:
$$
- \dfrac{\tilde{N}_k}{2} \left[ \text{tr}(\tilde{\mathbf{S}}_k\SIGMA_k^{-1}) + \log\det \SIGMA_k  \right] \quad \text{with}\quad \SIGMA_k \in \mathcal{C}^+\left( \GG_k \right),
$$
where
$$
\tilde{N}_k = N_k + \omega + V + 1, \qquad \tilde{\mathbf{S}}_k = \dfrac{1}{\tilde{N}_k} \left[ N_k \mathbf{S}_k + \mathbf{W} \right].
$$
The shape of the objective function corresponds to the one not regularized. Therefore, the same algorithm can be applied replacing $N_k$ and $\mathbf{S}_k$ with $\tilde{N}_k$ and $\tilde{\mathbf{S}}_k$.

\section{Initialization of the S-EM algorithm}
\label{appendix:b}
The S-EM algorithm requires two initialization steps: initialization of cluster allocations and initialization of the graph structure search. For the first task we use the Gaussian model-based hierarchical clustering approach of \cite{scrucca:init}, 
which has been shown to yield good starting points, be computationally efficient and work well in practice. For initialization of the graph structure search we use the following approach. Let $\mathbf{R}_k$ be the correlation matrix for component $k$, computed as:
$$
\mathbf{R}_k = \mathbf{U}_k\mathbf{S}_k\mathbf{U}_k,
$$
where $\mathbf{U}_k$ is a diagonal matrix whose elements are $\mathbf{S}_{k,[j,j]}^{-1/2}$ for $j=1,\ldots,V$, i.e. the within component sample standard deviations. A sound strategy is to initialize the search for the optimal association structure by looking at the most correlated variables. Therefore, we define the adjacency matrix $\A_k$ whose off-diagonal elements $a_{jhk}$ are given by:
$$
a_{jhk} = \begin{cases} 1 \quad \text{if}~~ \lvert r_{jhk} \lvert~~ \geq ~\rho,\\
		      0 \quad \text{otherwise}
	\end{cases}
$$
where $r_{jhk}$ is an off-diagonal element of $\mathbf{R}_k$ and $\rho$ is a threshold value. In practice, we define a vector of values for $\rho$ ranging from $0.4$ to $1$. For each value of $\rho$, the related adjacency matrix is derived and the corresponding sparse covariance matrix is estimated using the ICF algorithm. Then the different adjacency matrices are ranked according to their value of the objective function in \eqref{eq:5}. Subsequently the structure search starts from the adjacency matrix at the top of the rank.

\section{Details of simulation experiments}
\label{appendix:c}
This appendix section describes the various simulated data scenarios considered in Section~\ref{sim} of the paper. 

\textit{Scenario 1}:
In this setting we consider a structure with a single block of associated variables of size $\floor*{\frac{V}{2}}$. The groups are differentiated by the position of the block, top corner, center and bottom corner respectively. Figure~\ref{fig:sim:1} displays an example of such structure for $V=20$. To generate the covariance matrices, first we generate a $V\times V$ matrix with all entries equal to 0.9 and diagonal 1. Then we use it as input of the ICF algorithm to estimate the corresponding covariance matrix with the given structure.

\textit{Scenario 2}:
For this scenario, the graphs are generated at random from an Erd\H{o}s-R\'enyi model. The groups are characterized by different probabilities of connection, 0.3, 0.2 and 0.1 respectively. Figure~\ref{fig:sim:2} presents an example of a collection of structures of association for $V=20$. Starting from a $V\times V$ matrix with all entries equal to 0.9 and diagonal 1, we employ the ICF algorithm to estimate the corresponding sparse covariance matrix. In the simulated data experiment of Part III, we consider connection probabilities equal to 0.10, 0.05 and 0.03.

\textit{Scenario 3}:
This scenario is characterized by hubs, i.e. highly connected variables. Each cluster has $\frac{V}{2}$ such hubs. The graph structures and the corresponding covariance matrices are generated randomly using the R package \texttt{hglasso}. \citep{hglasso}. The three groups have different sparsity levels, respectively 0.7, 0.8 and 0.9. Figure~\ref{fig:sim:3} presents an example of this type of graphs for $V=20$. We point out that the method implemented in the package poses strict constraints on the covariance matrix and often some connected variables have weak correlations, making difficult to infer the association structure. 

\textit{Scenario 4}:
Here the groups have structures of different types: block diagonal, random connections and Toeplitz type. For the first group we consider a block diagonal matrix with blocks of size 5. Regarding the second, the graph is generated at random from an Erd\H{o}s-R\'enyi model with parameter 0.2. In both cases, we start from a $V\times V$ matrix with all entries equal to 0.9 and diagonal 1, and then we employ the ICF algorithm to estimate the corresponding sparse covariance matrices. For the Toeplitz matrix we take $\sigma_{j,\,j-1} = \sigma_{j-1,\,j} = 0.5$ for $j=2,\,\dots,\,V$. Figure~\ref{fig:sim:4} depicts an example of these graph configurations for $V=20$. In the simulated data experiment of Part III, we consider an Erd\H{o}s-R\'enyi model with parameter 0.05 and a block diagonal matrix with 5 blocks of size 20; the Toeplitz matrix is generated as before.

\section{A note on computing time}
\label{appendix:d}
In the simulated data experiment and illustrative examples we presented the computational time of our framework using as reference the computing time of the widely used software \texttt{mclust}. The software has more than twenty years history, is highly developed and the core functionalities are implemented in Fortran, for these reasons it is particularly efficient and fast. On the other hand, the code implementing our proposed method is written in pure R (which is known to be slower than compiled languages) and, although much care and effort have been put for an efficient implementation, it is the product of a shorter development time. Moreover, in our framework we are tackling the particularly complex problem of joint mixture and graphical model estimation: even for a relatively small size problem with $10$ variables and $2$ mixture components there are approximatively $7\times 10^{13}$ possible models. As expected, the runtime of our methodology is shown to be several orders of magnitude larger than \texttt{mclust}. Although computing time is a relevant variable to be taken into account in practice, here we argue that evaluating the effective runtime and speed of an algorithm or method is a very difficult task. This for multiple reasons: software implementation, modeling framework and purpose, computational resources, characteristics of the problem to be solved.  An intriguing discussion is in \cite{kriegel:2017} and references therein.

\bibliographystyle{apalike}
\bibliography{bibliography.bib}

\begin{thebibliography}{}

\bibitem[Amerine, 1953]{wine}
Amerine, M.~A. (1953).
\newblock The composition of wines.
\newblock {\em The Scientific Monthly}, 77(5):250--254.

\bibitem[Azizyan et~al., 2015]{azizyan:2015}
Azizyan, M., Singh, A., and Wasserman, L. (2015).
\newblock Efficient sparse clustering of high-dimensional non-spherical
  {G}aussian mixtures.
\newblock In {\em Artificial Intelligence and Statistics}, pages 37--45.

\bibitem[Baladandayuthapani et~al., 2014]{bala:2014}
Baladandayuthapani, V., Talluri, R., Ji, Y., Coombes, K.~R., Lu, Y., Hennessy,
  B.~T., Davies, M.~A., and Mallick, B.~K. (2014).
\newblock Bayesian sparse graphical models for classification with application
  to protein expression data.
\newblock {\em The Annals of Applied Statistics}, 8(3):1443--1468.

\bibitem[Banfield and Raftery, 1993]{banfield:raftery:1993}
Banfield, J.~D. and Raftery, A.~E. (1993).
\newblock Model-based {G}aussian and non-{G}aussian clustering.
\newblock {\em Biometrics}, 49(3):803--821.

\bibitem[Barber and Drton, 2015]{barber:2015}
Barber, R.~F. and Drton, M. (2015).
\newblock High-dimensional {I}sing model selection with {B}ayesian information
  criteria.
\newblock {\em Electronic Journal of Statistics}, 9(1):567--607.

\bibitem[Baudry and Celeux, 2015]{baudry:2015}
Baudry, J.-P. and Celeux, G. (2015).
\newblock {EM} for mixtures {I}nitialization requires special care.
\newblock {\em Statistics and Computing}, 25(4):713--726.

\bibitem[Bellman, 1957]{bellman:1957}
Bellman, R. (1957).
\newblock {\em Dynamic Programming}.
\newblock Princeton University Press.

\bibitem[Bien and Tibshirani, 2011]{bien:tibshirani:2011}
Bien, J. and Tibshirani, R.~J. (2011).
\newblock Sparse estimation of a covariance matrix.
\newblock {\em Biometrika}, 98(4):807--820.

\bibitem[Biernacki and Lourme, 2014]{biernacki:lourme:2014}
Biernacki, C. and Lourme, A. (2014).
\newblock Stable and visualizable {G}aussian parsimonious clustering models.
\newblock {\em Statistics and Computing}, 24(6):953--969.

\bibitem[Bollobas, 2001]{bollobas:2001}
Bollobas, B. (2001).
\newblock {\em Random {G}raphs}.
\newblock Cambridge University Press.

\bibitem[Bouveyron and Brunet, 2012]{bouveyron:2012}
Bouveyron, C. and Brunet, C. (2012).
\newblock Simultaneous model-based clustering and visualization in the fisher
  discriminative subspace.
\newblock {\em Statistics and Computing}, 22(1):301--324.

\bibitem[Bouveyron and Brunet-Saumard, 2014]{bouveyron:2014}
Bouveyron, C. and Brunet-Saumard, C. (2014).
\newblock Model-based clustering of high-dimensional data: A review.
\newblock {\em Computational Statistics \& Data Analysis}, 71:52--78.

\bibitem[Bozdogan, 2004]{bozdogan:2004}
Bozdogan, H. (2004).
\newblock Intelligent statistical data mining with information complexity and
  genetic algorithms.
\newblock {\em Statistical data mining and knowledge discovery}, pages 15--56.

\bibitem[Celeux and Govaert, 1995]{celeux:govaert:1995}
Celeux, G. and Govaert, G. (1995).
\newblock {G}aussian parsimonious clustering models.
\newblock {\em Pattern Recognition}, 28(5):781--793.

\bibitem[Chalmond, 2015]{chalmond:2015}
Chalmond, B. (2015).
\newblock A macro-{DAG} structure based mixture model.
\newblock {\em Statistical Methodology}, 25:99--118.

\bibitem[Chatterjee et~al., 1996]{chatterjee:1996}
Chatterjee, S., Laudato, M., and Lynch, L.~A. (1996).
\newblock Genetic algorithms and their statistical applications: an
  introduction.
\newblock {\em Computational Statistics \& Data Analysis}, 22(6):633--651.

\bibitem[Chaudhuri et~al., 2007]{chaudhuri:etal:2007}
Chaudhuri, S., Drton, M., and Richardson, T.~S. (2007).
\newblock Estimation of a covariance matrix with zeros.
\newblock {\em Biometrika}, 94(1):199--216.

\bibitem[Chen and Chen, 2008]{chen:2008}
Chen, J. and Chen, Z. (2008).
\newblock Extended {B}ayesian information criteria for model selection with
  large model spaces.
\newblock {\em Biometrika}, 95(3):759--771.

\bibitem[Ciuperca et~al., 2003]{ciuperca:2003}
Ciuperca, G., Ridolfi, A., and Idier, J. (2003).
\newblock Penalized maximum likelihood estimator for normal mixtures.
\newblock {\em Scandinavian Journal of Statistics}, 30(1):45--59.

\bibitem[Coomans et~al., 1983]{coomans:1983}
Coomans, D., Broeckaert, M., Jonckheer, M., and Massart, D. (1983).
\newblock Comparison of multivariate discriminant techniques for clinical data
  - {A}pplication to the thyroid functional state.
\newblock {\em Methods of Information Medicine}, 22:93--101.

\bibitem[Danaher et~al., 2014]{danaher:2014}
Danaher, P., Wang, P., and Witten, D.~M. (2014).
\newblock The joint graphical lasso for inverse covariance estimation across
  multiple classes.
\newblock {\em Journal of the Royal Statistical Society: Series B (Statistical
  Methodology)}, 76(2):373--397.

\bibitem[Dempster, 1972]{dempster:1972}
Dempster, A. (1972).
\newblock Covariance selection.
\newblock {\em Biometrics}, 28(1):157--175.

\bibitem[Dempster et~al., 1977]{dempster:etal:1977}
Dempster, A.~P., Laird, N.~M., and Rubin, D.~B. (1977).
\newblock Maximum likelihood from incomplete data via the {EM} algorithm.
\newblock {\em Journal of the Royal Statistical Society, Series B},
  39(1):1--38.

\bibitem[Drton and Maathuis, 2017]{drton:2017}
Drton, M. and Maathuis, M.~H. (2017).
\newblock Structure learning in graphical modeling.
\newblock {\em Annual Review of Statistics and Its Application}, 4(1):365--393.

\bibitem[Edwards, 2000]{edwards:2000}
Edwards, D. (2000).
\newblock {\em Introduction to Graphical Modelling}.
\newblock Springer-Verlag.

\bibitem[Erd{\H o}s and R{\'e}nyi, 1959]{erdos:1959}
Erd{\H o}s, P. and R{\'e}nyi, A. (1959).
\newblock On random graphs {I}.
\newblock {\em Publicationes Mathematicae (Debrecen)}, 6:290--297.

\bibitem[Erd{\H o}s and R{\'e}nyi, 1960]{erdos:1960}
Erd{\H o}s, P. and R{\'e}nyi, A. (1960).
\newblock On the evolution of random graphs.
\newblock {\em Publications of the Mathematical Institute of the Hungarian
  Academy of Sciences}, 5(1):17--60.

\bibitem[Forina et~al., 1986]{forina:1986}
Forina, M., Armanino, C., Castino, M., and Ubigli, M. (1986).
\newblock Multivariate data analysis as a discriminating method of the origin
  of wines.
\newblock {\em Vitis}, 25(3):189--201.

\bibitem[Foygel and Drton, 2010]{foygel:2010}
Foygel, R. and Drton, M. (2010).
\newblock Extended {B}ayesian information criteria for {G}aussian graphical
  models.
\newblock In {\em Advances in neural information processing systems}, pages
  604--612.

\bibitem[Fraley and Raftery, 2002]{fraley:raftery:2002}
Fraley, C. and Raftery, A.~E. (2002).
\newblock Model-based clustering, discriminant analysis and density estimation.
\newblock {\em Journal of the American Statistical Association}, 97:611--631.

\bibitem[Fraley and Raftery, 2005]{fraley:2005}
Fraley, C. and Raftery, A.~E. (2005).
\newblock Bayesian regularization for normal mixture estimation and model-based
  clustering.
\newblock Technical Report 486, Department of Statistics, University of
  Washington.

\bibitem[Fraley and Raftery, 2007]{fraley:2007}
Fraley, C. and Raftery, A.~E. (2007).
\newblock Bayesian regularization for normal mixture estimation and model-based
  clustering.
\newblock {\em Journal of Classification}, 24(2):155--181.

\bibitem[Friedman et~al., 2008]{friedman:etal:2008}
Friedman, J., Hastie, T., and Tibshirani, R. (2008).
\newblock Sparse inverse covariance estimation with the graphical lasso.
\newblock {\em Biostatistics}, 9(3):432--441.

\bibitem[Friedman, 1997]{friedman:1997}
Friedman, N. (1997).
\newblock Learning belief networks in the presence of missing values and hidden
  variables.
\newblock In Fisher, D., editor, {\em Proceedings of the Fourteenth
  International Conference on Machine Learning}, pages 125--133. Morgan
  Kaufmann.

\bibitem[Friedman, 1998]{friedman:1998}
Friedman, N. (1998).
\newblock The {B}ayesian structural {EM} algorithm.
\newblock In {\em Proceedings of the Fourteenth Conference on Uncertainty in
  Artificial Intelligence}, pages 129--138. Morgan Kaufmann.

\bibitem[Fr{\"u}hwirth-Schnatter, 2006]{fruhwirth:2006}
Fr{\"u}hwirth-Schnatter, S. (2006).
\newblock {\em Finite Mixture and Markov Switching Models}.
\newblock Springer Science \& Business Media.

\bibitem[Galimberti et~al., 2017]{galimberti:2017}
Galimberti, G., Manisi, A., and Soffritti, G. (2017).
\newblock Modelling the role of variables in model-based cluster analysis.
\newblock {\em Statistics and Computing}.

\bibitem[Galimberti and Soffritti, 2013]{galimberti:2013}
Galimberti, G. and Soffritti, G. (2013).
\newblock Using conditional independence for parsimonious model-based
  {G}aussian clustering.
\newblock {\em Statistics and Computing}, 23(5):625--638.

\bibitem[Gao et~al., 2016]{gao:2016}
Gao, C., Zhu, Y., Shen, X., and Pan, W. (2016).
\newblock Estimation of multiple networks in {G}aussian mixture models.
\newblock {\em Electronic Journal of Statistics}, 10(1):1133--1154.

\bibitem[Garber et~al., 2012]{thyro}
Garber, J., Cobin, R., Gharib, H., Hennessey, J., Klein, I., Mechanick, J.,
  Pessah-Pollack, R., Singer, P., and Woeber, K. (2012).
\newblock Clinical practice guidelines for hypothyroidism in adults:
  Cosponsored by the american association of clinical endocrinologists and the
  american thyroid association.
\newblock {\em Endocrine Practice}, 18(6):988--1028.

\bibitem[Goldberg, 1989]{goldberg:1989}
Goldberg, D. (1989).
\newblock {\em Genetic Algorithms in Search, Optimization, and Machine
  Learning}.
\newblock Addison-Wesley Publishing Company.

\bibitem[Green, 1990]{green:1990}
Green, P.~J. (1990).
\newblock On use of the {EM} for penalized likelihood estimation.
\newblock {\em Journal of the Royal Statistical Society. Series B
  (Methodological)}, pages 443--452.

\bibitem[Greenhalgh and Marshall, 2000]{greenhalgh:2000}
Greenhalgh, D. and Marshall, S. (2000).
\newblock Convergence criteria for genetic algorithms.
\newblock {\em SIAM Journal on Computing}, 30(1):269--282.

\bibitem[Guo et~al., 2011]{guo:2011}
Guo, J., Levina, E., Michailidis, G., and Zhu, J. (2011).
\newblock Joint estimation of multiple graphical models.
\newblock {\em Biometrika}, 98(1):1--15.

\bibitem[Harbertson and Spayd, 2006]{harbertson}
Harbertson, J.~F. and Spayd, S. (2006).
\newblock Measuring phenolics in the winery.
\newblock {\em American Journal of Enology and Viticulture}, 57(3):280--288.

\bibitem[Hoeting et~al., 1999]{hoeting:1999}
Hoeting, J.~A., Madigan, D., Raftery, A.~E., and Volinsky, C.~T. (1999).
\newblock Bayesian model averaging: a tutorial.
\newblock {\em Statistical Science}, 14(4):382--417.

\bibitem[Holland, 1992]{holland:1992}
Holland, J.~H. (1992).
\newblock Genetic algorithms.
\newblock {\em Scientific American}, 267(1):66--72.

\bibitem[Huang et~al., 2006]{huang:etal:2006}
Huang, J.~Z., Liu, N., Pourahmadi, M., and Liu, L. (2006).
\newblock Covariance matrix selection and estimation via penalised normal
  likelihood.
\newblock {\em Biometrika}, 93(1):85--98.

\bibitem[Hubert and Arabie, 1985]{hubert:arabie:1985}
Hubert, L. and Arabie, P. (1985).
\newblock Comparing partitions.
\newblock {\em Journal of Classification}, 2:193--218.

\bibitem[Kauermann, 1996]{kauermann:1996}
Kauermann, G. (1996).
\newblock On a dualization of graphical {G}aussian models.
\newblock {\em Scandinavian Journal of Statistics}, 23(1):105--116.

\bibitem[Koller and Friedman, 2009]{koller:2009}
Koller, D. and Friedman, N. (2009).
\newblock {\em Probabilistic Graphical Models: Principles and Techniques}.
\newblock The MIT Press.

\bibitem[Kriegel et~al., 2017]{kriegel:2017}
Kriegel, H.-P., Schubert, E., and Zimek, A. (2017).
\newblock The (black) art of runtime evaluation: {A}re we comparing algorithms
  or implementations?
\newblock {\em Knowledge and Information Systems}, 52(2):341--378.

\bibitem[Krishnamurthy, 2011]{krishnamurthy:2011}
Krishnamurthy, A. (2011).
\newblock High-dimensional clustering with sparse {G}aussian mixture models.
\newblock {\em Unpublished paper}.

\bibitem[Kumar et~al., 1977]{kumar:1977}
Kumar, M.~S., Safa, A.~M., Deodhar, S.~D., and .P., S.~O. (1977).
\newblock {The relationship of thyroid-stimulating hormone (TSH), thyroxine
  (T4), and triiodothyronine (T3) in primary thyroid failure}.
\newblock {\em American Journal of Clinical Pathology}, 68(6):747--751.

\bibitem[Lee and Xue, 2017]{lee:2017}
Lee, K.~H. and Xue, L. (2017).
\newblock Nonparametric finite mixture of {G}aussian graphical models.
\newblock {\em Technometrics}.

\bibitem[Lotsi and Wit, 2013]{lotsi:2013}
Lotsi, A. and Wit, E. (2013).
\newblock High dimensional sparse {G}aussian graphical mixture model.
\newblock {\em arXiv preprint arXiv:1308.3381}.

\bibitem[Ma and Michailidis, 2016]{ma:2016}
Ma, J. and Michailidis, G. (2016).
\newblock Joint structural estimation of multiple graphical models.
\newblock {\em Journal of Machine Learning Research}, 17(166):1--48.

\bibitem[Madigan and Raftery, 1994]{madigan:1992}
Madigan, D. and Raftery, A.~E. (1994).
\newblock Model selection and accounting for model uncertainty in graphical
  models using {O}ccam's window.
\newblock {\em Journal of the American Statistical Association},
  89(428):1535--1546.

\bibitem[Malsiner-Walli et~al., 2016]{malsiner:etal:2016}
Malsiner-Walli, G., Fr{\"u}hwirth-Schnatter, S., and Gr{\"u}n, B. (2016).
\newblock Model-based clustering based on sparse finite {G}aussian mixtures.
\newblock {\em Statistics and Computing}, 26(1):303--324.

\bibitem[Mart\'{i}nez and Vitria, 2000]{martinez:2000}
Mart\'{i}nez, A.~M. and Vitria, J. (2000).
\newblock Learning mixture models using a genetic version of the {EM}
  algorithm.
\newblock {\em Pattern Recognition Letters}, 21(8):759 -- 769.

\bibitem[McLachlan and Peel, 2000]{mclachlan:peel:2000}
McLachlan, G. and Peel, D. (2000).
\newblock {\em Finite Mixture Models}.
\newblock Wiley.

\bibitem[McLachlan and Rathnayake, 2014]{mcLachlan:rathnayake:2014}
McLachlan, G.~J. and Rathnayake, S. (2014).
\newblock On the number of components in a {G}aussian mixture model.
\newblock {\em Wiley Interdisciplinary Reviews: Data Mining and Knowledge
  Discovery}, 4(5):341--355.

\bibitem[McNicholas and Murphy, 2008]{mcnicholas:murphy:2008}
McNicholas, D.~P. and Murphy, T.~B. (2008).
\newblock Parsimonious {G}aussian mixture models.
\newblock {\em Statistics and Computing}, 18(3):285--296.

\bibitem[McNicholas, 2016]{mcnicholas:2016}
McNicholas, P.~D. (2016).
\newblock Model-based clustering.
\newblock {\em Journal of Classification}, 33(3):331--373.

\bibitem[Miller, 2002]{miller:2002}
Miller, A. (2002).
\newblock {\em Subset Selection in Regression}.
\newblock Chapman \& Hall/CRC.

\bibitem[Mohan et~al., 2012]{mohan:2012}
Mohan, K., Chung, M., Han, S., Witten, D., Lee, S.-i., and Fazel, M. (2012).
\newblock Structured learning of {G}aussian graphical models.
\newblock In Pereira, F., Burges, C. J.~C., Bottou, L., and Weinberger, K.~Q.,
  editors, {\em Advances in Neural Information Processing Systems 25}, pages
  620--628.

\bibitem[Mohan et~al., 2014]{mohan:2014}
Mohan, K., London, P., Fazel, M., Witten, D., and Lee, S.-I. (2014).
\newblock Node-based learning of multiple {G}aussian graphical models.
\newblock {\em The Journal of Machine Learning Research}, 15(1):445--488.

\bibitem[Pan and Shen, 2007]{pan:shen:2007}
Pan, W. and Shen, X. (2007).
\newblock Penalized model-based clustering with application to variable
  selection.
\newblock {\em Journal of Machine Learning Research}, 8:1145--1164.

\bibitem[Pan et~al., 2006]{pan:2006}
Pan, W., Shen, X., Jiang, A., and Hebbel, R.~P. (2006).
\newblock Semi-supervised learning via penalized mixture model with application
  to microarray sample classification.
\newblock {\em Bioinformatics}, 22(19):2388--2395.

\bibitem[Pernkopf and Bouchaffra, 2005]{pernkopf:2005}
Pernkopf, F. and Bouchaffra, D. (2005).
\newblock Genetic-based {EM} algorithm for learning {G}aussian mixture models.
\newblock {\em IEEE Transactions on Pattern Analysis and Machine Intelligence},
  27(8):1344--1348.

\bibitem[Peterson et~al., 2015]{peterson:2015}
Peterson, C., Stingo, F.~C., and Vannucci, M. (2015).
\newblock Bayesian inference of multiple {G}aussian graphical models.
\newblock {\em Journal of the American Statistical Association},
  110(509):159--174.

\bibitem[Poli and Roverato, 1998]{poli:1998}
Poli, I. and Roverato, A. (1998).
\newblock A genetic algorithm for graphical model selection.
\newblock {\em Journal of the Italian Statistical Society}, 7(2):197--208.

\bibitem[Pourahmadi, 2011]{pourahmadi:2011}
Pourahmadi, M. (2011).
\newblock Covariance estimation: The {GLM} and regularization perspectives.
\newblock {\em Statistical Science}, 26(3):369--387.

\bibitem[{R Core Team}, 2017]{R}
{R Core Team} (2017).
\newblock {\em R: A Language and Environment for Statistical Computing}.
\newblock R Foundation for Statistical Computing, Vienna, Austria.

\bibitem[Richardson and Spirtes, 2002]{richardson:spirtes:2002}
Richardson, T. and Spirtes, P. (2002).
\newblock Ancestral graph markov models.
\newblock {\em The Annals of Statistics}, 30(4):962--1030.

\bibitem[Rodr\'{i}guez et~al., 2011]{rodriguez:2011}
Rodr\'{i}guez, A., Lenkoski, A., and Dobra, A. (2011).
\newblock Sparse covariance estimation in heterogeneous samples.
\newblock {\em Electronic Journal of Statistics}, 5:981--1014.

\bibitem[Rothman, 2012]{rothman:2012}
Rothman, A.~J. (2012).
\newblock Positive definite estimators of large covariance matrices.
\newblock {\em Biometrika}, 99(3):733--740.

\bibitem[Roverato, 2002]{roverato:2002}
Roverato, A. (2002).
\newblock Hyper inverse {W}ishart distribution for non-decomposable graphs and
  its application to {B}ayesian inference for {G}aussian graphical models.
\newblock {\em Scandinavian Journal of Statistics}, 29(3):391--411.

\bibitem[Roverato and Paterlini, 2004]{roverato:2004}
Roverato, A. and Paterlini, S. (2004).
\newblock Technological modelling for graphical models: {A}n approach based on
  genetic algorithms.
\newblock {\em Computational Statistics \& Data Analysis}, 47(2):323--337.

\bibitem[Ruan et~al., 2011]{ruan:2011}
Ruan, L., Yuan, M., and Zou, H. (2011).
\newblock Regularized parameter estimation in high-dimensional {G}aussian
  mixture models.
\newblock {\em Neural Computation}, 23(6):1605--1622.

\bibitem[Schwarz, 1978]{schwarz:1978}
Schwarz, G. (1978).
\newblock Estimating the dimension of a model.
\newblock {\em Annals of Statistics}, 6(2):461--464.

\bibitem[Scrucca, 2013]{scrucca:2013}
Scrucca, L. (2013).
\newblock {GA}: {A} package for genetic algorithms in {R}.
\newblock {\em Journal of Statistical Software, Articles}, 53(4):1--37.

\bibitem[Scrucca, 2017]{scrucca:2017}
Scrucca, L. (2017).
\newblock {On Some Extensions to GA Package: Hybrid Optimisation,
  Parallelisation and Islands Evolution}.
\newblock {\em {The R Journal}}, 9(1):187--206.

\bibitem[Scrucca et~al., 2016]{scrucca:etal:2016}
Scrucca, L., Fop, M., Murphy, T.~B., and Raftery, A.~E. (2016).
\newblock mclust 5: {C}lustering, classification and density estimation using
  {G}aussian finite mixture models.
\newblock {\em {The R Journal}}, 8(1):289--317.

\bibitem[Scrucca and Raftery, 2015]{scrucca:init}
Scrucca, L. and Raftery, A.~E. (2015).
\newblock Improved initialisation of model-based clustering using {G}aussian
  hierarchical partitions.
\newblock {\em Advances in Data Analysis and Classification}, 9(4):447--460.

\bibitem[Sharapov and Lapshin, 2006]{sharapov:2006}
Sharapov, R.~R. and Lapshin, A.~V. (2006).
\newblock Convergence of genetic algorithms.
\newblock {\em Pattern Recognition and Image Analysis}, 16(3):392--397.

\bibitem[Shen and Ye, 2002]{shen:2002}
Shen, X. and Ye, J. (2002).
\newblock Adaptive model selection.
\newblock {\em Journal of the American Statistical Association},
  97(457):210--221.

\bibitem[Talluri et~al., 2014]{talluri:2014}
Talluri, R., Baladandayuthapani, V., and Mallick, B.~K. (2014).
\newblock Bayesian sparse graphical models and their mixtures.
\newblock {\em Stat}, 3(1):109--125.

\bibitem[Tan, 2014]{hglasso}
Tan, K.~M. (2014).
\newblock {\em hglasso: Learning graphical models with hubs}.
\newblock R package version 1.2.

\bibitem[Thiesson et~al., 1997]{thiesson::1997}
Thiesson, B., Meek, C., Chickering, D.~M., and Heckerman, D. (1997).
\newblock Learning mixtures of {DAG} models.
\newblock In {\em Proceedings of the Fourteenth conference on Uncertainty in
  artificial intelligence}, pages 504--513.

\bibitem[Titterington et~al., 1985]{titterington:1985}
Titterington, D., Smith, A., and Makov, U. (1985).
\newblock {\em Statistical analysis of finite mixture distributions}.
\newblock Wiley.

\bibitem[Wang, 2015]{wang:2015}
Wang, H. (2015).
\newblock Scaling it up: {S}tochastic search structure learning in graphical
  models.
\newblock {\em Bayesian Analysis}, 10(2):351--377.

\bibitem[Wermuth et~al., 2006]{wermuth:etal:2006}
Wermuth, N., Cox, D., and Marchetti, G.~M. (2006).
\newblock Covariance chains.
\newblock {\em Bernoulli}, 12(5):841--862.

\bibitem[Whittaker, 1990]{whittaker:1990}
Whittaker, J. (1990).
\newblock {\em Graphical Models in Applied Multivariate Statistics}.
\newblock Wiley.

\bibitem[Wiegand, 2010]{wiegand:2010}
Wiegand, R.~E. (2010).
\newblock Performance of using multiple stepwise algorithms for variable
  selection.
\newblock {\em Statistics in Medicine}, 29(15):1647--1659.

\bibitem[Wu, 1983]{wu:1983}
Wu, C. F.~J. (1983).
\newblock On the convergence properties of the {EM} algorithm.
\newblock {\em The Annals of Statistics}, 11(1):95--103.

\bibitem[Xie et~al., 2008]{xie:2008}
Xie, B., Pan, W., and Shen, X. (2008).
\newblock Variable selection in penalized model-based clustering via
  regularization on grouped parameters.
\newblock {\em Biometrics}, 64(3):921--930.

\bibitem[Yuan and Lin, 2007]{yuan:2007}
Yuan, M. and Lin, Y. (2007).
\newblock Model selection and estimation in the {G}aussian graphical model.
\newblock {\em Biometrika}, 94(1):19--35.

\bibitem[Zhou et~al., 2009]{zhou:2009}
Zhou, H., Pan, W., and Shen, X. (2009).
\newblock Penalized model-based clustering with unconstrained covariance
  matrices.
\newblock {\em Electronic Journal of Statistics}, 3:1473--1496.

\bibitem[Zhou et~al., 2011]{zhou:etal:2011}
Zhou, S., Rütimann, P., Xu, M., and Bühlmann, P. (2011).
\newblock High-dimensional covariance estimation based on {G}aussian graphical
  models.
\newblock {\em Journal of Machine Learning Research}, 12:2975--3026.

\bibitem[Zhu et~al., 2014]{zhu:2014}
Zhu, Y., Shen, X., and Pan, W. (2014).
\newblock Structural pursuit over multiple undirected graphs.
\newblock {\em Journal of the American Statistical Association},
  109(508):1683--1696.

\bibitem[Zou et~al., 2007]{zou:2007}
Zou, H., Hastie, T., and Tibshirani, R. (2007).
\newblock On the ``degrees of freedom'' of the lasso.
\newblock {\em The Annals of Statistics}, 35(5):2173--2192.

\end{thebibliography}

\end{document}